\documentclass[11pt]{article}  
\usepackage{graphicx}
\usepackage[a4paper,left=3cm,right=3.cm, bottom=2.cm, top=2.0cm]{geometry}
\usepackage{amssymb}
\usepackage{amsmath}
\usepackage{multirow}
\usepackage{nicefrac}
\usepackage{csquotes}
\usepackage{bm}
\usepackage[margin=1pt,font=small,labelfont=bf]{caption}
\usepackage[dvipsnames]{xcolor}
\definecolor{citecolor}{RGB}{128,0,32}
\definecolor{headcolor}{RGB}{128,128,128}
\usepackage[unicode,hyperfootnotes=false,breaklinks=true,colorlinks=true,allcolors=citecolor]{hyperref}
\usepackage{setspace}
\usepackage{fancyhdr}
\usepackage{titlesec}
\usepackage{algorithm}
\usepackage{algpseudocode}
\usepackage{etoolbox}

\usepackage{libertinus}
\usepackage{tabularx}
\usepackage{datetime} 
\usepackage{etoolbox}
\usepackage{subcaption} 
\patchcmd{\linenumber}{\hb@xt@}{\hbox}{}{}
\usepackage[ giveninits=true, style=authoryear, backend=biber, maxcitenames=2, maxbibnames=99,
  hyperref=true, backref=false, sorting=nyt, uniquename=init, autolang=hyphen,
  dashed=false]{biblatex}

\AtEveryCite{\it \color{citecolor}} \AtEveryBibitem{\clearfield{month}}
\AtEveryBibitem{\clearfield{issn}}
\renewbibmacro{in:}{}

\let\citet\textcite
\let\citep\parencite
\graphicspath{{./figs/}}
\DeclareGraphicsExtensions{.jpg}
\newdateformat{usvardate}{\shortmonthname[\THEMONTH]. \THEDAY, \THEYEAR}
\newcommand{\lastmodified}{%
  \textcolor{headcolor}{\scriptsize \usvardate\today{} at \currenttime} }
\fancypagestyle{custom}
{      
\fancyhead[L]{}
\fancyhead[C]{\textcolor{headcolor}{\scriptsize submitted to \textit{Tectonophysics}}}
\fancyhead[R]{}
\fancyfoot[C]{\textcolor{gray}{\thepage}}
\fancyfoot[R]{\lastmodified}
\fancyfoot[L]{}
}
\renewcommand\thesection{\arabic{section}} \titlespacing\section{0pt}{12pt plus 4pt minus 2pt}{0pt
plus 2pt minus 2pt}

\begin{document}

\pagestyle{custom}
\begin{center}
\LARGE {\bf 3-D thermomechanical geodynamic modelling of heating and volcanism in the Massif Central (France) and Eifel Volcanic Region (Germany)\\[12pt]}

\normalsize
Ekeabino Momoh$^{1,2,\P}$,
Stephen Tait$^{1,2}$,
Harsha S. Bhat$^{3}$
\\[12pt]

\begin{enumerate}
	\scriptsize
	\setlength\itemsep{-5pt}
  	\item {Geosciences Environnement Toulouse - GET, Universit\'{e} Toulouse III - Paul Sabatier,  CNRS UMR 5563, {31400} Toulouse, France}
    \item {Institut de Physique du Globe de Paris, UMR 7154 Universite de Paris {75005} Paris, France}
    \item {Laboratoire de Géologie, École Normale Supérieure, CNRS UMR 8538, PSL Research University, 75005 Paris, France}    
\end{enumerate}

\let\thefootnote\relax\footnotetext{$\P$ Corresponding author: \texttt{ekemomoh@gmail.com}}
\end{center}

\section*{Abstract}
\small
The French Massif Central and the Eifel Volcanic Field represent two of the enigmatic features in France and Germany, respectively. Both areas have been affected by Cenozoic volcanism whose origin is still debated.  Using 3-D numerical geodynamic modelling of recently developed solid-mechanical constitutive laws including viscoelastic-viscoplastic behaviour, we explored a new hypothesis to highlight the contribution of localised deformational (inelastic) heating towards the evolution of localised volcanism as being due to far-field-induced compressive stresses from the Africa-Eurasia convergence. Our simulations show that variations in crustal thicknesses can help localise inelastic strain near the Moho of relatively thicker crust, which leads to enhanced heating. Localised viscoplastic deformation above the brittle-ductile transition in the crust also formed conjugate deformation bands, resulting in localised changes in the topography which can be compared to present-day topography. The first-order picture that emerges from our simulations shows an association of elevated topography and tectonic structures that localise deformation, with a geothermal anomaly sufficient to explain the elevated heat flow and cause a potential localised zone for partial melting in hydrated or CO\textsubscript{2}-bearing mantle below the Moho.

\section{Introduction, Geological Context and Problem Statement}
\label{introduction}
\subsection{The French Massif Central, France}
\begin{figure}[t!]
\centering{\includegraphics[width=0.8\textwidth]{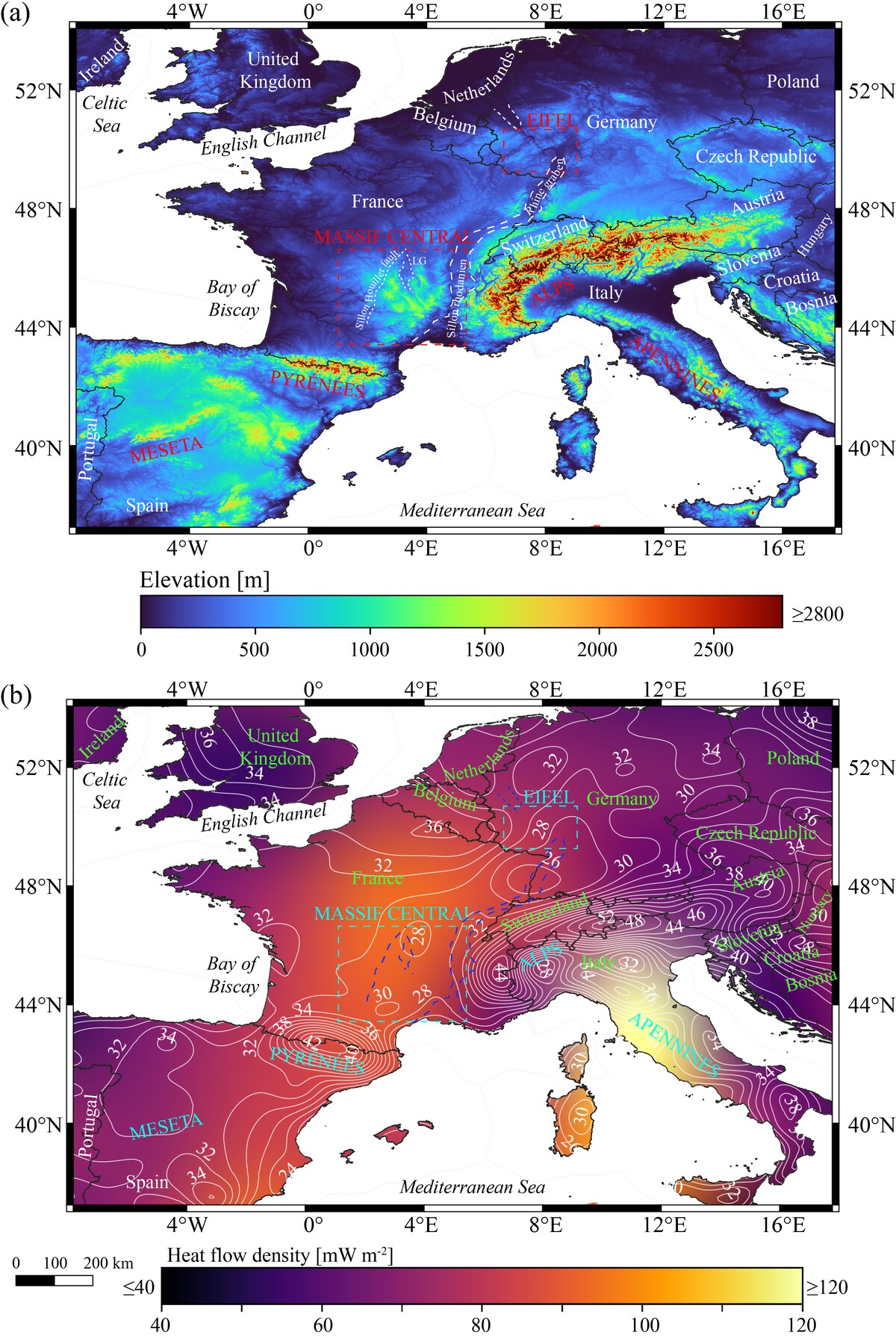}}
\caption{(a) Digital elevation map for part of Europe encompassing the highlands of Massif Central (red dashed rectangle), Pyrenees, Alps, and Apennines. The white dashed lines outline the Sillon rhodanien, which is the southern tip of the European Cenozoic Rift System (ECRIS) and separates the French Massif Central (FMC) from the Alps. The white dashed lines within the French Massif Central outline the Limagne Graben, also part of the ECRIS; while the other white dashed line outlines the surface expression of the lithospheric-scale Sillon Houiller strike-slip fault, an inheritance from the Variscan. The white dashed line North of the Alps, trending SSW-NNE, is the trace of the Upper Rhine Graben, while the white dashed lines trending NNW-SSE outline the Lower Rhine Graben, with the Eifel Volcanic Fields (shown in red dashed rectangle) forming an intersection between the Cenozoic grabens. (b) Heat flow density map overlain with contours of estimated crustal thicknesses in km. The blue lines represent the trace of the Sillon Houiller fault and the Limagne graben area in the Massif Central and the Rhine graben system in the Eifel area. The digital elevation model data were obtained from \url{https://www.mapsforeurope.org/access-data}. The heat flow data were obtained from \cite{davies2013global}, while the crustal thickness data were obtained from \cite{grad2009moho}. The map shown here has been created using the Free and Open Source QGIS \citep{QGIS_software}.}
\label{momoh_map}
\end{figure}

\begin{figure}[htbp]
\centering{\includegraphics[width=\textwidth]{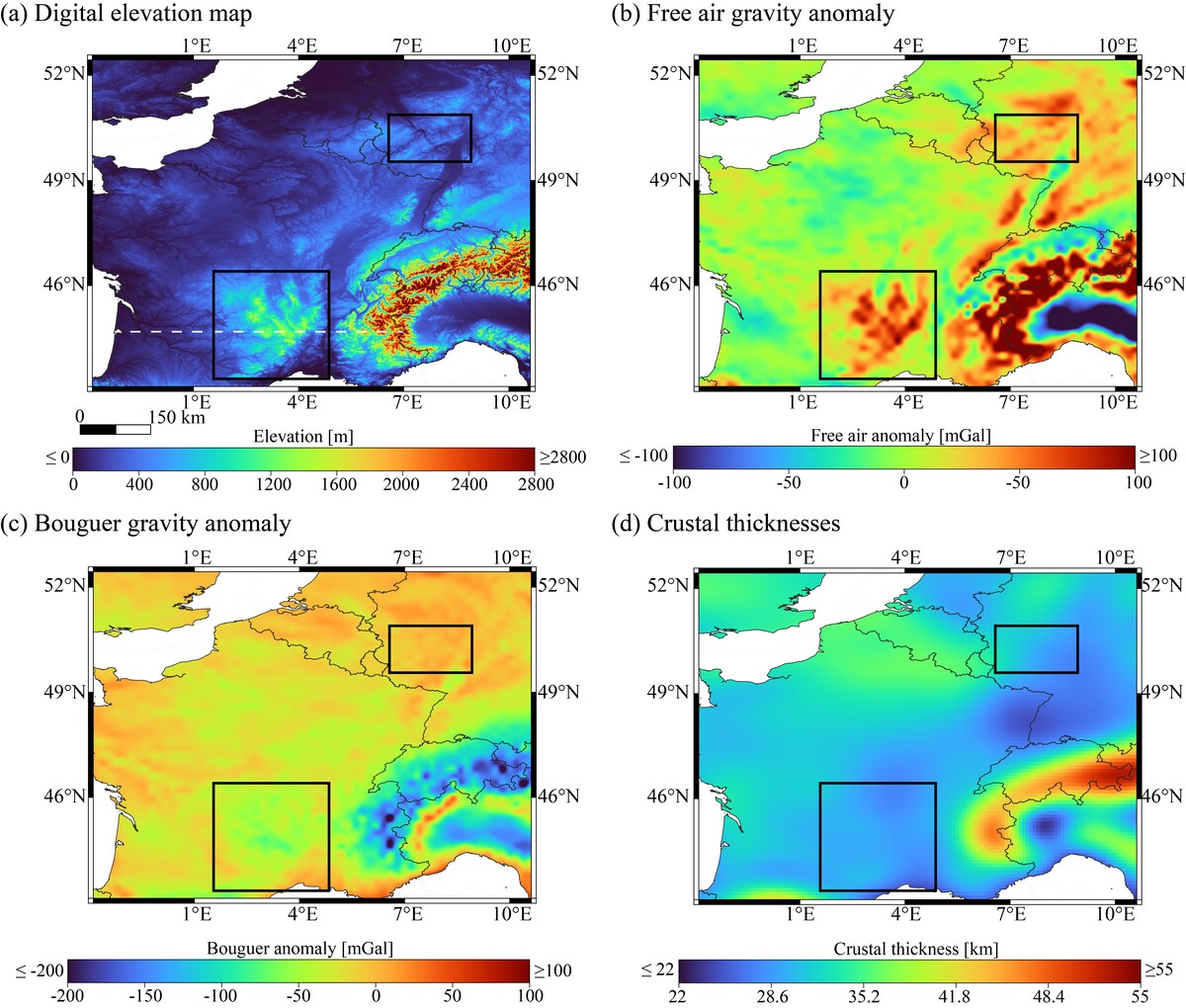}}
\caption{Zoom on the French Massif Central area (black rectangle in the South) and Eifel Volcanic area (black rectangle in the Northeast): (a) Digital elevation map, (b) free-air gravity anomaly, (c) Bouguer gravity anomaly, and (d) crustal thickness. The free air and Bouguer gravity anomaly corrections have been obtained from the International Gravimetric Bureau \citep{poutanen2020geodesist}. The line in (a) shows the path from latitude 44.74\textsuperscript{${\circ}$}N and longitude 1.344.74\textsuperscript{${\circ}$}W to longitude 7\textsuperscript{${\circ}$}E for extracts of elevation, gravity and crustal thickness in \autoref{momoh_map_profiles}.}
\label{momoh_map_gravs_heat_flow}
\end{figure}

The French Massif Central (FMC), in the southeast of France (\autoref{momoh_map} and \autoref{momoh_map_gravs_heat_flow}), represents one of the prominent geological features in Metropolitan France, where outcrops of Variscan and plutonic rocks are not uncommon \citep{faure2009review}. The FMC is a relic of the Variscan or Hercynian Orogeny and forms the central part of the European Hercynian belt \citep{zeyen1997refraction}. The area was an ocean ${\sim}$450 million years ago. The Hercynian orogeny was a mountain-building event that began during the continent-continent collision between Laurussia and Gondwana to form the supercontinent, Pangea, during the mid-Palaeozoic \citep{lefort1981kinematic}. The climax of the Hercynian orogeny was characterised by low-pressure, high-temperature metarmorphism and abundant magmatism \citep{lucazeau1984interpretation}. The Hercynian mountains collapsed, leading to the nucleation of the Sillon Houiller strike-slip fault, which now demarcates the Western non-volcanic (Limousin) region from the Eastern volcanic Massif Central region. The remnants of the Hercynian mountains underwent prolonged erosion to a flat plain, followed by sea-level rise during the Jurassic and Cretaceous.

After a long period of stability during the Mesozoic, the evolutionary history was influenced by the Pyren{\'e}an orogeny from about 80 million years ago and the Alpine orogeny from around 65 million years ago. This was followed by the formation of the European Cenozoic Rift System  (ECRIS), which was nucleated by the reactivation of the pre-existing Variscan fracture network, leading to the formation of numerous grabens and half-grabens. These include the Rhine Graben system, Limagne Graben, Forez Graben, and Sillon rhodanien, which separates the FMC from the Alps (\autoref{momoh_map}). The Limagne Graben is prominent in the Northern segment of the massif \citep{werling1997thermal}, and along with the Rhine Graben, was formed during the large-scale rifting in Europe before the onset of the most recent volcanism \citep{michon2000crustal,dezes2004evolution}. Simultaneously with the graben and half-graben formation, there was an uplift during the Oligocene, with little volcanism. However, from 14 million years ago, volcanic eruptions have interspersed the FMC, forming, among others, a North-South trending Cha\^ine des Puys, and Lac Pavin, representing the youngest volcano (now represented by a crater) formed by explosive volcanism due to the interaction between magma and water (phreato-magmatism) some 6,700 years ago. Continuous uplift of the FMC has been attributed to thermal thinning of the lithosphere due to plumes rising in the upper mantle \citep{dezes2004evolution}.  

\begin{figure}[htbp]
\centering{\includegraphics[width=\textwidth]{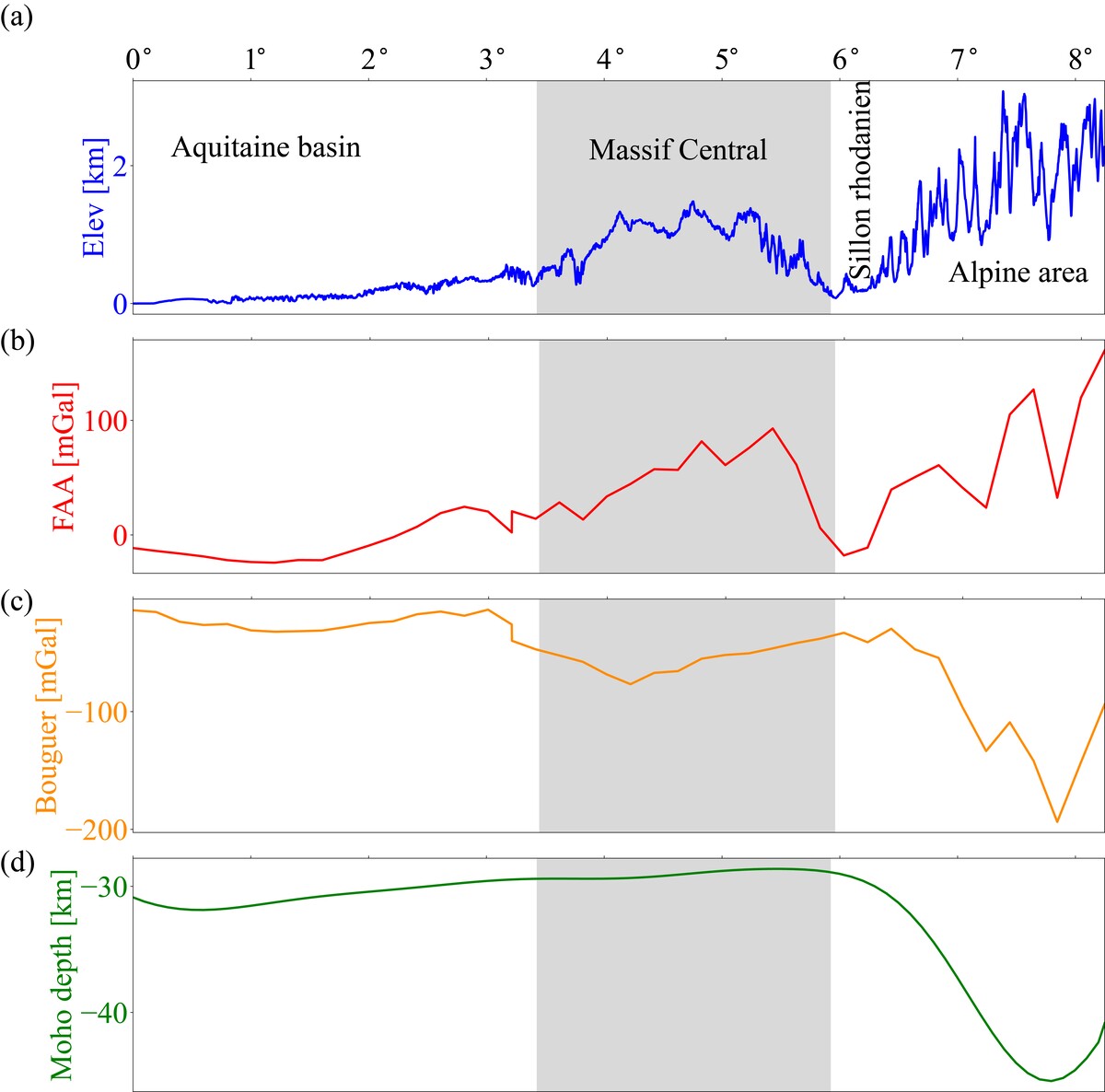}}
\caption{Profiles extracted along latitude 44.74\textsuperscript{${\circ}$}N and longitude 1.3\textsuperscript{${\circ}$}W to longitude 7\textsuperscript{${\circ}$}E  through the central part of the French Massif Central. (a) Elevation, (b) free-air gravity anomaly, (c) Bouguer gravity anomaly, and (d) Moho depth.}
\label{momoh_map_profiles}
\end{figure}

Towards understanding the source of intraplate Cenozoic volcanism in Europe, seismic tomography investigations were carried out in several volcanic provinces, e.g., \cite{granet1995massif,granet1995imaging,ritter2000teleseismic,ritter2001mantle,budweg2006eifel}. From the results of \textit{P}-wave tomographic velocity models derived from teleseismic events of the FMC some thirty years ago, the velocity structure showed a low-velocity anomaly from 70 to 120 km beneath the volcanic areas  \citep{granet1995imaging,sobolev1997temperature,sobolev1997upper} and 100 to 200 km in diameter from 100 km depth \citep{granet1995massif,barth2007crustal}. Wider aperture coverage helped to identify a low-velocity anomaly over a depth range of 50 to 325 km underneath the FMC, separated from a faster Western velocity anomaly by the Sillon Houiller fault \citep{chevrot2014high}. 
Seismic tomography has also revealed variations in crustal thicknesses in the French Massif Central (\autoref{momoh_map}b, \autoref{momoh_map_gravs_heat_flow}d, and \autoref{momoh_map_profiles}d). Beneath some large volcanic edifices like C{\'e}zalier in the FMC, slightly thicker crust has been observed and attributed to cooling magma and heat from associated hot plume mantle material \citep{zeyen1997refraction}, while the volcanism in the Limagne area is associated with thinner crust which has been interpreted as due to entrapment of basaltic magma under the crust \citep{lucazeau1984interpretation}. The lithosphere-asthenosphere boundary shallows to between 70 and 80 km beneath the Limagne graben and 50 to 60 km beneath the volcanic edifices \citep{sobolev1997upper,plomerova2000temporary}.  
The Bouguer gravity anomaly  (\autoref{momoh_map_gravs_heat_flow}c and \autoref{momoh_map_profiles}c), was interpreted as being due to a low-density underlying material from anomalously high temperatures, possibly due to radioactive crustal enrichment and/or tectonic activity, or in the context of convective motions in the upper mantle, giving rise to a low-density underlying material \citep{vasseur1980trend}.
High surface heat flow data (${>}$100 mW m\textsuperscript{-2}) over a distance of 300 km (\autoref{momoh_map}b) had earlier been interpreted in the context of a thermal mantle diapir with a width of 40 km and a nucleating depth of 150 km  \citep{lucazeau1984interpretation,lucazeau1989heat}. The areas of high heat flow (${>\mathrm{80 mW/m^2}}$) coincided with a thin to medium (50 to 90 km) lithosphere (\autoref{momoh_map}, \cite{muller1992regional}). As we will argue here, however, the observation of thin lithosphere, i.e., zones where isotherms dome up towards the shallower levels, is important but may not necessarily imply the action of a mantle plume. A mechanism of in situ heating may also potentially explain the observations.

\cite{michon2001evolution} discussed three distinct magmatic sequences associated with the FMC: (1) Pre-rift magmatism with extremely small volcanic events in the Northern FMC likely resulting from bending of the European lithosphere predating the Alpine orogeny, (2) Syn-rift magmatism during the late Eocene to early Miocene in the Northern FMC associated with crustal thinning in the Limagne Graben, for example; and formation of North-South half-grabens without magmatism in the South of the FMC, and (3) The major post-rift magmatism starting from the South of the FMC around 14 million years ago and North of the FMC 5.5 milllion years ago, unassociated with extensional tectonics, and were said to be concomitant with two periods of uplift. Geochemical models suggest that the origin of the latest eruptions was the base of the continental crust with partial melting less than 10\% \citep{lucazeau1984interpretation}. In nature, uplift can be caused by lithospheric-scale buckling due to compression, or local uplift can be attributed to lithospheric thinning, which may result from rising hot, buoyant material in the underlying ductile mantle \citep{cloetingh1999lithosphere,saunders2007regional}.
The nature of the FMC Cenozoic volcanism reflects fissures or cracks, an indication of tectonic-related volcanism \citep{granet1995massif}. 

The low-velocity anomalies from teleseismic tomographic images, negative Bouguer gravity anomaly, thermal modelling incorporating mineral physics and surface heat flow data have been interpreted as due to the presence of a low-density, hot, thermally and chemically heterogeneous mantle plume during Tertiary times now in a cooling phase \citep{granet1995massif,granet1995imaging,sobolev1997temperature}.  The presence of a hot plume that causes lithospheric thinning as the isotherms are raised towards the surface, as has been suggested, implies a process of thermal conduction and mechanical erosion of the lithosphere, both of which take some time.

The current topographic expression of the FMC is younger than the adjacent Pyrenees and Alps, which terminated collision some 50 million years ago. The FMC is separated from the Alps by the  Sillon rhodanien cleft, which runs from the Northeast of France and terminates on the South, near the Pyrenees. The Sillon rhodanien is connected to the Upper Rhine Graben in the Northeast of France through the Bresse-Rhine strike-slip system (\autoref{momoh_map}). The geologic history of the FMC in the last 50 million years was said to have been influenced by compressional tectonics, possibly from far-field stresses due to the Pyrenees and the Alps or indirectly, from the Africa-Eurasia convergence, which led to regional uplift.  Cenozoic volcanism persisted from 35 million years to about 6 thousand years ago, although with a clear maximum in magmatic flux during the post-rift phase (less than 15 million years ago). The Alpine orogeny, and the collision of the Iberia microplate and the SouthWestern segment of Eurasia (forming the Pyrenees) induced far-field stresses that likely influenced the FMC,  as these were the major tectonic zones in the proximity of the FMC. Most of the dormant volcanic edifices are atop domally uplifted Hercynian basement, with Cenozoic volcanic phases associated with intraplate compressive stress relaxation of the Alpine foreland \citep{wilson1999tertiary}. While the Alpine orogeny and Pyrenean orogeny were consequences of the broader Africa-Eurasia convergence as the stresses therefrom were absorbed by the mountain chains, the impact of the residual stresses from the Africa-Eurasia convergence deserve to be investigated.  

\subsection{Eifel Region, Germany}
The Eifel region in Germany is part of the Hercynian Rheinish Massif found at the junction of the Lower Rhine Graben and Upper Rhine Graben (\autoref{momoh_map} and \autoref{momoh_map_gravs_heat_flow}). Presently, it represents the youngest volcanic region in Central Europe, mostly represented by peneplains \citep{ritter2000teleseismic,guillou2007deciphering}. Similar to the FMC, the Rheinish Massif volcanism formed after the Rhine Graben rifting with major volcanic events concomitant with the updoming of the massif, and volcanism shifting to the Eifel region during the Quaternary \citep{walker2005shear,schmincke2007quaternary}. Uplift can be accomplished by upwelling of hotter mantle material \citep{schmincke1983quaternary}, or by compression. 

The crustal thickness of the Eifel region averages 30 km (27 to 33 km) with a local crustal thinning beneath the Eifel volcanic provinces \citep{budweg2006eifel,seiberlich2013topography}.  Lithospheric thinning to 41${\pm}$5 km was also said to have been observed in the Eifel region, attributed to thermal erosion due to an upwelling mantle plume \citep{seiberlich2013topography}. Volcanic episodes had earlier occurred in the Cretaceous and Tertiary in the East and West Eifel, followed by Quaternary volcanism whose latest episode was about 11,000 years ago \citep{budweg2006eifel,schmincke1983quaternary,schmincke2007quaternary}. Intraplate volcanism had been attributed to a major European hotspot, similar to the FMC. Receiver function analyses of data collected from the Eifel region attributed a low-velocity anomaly with 100 km diameter extending from a depth of 60 to 90 km under West Eifel to the existence of a mantle plume feeding the Eifel volcanism \citep{walker2005shear,keyser20023d,budweg2006eifel}. Similar to the FMC, the Lower Rhine Graben cut through the Eifel region with average heat flow ${\ge\mathrm{80mW m^{-2}}}$ (\autoref{momoh_map}).

The orientation of maximum compressive stresses in Western Europe suggests that the Alps may be a source for intraplate compression, and correlations between switches in the regional stress field and volcanism gave a hint that volcanism may be associated with an intraplate compressive stress state \citep{grunthal1992recent,wilson1999tertiary,walker2005shear}.
However, data are insufficient to reconcile distinct volcanic activity with tectonic phases \citep{schmincke1983quaternary}.

Deep (10 to 40 km) low-frequency (2-8 Hz) events have been reported for beneath Laacher See Volcano in the East Eifel, suggesting an active magmatic system \citep{hensch2019deep}. The low frequency-events originate below the center of what was imaged as a low-velocity gradient layer, and followed a finger-like path upward to the youngest volcanic center in East Eifel with indications of a low-velocity layer within the mantle (30 to 36 km depth) encompassing all volcanic fields within the Eifel region; and within the lower crust (6 to 10 km) interpreted as an upper mantle and lower-crustal magmatic reservoir which can focus partial melt or magmatic fluids presently \citep{dahm2020seismological}. From this observation of two low-velocity anomalies both in the lower crust and upper mantle and a channel of deep low-frequency events linking them and lack of observations of low-velocity intervening layer, it may not be unreasonable for the deep low frequency events to be of tectonic origin, for example due to compression in the Eifel region, which can lead to heat production that may even lead to partial melting, as explored here. If this alternative view is admissible, explanations are needed as to the origin and location of the partial melt as interpreted by \cite{dahm2020seismological}. Analyses of Global Positioning Satellite (GPS) data in the Eifel region suggested a maximum uplift of 1 mm/year with a compressive halo defined by radial shortening, without a similar uplift in the FMC, even though radial shortening was reported \citep{kreemer2020geodetic}. The presence of an uplift, along with horizontal extension at the apex and radial shortening surrounding it; and adjacent seismically active zones, led previous workers to suggest that the source could be a buoyant mantle plume  \citep{kreemer2020geodetic,dahm2020seismological,silverii2023lithospheric}.  

Recently, seismic reflection reprocessing of legacy seismic data (acquired in 1987) on the Western Eifel region indicated a cluster of intracrustal bright sub-parallel seismic reflectors from 14.5 km to 17.5 km; reducing in strength and extending to the Moho (${\sim}$31 km) which were attributed to elongated melt sills or fluid or melt from the mantle in the context of a magmatic underplating model \citep{eickhoff2024seismic,eickhoff2025seismic}. This suite of reflectors demonstrated a negative polarity switch, which could indicate a reduction in seismic \textit{P}-wave velocity and/or density; proxies for partial melt or magmatic fluids \citep{eickhoff2024seismic}. What was also observed was that some of these reflectors were clustered beneath local fault zones. A basal detachment fault extending over 80 km to a depth of 10 km with surface expressions of thrust fault reflectors, imaged to terminate on the extensive detachment fault, gave additional indication of compressional tectonics \citep{eickhoff2025seismic}. The presence of additional reflectors was attributed to possible magmatic incursions overprinting former tectonic zones, since wide damage zones may lead to increased porosity and magmatic fluid incursions.

The signature of compressional tectonics is evidenced by the data sources discussed above. Whether a rising mantle plume is responsible for observed uplift and/or compression from far-field stresses due to the Alpine orogeny or Variscan deformation front, and the contribution to the surface morphology and thermal state, is an issue worth investigating. 

\subsection{Current Challenges and Problem Statement}
Heat flow in France and Germany ranges between 45 to 176 ${\mathrm{mW m^{-2}}}$ and 21 to 172 ${\mathrm{mW m^{-2}}}$, respectively \citep{vcermak1979heat}. While the volcanic flux is insufficient to reconcile the surface heat flow data, a common hypothesis for the FMC and Eifel volcanic Region to explain volcanism and anomalous heat flow is the invocation of a low-velocity columnar body, which has been attributed to a mantle plume with a plume head at ${<}$ 100 km. This series of sub-vertical low-velocity anomalies has been inferred to be \enquote{baby plumes} within the European upper mantle associated with areas of recent volcanism (\cite{granet1995massif,granet1995imaging,ritter2001mantle}. The timing of Cenozoic volcanism is summarised in Table \ref{volcanism}. Other studies proposed even larger lower-mantle plumes at depths of 660 to 2800 km, hypothesised to feed the smaller overlying baby plumes \citep{goes1999lower,dezes2004evolution}. 

{\begin{table*}
\begin{center}
\caption{Historicity of Cenozoic volcanism in the French Massif Central and Eifel Volcanic Area. Adapted from \cite{granet1995massif,granet1995imaging,michon2001evolution}}.
\begin{tabular} {@{}lcc}  
\hline
Event& Period (Myrs) & Areas\\
\hline
\multicolumn{3}{l}{\textbf{French Massif Central}}\\
\hspace{2mm}Pre-rift volcanism& 65 to 35 & Bourgogne, Causses, Forez, Menat,  \\ 
\hspace{2mm}Post-rift volcanism &&\\
\hspace{4mm}First burst & 35 to 23 & Causses, Limagne, Forez\\
&&\\
\hspace{4mm}Second burst & 7.5 to 4 & Cantal, C{\'e}zalier, Coirons, Velay, Aubrac, Escandorgue \\ 
&&\\
\hspace{4mm}Third burst & 3 to 0.3 & Limagne, Mont Dore, Dev{\'e}s, Cha{\^i}ne de Puys\\ 
\hline
\multicolumn{3}{l}{\textbf{Eifel Volcanic Fields}}\\
\hspace{2mm}Quaternary volcanism& 0.7 to 0.011 & West Eifel and East Eifel \\ 
\hline
\label{volcanism}
\end{tabular}
\end{center}
\end{table*}}

Despite its popularity, the baby plume hypothesis has also attracted controversy. For example, elongated vertical low-velocity anomalies were not imaged in some other areas with Tertiary volcanism, like the Bohemian Massif  \citep{plomerova2016cenozoic}. Additionally, the existence of such a baby plume would allude to a hot spot in the FMC and Eifel, due to the vertical upwelling and horizontal plume ponding as observed in other areas with mantle plumes, like the Reunion Islands  \citep{dongmo2023imaging}. While the volcanic flux in the FMC and Eifel is not sufficiently high to reconcile with the surface heat flow, the surface heat flow is not sufficiently high to be related to the plume concept as observed in other areas. Continental flood basalts, as would be expected for the early stages of a mantle plume \citep{richards1989flood,hill1993mantle,white1995mantle}, for example, are also absent. Regional studies using full waveform inversion of body waves and surface waves (incorporating both amplitude and phase information of recorded arrivals) from three-component sensors showed a broad low-velocity zone from 80 km to 250 km depth beneath the Massif Central and Southwards towards Africa, but an absence of sub-vertical low-velocity anomalies beneath the Massif Central that would have lent credence to the mantle plume hypothesis  \citep{fichtner2015crust}. It was suggested that body-wave traveltime tomographic results that had been interpreted as due to mantle plumes, \citep{granet1995massif,goes1999lower,ritter2001mantle}, may be artefacts due to vertical smearing in the upper mantle \citep{fichtner2015crust}. What was observed beneath the French Massif Central by \cite{fichtner2015crust} was a roundish low-velocity anomaly down to 250 km depth, similar to a low \textit{P-}wave velocity anomaly reported by \cite{chevrot2014high}. The consensus thus far is a heterogeneous (possibly hot) mantle beneath the FMC and Eifel, but not conclusively the presence of a mantle plume. 

Towards understanding the structure of the crust and upper mantle, and characterising the possible volcanic systems that may be sources for intermittent Cenozoic volcanism and sources of heat for geothermal exploration, was the motivation for deploying a temporary network of seismic stations in the FMC from 2023 to 2027 \citep{aubert2025maciv}. This underscores the importance of the FMC both for scientific interests and stakeholders in the energy transition conversations. One of the current results from the MACIV data complementing previous records is the discovery of low-frequency earthquakes at the crust-mantle boundary beneath a few of the most recent FMC volcanoes (the Puy de D{\^o}me and Pariou volcanic edifices), which would indicate that the deep magmatic systems may still be active either, speculatively,  as fresh melt is supplied from the mantle or cooling of melt bodies or both \citep{shapiro2025deep}.

Central Europe, of which the FMC is a component, has been undergoing active deformation since the Cretaceous and throughout the Cenozoic \citep{hetenyi2009anomalously}, and it has been suggested that the present crustal architecture is an outcome of recent lithospheric deformation that overprinted the earlier Caledonian and Variscan crustal domains \citep{ziegler2006crustal}. Geological studies of the Alpine fold belt suggested that there was a change from transform motion to convergence in the Western Mediterranean, which subjected the FMC to compression and uplift leading to shear-induced melting, melt ponding, and migration through fault conduits \citep{chesworth1975mantle}. Despite the active tectonic processes, little attention has been paid to additional sources of heat production due to long-term tectonic deformation. Areas of tectonic deformation, especially fault junctions, have indicated high heat flow, which is usually inexplicable by volcanic flux alone, especially when volcanism is sparse. 

Our contribution here is to use 3-D numerical geodynamic modelling \citep{momoh2025volumetric} to investigate possible conditions whereby localisation of deformation can lead to localised heating and thus contribute to high heat flow in the FMC and Eifel region. Another preliminary study addressing volcanism in Tibet and Anatolia looked at whether heat flow from deformational work could contribute to mantle and crustal melting under different considerations, where deformation at incompatible junctions, like fault intersections, is distributed off the fault to surrounding regions where the stress is high enough and conductive heat loss is minimal, leading to temperatures that may overcome the solidus \citep{deves2014strain}. Here, we aim at a more complete description of the energy dissipation processes by including a full constitutive update of the solid materials at each time step as intensive parameters evolve. We organize our presentation as follows: we give an overview of our constitutive laws, present a simple 3-D example for crustal dynamics to demonstrate the possibility of heat production due to deformation; thereafter, we present the results of 3-D simulations of the FMC and Eifel regions, we then evaluate the amount of partial melt that can be possible due to temperature rise from deformational heating, and discuss the outcomes of our investigations.
\section{Methodology}
\label{methodology}
\subsection{Thermomechanical conservation laws}
We consider a viscoelastic-viscoplastic rheology assuming a non-associative Drucker-Prager plastic yield criterion and infinitesimal strain theory for a deformable solid. We assume heat production is solely from mechanical deformations, accounting for a combination of shear heating and dilatant heating,  whereby the latter arises from dilatant plasticity. The full description of our conservation laws, constitutive updates, and consistent tangent modulus for the adaptive time-stepping scheme adopted here has been given in \cite{momoh2025volumetric}. These updates are implemented in the ABAQUS finite element solver. We only state the equations we have utilised here and extensions from our previous work. The conservation of linear momentum is given by:
\begin{equation}
\dfrac{\partial{{\sigma}_{ij}}}{\partial{x}_j} +f_i=\rho\dfrac{\partial^2{u_i}}{\partial{t}^2}.
\label{conservation_equations}
\end{equation}
\noindent${\sigma_{ij}}$ represents components of the Cauchy stress tensor, ${f_i}={\rho g}$ represents body forces (here represented by the weight); ${\rho}$ is density, ${g}$ is gravitational acceleration, ${x_j}$  represents spatial variables in the Cartesian coordinates; ${{u_i}}$ are the components of displacement. In our approach, we account for small accelerations in the system. The mass balance in Lagrangian framework is given as:

\begin{equation} 
\rho_\mathrm{R}=\mathrm{det}\left(\delta_{ij}+\dfrac{\partial u_i}{\partial x_j}\right)\rho,
\label{eqn:massb}
\end{equation} 

where ${\rho_\mathrm{R}}$ and ${\rho}$ represent the mass densities in undeformed and deformed configurations, respectively. The determinant of the Jacobian maps the density from the undeformed to the deformed configuration. The heat generation and conductive transport are incorporated in the energy conservation equation as:

\begin{equation} 
\dfrac{\partial{T}}{\partial{t}}=\nabla\cdot(\alpha_\textrm{th}{\nabla T})+\beta\dfrac{{\sigma_{ij}}\left(\dot\varepsilon^\textrm{v}_{ij}+\dot\varepsilon^\textrm{vp}_{ij}\right)}{\rho C_\mathrm{p}} + H_\mathrm{crust}-\dfrac{C_\mathrm{L}}{C_\mathrm{p}}\dfrac{\partial}{\partial t}F(p,T),
\label{eqn:thermal}
\end{equation} 
where ${{C_\mathrm{P}\;(\mathrm{J\;kg^{-1}\; K^{-1}})}}$ is the specific heat capacity and ${\alpha_\textrm{th}\;\mathrm{(m^{2}s^{-1})}}$ is the thermal diffusivity, i.e., thermal conductivity divided by the product of density and specific heat capacity; (${\lambda_\mathrm{th}/\rho C_\mathrm{P}}$). ${{{\dot\varepsilon}}^{\textrm{v}}_{ij}}$ and ${{{\dot\varepsilon}}^{\textrm{vp}}_{ij}}$ represent the deformational contribution to strain rate from creep and viscoplastic deformations, $\sigma_{ij}\left(\dot{\varepsilon}^\textrm{v}_{ij}+\dot{\varepsilon}^\textrm{vp}_{ij}\right)$ is the contribution to heating from irreversible deformational work (heat generation rate in ${ \mathrm{W\;m^{-3}}}$), with ${\beta}$, the Taylor-Quinney coefficient which quantifies the proportion of deformational work which is converted to heat.  ${H_\mathrm{crust}}$ represents radiogenic heat production applied for the crust and weak zones. ${{C\mathrm{_L\;(kJ\;kg^{-1})}}}$ is the latent heat where the ${F (p, T)}$ represents the melt fraction, which is a function of temperature and pressure, only activated when partial melt is generated (discussed later in this section). When the partial melt fraction increases, the term absorbs heat into the melt, while heat is released during crystallisation. Thermal \enquote{steady-state} occurs when the heat production terms in Equation \ref{eqn:thermal} are balanced by the diffusion term. The dominant heat-production term is deformational heating. The density is allowed to depend on temperature change with respect to a reference temperature, e.g., \cite{gerya2004thermomechanical,afonso2005thermal,thielmann2012shear,bessat2020stress}. Dissipative heating (shear heating) transforms inelastic deformation to heat, and its impact on lithospheric-scale localisation has been studied for subduction initiation \citep{thielmann2012shear}. Other studies have also demonstrated the conversion of inelastic work to heat at crustal and lithospheric scales \citep{regenauer1998rapid,leloup1999shear,jaquet2018spontaneous}. \cite{momoh2025volumetric} explored the role of plastic compressibility in the heat budget and showed that it can be important on lithospheric-scale deformation. While viscous shear heating has been considered for the Lepontine Dome in the Central Alps and argued to lead to equal or exceed heat generation due to radiogenic heat production thereby making it an important contributor to the energy budget \citep{burg2005role}, inelastic dissipative heating of the upper mantle causing a thermal anomaly and uplift has not, to our knowlegde,  been quantitatively considered for the Massif Central and Eifel region. 
\subsection{Deformational constitutive laws}
Our formulation relies on a small strain assumption at each iterative step of the numerical calculation \citep{lemaitre1994mechanics,bower2009applied,de2011computational,de2012nonlinear}. We compute the stress evolution, creep and plastic flow relying on an additive decomposition of the total strain rate tensor into elastic, viscous, and viscoplastic strain rates, represented, respectively, by superscripts \enquote{$\textrm{e}$}, \enquote{$\textrm{v}$} and \enquote{${\textrm{vp}}$} below:
\begin{equation}
{\dot\varepsilon}_{ij} = {\dot\varepsilon}^\textrm{e}_{ij}+{\dot\varepsilon}^\textrm{v}_{ij}+{\dot\varepsilon}^\textrm{vp}_{ij},
\label{rateform}
\end{equation}
\begin{equation}
{\dot\varepsilon}_{ij} = {\dot\varepsilon}^\textrm{e}_{ij}+{\dot\gamma^\textrm{v}}\dfrac{\partial\Phi^\textrm{v}_\textrm{F}}{\partial s_{ij}}+{\dot\gamma^\textrm{vp}}\dfrac{\partial\Phi^\textrm{vp}_\textrm{F}}{\partial\sigma_{ij}}.
\label{decomposition0}
\end{equation}
${\Phi^\textrm{v}_\textrm{F}}$ and ${\Phi^\textrm{vp}_\textrm{F}}$ are viscous (creep) and viscoplastic flow potentials, respectively, and ${{\dot\gamma}^\textrm{v}}$ and ${{\dot\gamma}^\textrm{vp}}$ are viscous and viscoplastic multipliers in rate form; ${s_{ij}}$ represent the deviatoric stress tensors with ${{s_{ij}=\sigma_{ij}}-\sigma_{kk}/3}$. The quantities coupled with the viscous and viscoplastic multipliers in Equation \ref{decomposition0} above indicate flow directions for viscous creep and viscoplastic flow, respectively.
\subsubsection{Elastic rheology}
For the elastic rheology, we consider an isotropic solid characterised by its Young's modulus, $E$, and Poisson's ratio, $\nu$, incorporated into an elastic stiffness tensor, ${C_{ijkl}^\textrm{e}}$. The elastic stress state is estimated from the conventional Hooke's law relating stress and strain via the stiffness tensor: ${{\sigma}_{ij}^\textrm{e}=C_{ijkl}^\textrm{e}({\varepsilon}_{kl}^\textrm{e}-\alpha_\mathrm{ex}\Delta T\delta_{kl}),}$, where we have also accounted for thermoelastic strains \citep{turcotte2002geodynamics,bower2009applied,gere2009mechanics,jacquey2020multiphysics}. 
${\alpha_\mathrm{ex}}$ represents the coefficient of thermal expansion.
\subsubsection{Creep rheology}
For the creep rheology, we defined a creep potential function, which is essentially a non-zero elastic stress state. For computing creep strain rates, we proceed by defining:
\begin{equation}
\dot{\varepsilon}_{ij}^\textrm{v} = \dot{\gamma^\textrm{v}}({\sigma_{ij}^\textrm{e}},T)\dfrac{\partial{\Phi_\textrm{F}^\textrm{v}}}{\partial{s}_{ij}^\textrm{e}},
\label{viscousflow}
\end{equation}
with ${\dot{\gamma^\textrm{v}}}$, incorporating a non-negative creep multiplier describing a magnitude of creep flow in rate form and ${{\partial{\Phi_\textrm{F}^\textrm{v}}}/{\partial{s}_{ij}^\textrm{e}}}$ specifying the direction of creep deformation taking account of only deviatoric stresses. $\Phi_\textrm{F}^\textrm{v} = \sqrt{J_\textrm{II}^\textrm{e}}$ with ${{J_\textrm{II}^\textrm{e}}}$ representing the invariant of the deviatoric elastic stress tensor, is the viscous flow potential whose direction is given by: 
\begin{equation}
 \dfrac{\partial{\Phi_\textrm{F}^\textrm{v}}}{\partial{s}_{ij}^\textrm{e}}=\dfrac{{s}_{ij}^\textrm{e}}{2\sqrt{J_\textrm{II}^\textrm{e}(s_{ij}^\textrm{e})}}.
\end{equation}
For computing the amplitude of creep flow, we utilised the power-law flow form:
\begin{equation}
{\dot{\gamma}^\textrm{v}({\sigma_{ij}^\textrm{e}},T)=\left({\Phi^\textrm{v}}\right)^mf(T).}
\label{functional_form}
\end{equation}
Where $m$ represents the power law exponent. For the temperature dependence, we utilised the functional form for dislocation creep: 
\begin{equation}
f(T)=A{\textrm e}^{-\frac{E_a}{RT}}.
\label{disloc}
\end{equation}
$E_a$, $R$, $T$, and $A$ are the activation energy, molecular gas constant,  absolute temperature (in Kelvin, K), and pre-exponential factor, respectively. During creep deformation, we defined a residual creep potential function, which is a non-linear equation whose root is to be determined. For computational purposes, we decompose the quantities in rate form to incremental form as we solve the non-linear equations within a time step. For solving the non-linear equation, we used Newton-Raphson iterations, which is a useful numerical scheme for finding the roots of non-linear equations:
\begin{equation}
{\tilde\Phi^\textrm{v}}{({\sigma_{ij}^\textrm{e}},T)=\left({\Phi^\textrm{v}}\right)^mf(T)-\dfrac{\Delta{\gamma}^\textrm{v}}{\Delta{t}}}=0,
\label{viscous_functional_form}
\end{equation}
where ${\Delta{t}}$ is the time increment at a step, introduced to switch from rate form to incremental form for computational purposes; ${{\Phi^\textrm{v}}=\sqrt{J_\textrm{II}^\textrm{e}(s_{ij}^\textrm{e, trial})}-G\Delta\gamma^\textrm{v}}$. In finding the root of Equation \ref{viscous_functional_form}, we require the derivative of ${{\tilde\Phi^\textrm{v}}}$ with respect to ${\Delta\gamma^\textrm{v}}$:
\begin{equation}
\dfrac{{\textrm{d}\tilde\Phi^\textrm{v}}}{{\textrm{d}\Delta\gamma^\textrm{v}}}={=-mG\left({\Phi^\textrm{v}}\right)^{m-1}f(T)\Delta{t}-1}=0.
\label{viscous_functional_form_NR}
\end{equation} 

Once the creep multiplier is obtained, we update stress components as follows:
\begin{equation}
s_{ij}^\textrm{v}=\left(1-\dfrac{G\Delta\gamma^\textrm{v}}{\sqrt{J_\textrm{II}}}\right)s_{ij}^\textrm{e}.
\end{equation}
The correction to the deviatoric stresses due to creep flow can be interpreted as a reduction of the elastic trial state. Since the major control for creep flow is the temperature, the correction to creep flow becomes more dominant with increasing temperature. Since we assume that ductile creep is incompressible flow, we do not update the pressure.

The creep strain increments are updated as follows:
\begin{equation}
\varepsilon_{ij}^\textrm{v}=\Delta\gamma^\textrm{v}\dfrac{s_{ij}^\textrm{e}}{\sqrt{J_\textrm{II}\left({s_{ij}^\textrm{e}}\right)}}.
\end{equation}
\subsubsection{Viscoplastic rheology}
After implementing a correction to the elastic stress state, we subject the (creep-updated) stress state to a test of plasticity. Depending on the combination of stress states, plastic yielding is defined by a plastic potential function, described here by the pressure-sensitive Drucker-Prager plastic yield criterion given by \citep{drucker1952soil}:
\begin{equation} 
\Phi_{Y}({\sigma}_{ij}, c) = \sqrt{J_\textrm{II}^\textrm{v}}+\alpha_1P-\alpha_2c \ge 0.
\end{equation}
Where $J_\textrm{II}^\textrm{v}=s_{ij}^\textrm{v}s_{ij}^\textrm{v}/2$ represents the second invariant of the post-creep deviatoric stress tensor describing a magnitude of deviatoric stresses; ${c}$ is the cohesion which may depend on the viscoplastic strain history (to be discussed later in this section); ${\alpha_1}$ and ${\alpha_2}$ are material-dependent constants which are given functions of the internal friction angle (${\varphi}$) as follows:
\begin{equation} 
\alpha_1=\dfrac{6 \sin{\varphi}}{\sqrt{3}(3-\sin\varphi)}, \alpha_2=\dfrac{6\cos{\varphi}}{\sqrt{3}(3-\sin\varphi)}.
\end{equation}
As we demonstrated in the case of creep flow, we define a residual non-linear plastic function including an amplitude and flow direction:
\begin{equation}
\dot{\varepsilon}_{ij}^\textrm{vp} = \dot{\gamma}({\sigma_{ij}},T)\dfrac{\partial{\Phi_{F}}}{\partial{\sigma}_{ij}}
\label{viscoflow}
\end{equation}
$\Phi_{F} = \sqrt{J_\textrm{II}} + \alpha_3 P$ is the viscoplastic flow potential and ${\alpha_3=6\sin{\psi}/ (\sqrt{3}(3-sin\;{\psi}))}$, ${\psi}$ is the dilatancy angle which may be equal or less than the friction angle \citep{vermeer1984non}. 
\begin{equation}
 \dfrac{\partial{\Phi_{F}}}{\partial{\sigma}_{ij}}=\dfrac{{s}_{ij}}{2\sqrt{J_\textrm{II}}}+ \dfrac{\alpha_3}{3}\delta_{ij}.
\end{equation}
The equation above indicates that there is a deviatoric and volumetric (dilatant) contribution to the direction of viscoplastic flow. Non-zero dilatancy removes the assumption of an incompressible material, i.e., one for which pressure changes due to viscoplastic deformation do not result in a net volume change in the material \citep{turcotte2002geodynamics}. The presence of volumetric inelastic strain thus provides a direct and useful way of visualising the lithosphere in the results.

In solving for the viscoplastic multiplier, we utilise a Newton-Raphson iteration for finding the roots of a nonlinear residual viscoplastic equation. We define the residual viscoplastic flow functions as follows:
\begin{equation}
\tilde{\Phi}^\textrm{vp}_\textrm{R}{({\sigma_{ij}^\textrm{v}},T)=\dfrac{\Delta{t}}{\mu}\left({\dfrac{\Phi_\textrm{R}^\textrm{DP}}{\Phi_{0}}}\right)^m-\Delta{\gamma}^\textrm{vp}}=0.
\label{residual_functionals_vp}
\end{equation}
Where ${{\Phi_\textrm{R}^\textrm{DP}}= \Phi_\textrm{Y}^\textrm{DP}-c_1\Delta\gamma^\textrm{vp}}$, with ${c_1 = (G+\alpha_1 \alpha_3 K +\alpha_2^\textrm{2} H)}$, where ${H}$ is the hardening modulus of cohesion hardening if it is included (to be discussed later in this section); ${\Phi_\textrm{0}=c_0}$, and ${\mu}$ is a viscosity- or strain-rate related term (we have used the latter here). ${m}$ is the stress sensitivity term (${\ge 1}$). In finding the unknown ${\Delta\gamma^\textrm{vp}}$  by Newton-Raphson iterations as for the creep case, the derivative of the residual viscoplastic potential function required for these iterations is given by:
\begin{equation}
\dfrac{\textrm{d}{\tilde\Phi_\textrm{R}^\textrm{vp}}}{\textrm{d}\Delta\gamma^\textrm{vp}} = -\dfrac{c_{1}m\Delta{t}}{c_0 \mu}\left(\dfrac{\Phi_\textrm{Y}-c_{1}\Delta\gamma^\textrm{vp}}{c_0}\right)^{m-1} - 1
\label{residual_functionals_vp_derive}
\end{equation}
Once the quantity ${\Delta\gamma^\textrm{vp}}$ is obtained, the deformation state due to viscoplastic flow is updated as follows:
\begin{equation}
\Delta{\varepsilon}^\textrm{vp}_{ij}  = \Delta{\gamma^\textrm{vp}}\left(\underbrace{\dfrac{{s}^{\textrm{v}}_{ij}}{2\sqrt{J_\textrm{II}^\textrm{v}({s}^{\textrm{v}}_{ij})}}}_\textrm{deviatoric}+ \underbrace{\dfrac{\alpha_3}{3}\delta_{ij}}_\textrm{volumetric}\right).
\label{vpupdate}
\end{equation}
Equation \ref{vpupdate} contains both deviatoric and volumetric contributions, where the volumetric contributions come from dilatant plastic flow. The deviatoric stresses are updated as:

\begin{equation}
{{{s}_{ij}^\textrm{vp} ={\left(1-\dfrac{G\Delta{\gamma^\textrm{vp}}}{\sqrt{J_\textrm{II}^\textrm{v}({s}^{\textrm{v}}_{ij})}}\right)}{s}_{ij}^{\textrm{v}}}},
\label{deviatoric_update}
\end{equation}
which is a correction of the previous creep-updated deviatoric stress. 

And the pressure is also corrected as a function of dilatant plastic flow:
\begin{equation}
 P=P^{\textrm{e}}-\underbrace{{\alpha_3}\Delta{\gamma^\textrm{vp}}K}_\textrm{volumetric}.
\label{pressure_update}
\end{equation}
Where ${K}$ is the bulk modulus.

The updated stress tensor is given by:

\begin{equation}
\sigma_{ij}=s_{ij}^\textrm{vp}+P,
\label{stress_update_vp}
\end{equation}
\subsubsection{Model for partial melting}
To account for partial melting if the temperature and pressure conditions are such that it is possible in our simulations, we consider a melting model for anhydrous peridotite which follows the power-law melting formulation of \citep{hirschmann2000mantle,katz2003new}, given by:
\begin{equation}
F(p,T) = \left(\dfrac{T-T_{\textrm{solidus}}}{T_{\textrm{liquidus}}-T_{\textrm{solidus}}}\right)^{1.5} \; \textrm{for} \; T_{\textrm{solidus}}< T < T_{\textrm{liquidus}}
\end{equation} 
where ${F(p, T)}$ represents the degree of melt as a function of temperature (in ${{\textrm{T}}}$) and pressure (in ${\textrm{GPa}}$). We assume batch melting, i.e., one for which an ascending mantle melts without an instantaneous melt extraction, which therefore does not take account of the melting history in our model or requires a modification of the description of the melted units. The ${T_{\textrm{solidus}}}$ and ${T_{\textrm{liquidus}}}$ follow the form proposed by \citep{hirschmann2000mantle}:
\begin{equation}
\begin{split}
T_{\textrm{solidus}} =A_1P + A_2P + A_3P^2 \\
T_{\textrm{liquidus}} =B_1P + B_2P + B_3P^2,	
\end{split}
\end{equation}
where ${A_1}$, ${A_2}$, and ${A_3}$ = 1085.7${^\circ}$C GPa\textsuperscript{-1}, 132.9${^\circ}$C GPa\textsuperscript{-1} and -5.1${^\circ}$C GPa\textsuperscript{-1}, respectively; and ${B_1}$, ${B_2}$, and ${B_3}$ = 1475${^\circ}$C GPa\textsuperscript{-1}, 80${^\circ}$C GPa\textsuperscript{-1} and -3.2${^\circ}$C GPa\textsuperscript{-1} \citep{hirschmann2000mantle,katz2003new}. The pressure for estimating the melt fraction is given in GPa. Such parameterisation for sublithospheric batch melting for pressures less than 13 GPa has been incorporated in other numerical modelling codes \citep{dannberg2019new,njinju2021lithospheric,njinju2023instantaneous,ball2022coupled}. We acknowledge that a certain thickness of continental mantle \enquote{root} immediately beneath the crust may well contain enough volatiles (for example, present in phases such as pargasitic amphibole, sometimes referred to as metasomatised mantle) that its solidus may be considerably below that of anhydrous peridotite. This is obviously highly relevant to the genesis of small amounts of partial melts and their compositions and should be explored in future work. However, in this first set of calculations, it seems reasonable to avoid this secondary complexity and use the anhydrous case. We revisit this point in the discussion following the presentation of our results. 

\subsubsection{Friction and Cohesion Hardening}
We incorporate two hardening laws in
our constitutive formulation. First is a linear cohesion hardening, which depends on the deformation history, and the second is friction hardening. The cohesion hardening is given by:
\begin{equation}
c=c_0+H\bar\varepsilon^\mathrm{vp}.
\end{equation}
Where ${c}$ is the updated cohesion, ${c_0}$ is the initial cohesion, ${H}$ is the hardening modulus, and ${\bar\varepsilon^\mathrm{vp}}$ is a quantity that records the deformation history during viscoplastic deformation, which can be seen as accumulated viscoplastic strain: ${\dot{\bar\varepsilon}^\mathrm{vp}}=\alpha_2\dot{\gamma}$, or in incremental form, ${\Delta{\bar\varepsilon}^\mathrm{vp}}=\alpha_2\Delta{\gamma}$. The hardening modulus can be interpreted as the slope of the stress-strain curve during plasticity, i.e., an incremental yield surface as a function of the viscoplastic deformation history.
\begin{equation}
c(\bar{\varepsilon}^\textrm{vp})=  c_0 + H\bar{\varepsilon}_{n+1}^\textrm{vp} = c_0 + H(\bar{\varepsilon}_{n}^\textrm{vp} + \Delta{\bar{\varepsilon}^\textrm{vp}})= c_0 + H(\bar{\varepsilon}_{n}^\textrm{vp}+\alpha_2\Delta{\gamma}^\mathrm{vp}).
\end{equation}

For frictional hardening, we use the form of \cite{leroy1989finite,leroy1990finite} which incorporates a monotonically increasing friction angle with the inelastic deformation:
\begin{equation}
sin\;\varphi=sin\;\varphi_\textrm{i}+\dfrac{2\left(sin\; \varphi_\textrm{f}-sin\;\varphi_\textrm{i}\right)\sqrt{\varepsilon_\textrm{eff}^\textrm{vp}\varepsilon_\textrm{crit}^\textrm{vp}}}{\varepsilon_\textrm{eff}^\textrm{vp}+\varepsilon_\textrm{crit}^\textrm{vp}}.
\end{equation}
Where the effective friction angle ${\varphi}$ increases from an initial friction angle ${\varphi_\textrm{i}}$ to a final friction angle ${\varphi_\textrm{f}}$, where the final friction angle is attained at a critical strain  ${\varepsilon_\textrm{crit}^\mathrm{vp}}$. The effective viscoplastic strain, which is a measure of the accumulated viscoplastic deformation, is given by:
\begin{equation}
\varepsilon_\textrm{eff}^\textrm{vp}=\sqrt{\dfrac{2}{3}\varepsilon_{ij}^\textrm{vp}\varepsilon_{ij}^\textrm{vp}}.
\end{equation}
\subsection{Numerical solver}
We solved the conservation laws using Abaqus\textsuperscript{\textcopyright}, which is an engineering-optimised finite element solver built to handle various engineering problems. Although there exists a wide array of mechanical, thermal, and electrical material libraries, the unique scripting interface within  Abaqus\textsuperscript{\textcopyright} allows customisation of any constitutive laws through user subroutines. We transcribed our deformational constitutive laws into a customised Abaqus\textsuperscript{\textcopyright} user material mechanical subroutine to solve for the mechanical deformations and define the heat source, and a customised heat transfer subroutine to solve the thermal balance. To account for variability of material properties due to evolving intensive parameters, we utilised user-defined field subroutines, for example, to handle the density changes due to temperature changes during the calculations.

Abaqus\textsuperscript{\textcopyright} has a suite of element libraries that can be utilised, but for our purposes, involving mechanical and thermal considerations, we have utilised the elements containing both displacement (fundamental variable) and temperature degrees of freedom. For all simulations, we have used a 4-node continuum solid displacement and thermally coupled tetrahedron, with three linear displacement and one temperature degrees of freedom (C3D4T). This element library provides versatility in the mutual exchange of information during the analysis. It allows the heat generation rate to be defined in the mechanical subroutine after stresses and inelastic strain rates have been calculated, and then passed on to the thermal subroutine, which solves the diffusion problem, updates the temperature, which is in turn used by the mechanical subroutine to calculate stresses in the subsequent time step. In computational plasticity, a critical aspect is time stepping. We have utilised an adaptive time-stepping scheme whereby the time step used depends on the instantaneous deformation state. This allows for implementation efficiency and numerical convergence of the global equations.

\section{Crustal Dynamics Example}
\subsection{Model setup}
In a previous study, we carried out 2-D numerical benchmarking experiments to address deformation problems under plane strain conditions \citep{momoh2025volumetric}. The versatility of Abaqus\textsuperscript{\textcopyright} allows the switch from 2-D (plane strain/axisymmetric), plane stress, to full 3-D problems depending on the problem setup, i.e., the number of direct (normal) tensor and shear tensor components, and the geometry.

\begin{figure}[htbp]
\centering{\includegraphics[width=\textwidth]{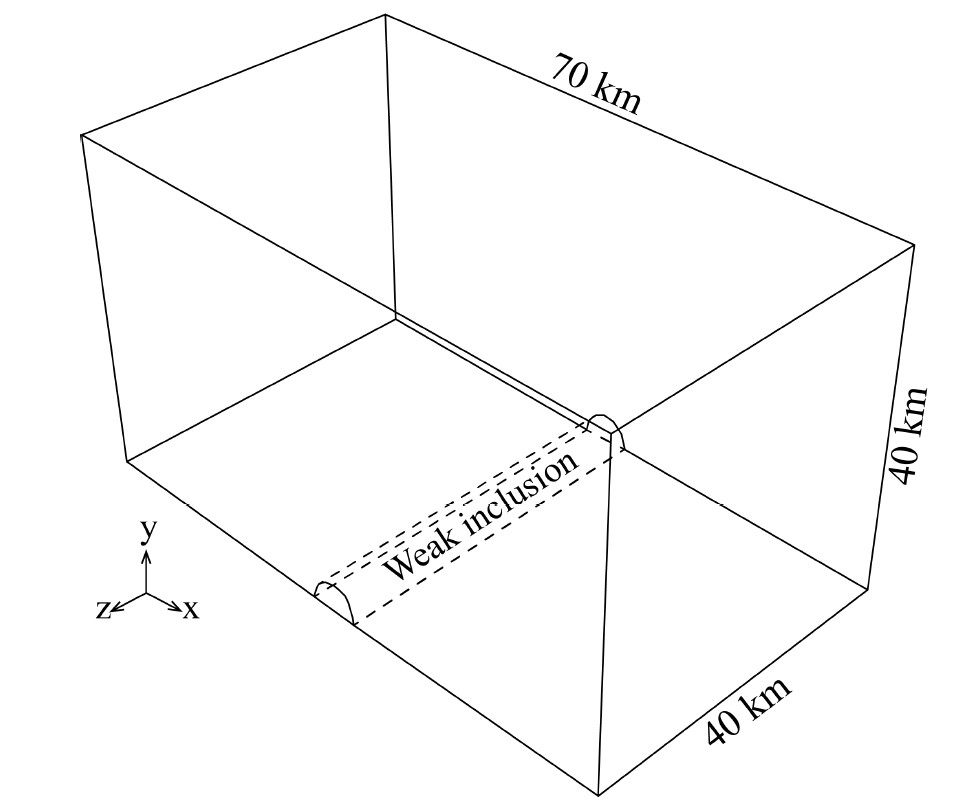}}
\caption{3-D input model characterized by a 70 km ${\times}$ 40 km ${\times}$ 40 km crust in the x, y and z directions, respectively; with a semi-circular weak inclusion at the bottom. The setup is used to investigate visco-elastic and elasto-viscoplastic deformation.}
\label{momoh1}
\end{figure}

To demonstrate heat production within deformation zones in 3-D, we present a simple 3-D test case by extending the 2-D setup already discussed in previous studies \citep{duretz2014physics,momoh2025volumetric}. The geometry is shown in \autoref{momoh1}, composed of a 40 km-thick crust with a weak semi-circular inclusion at the bottom. The rheological properties are shown in Table \ref{duretz2014table1}. As the rheological switch is controlled by either temperature or the plastic yield criterion, we have investigated two rheological cases: (1)  a warm case, with an isothermal initial temperature of 673 K for a viscoelastic rheology, and (2) a cold case with an initial temperature of 473 K for a viscoplastic rheology. In the latter case, once the temperature rises sufficiently to accommodate creep behaviour, the rheology can switch from viscoplastic to viscoelastic, which can lead to stress relaxation that may inhibit plastic yielding. We did not include gravity in these test cases and ran the simulations for 5 ${\times}$ 10\textsuperscript{12} s. For this exercise, we have imposed decreasing compressional velocities on the lateral boundaries and increasing extensional velocities on the top to maintain a constant boundary strain rate of 5 ${\times}$ 10\textsuperscript{-14}s\textsuperscript{-1}.
{{\begin{table*}
\begin{center}
\caption{Rheological parameters for 3-D crustal dynamics example. ${E\;=\;}$Young's modulus, ${\nu\;=\;}$Poisson's ratio, ${E_a\;=\;}$dislocation creep pre-exponential multiplier, ${m\;=\;}$dislocation creep exponent, ${E_a\;=\;}$dislocation creep activation energy, ${\alpha_\textrm{th}\;=\;}$thermal diffusivity, ${C_\mathrm{P}\;=\;}$specific heat capacity, and ${\rho\;=\;}$density. For the cold case (viscoplastic) simulations, we have used additional parameters of 10\textsuperscript{15} s for the inverse of the relative rate of viscoplastic strain (${\mu}$), 1 MPa for the cohesion and 250 MPa for hardening moduli for the two materials; constant friction angles of 25${^\circ}$ for the matrix and 20${^\circ}$ for the weak inclusion, with dilatancy angles of 10${^\circ}$ and 5${^\circ}$ for the matrix and weak inclusions, respectively. The parameters for the creep laws have been taken from \cite{duretz2014physics}, while the additional elasticity and plasticity parameters have been drawn from \cite{momoh2025volumetric}. We used the creep parameters for the matrix and weak inclusions for creep in order to focus on the plastic deformations. These simulations include density variations with temperature.}
{\footnotesize
\begin{tabular}{@{}ccccccccc}
\hline
Material&${E\;}$ (GPa)&${\nu}$& ${A}$ (Pa${^{-m}}$s${^\textrm{-1})}$ & ${m}$ & $E_a\; $ (kJ mol${^\textrm{-1})}$& ${\alpha_{th}}$ (m${^\textrm{2}}$s${^\textrm{-1}}$) & ${C_\mathrm{P}}$ (J kg${^\textrm{-1}}$K${^\textrm{-1}}$) &${\rho}$ (kgm${^\textrm{-3}}$)\\
\hline
Matrix&25&0.25&3.20${\times}$10${^\textrm{-20}}$&3&276&${8.82\times{10^{-7}}}$ &1050&2700\\
Inclusion&25&0.25&3.16${\times}$10${^\textrm{-26}}$&3.3&186&${8.82\times{10^{-7}}}$ &1050&2700\\
\hline
\end{tabular}
}
\label{duretz2014table1}
\end{center}
\end{table*}}}

\begin{figure}[htbp]
\centering{\includegraphics[width=0.75\textwidth]{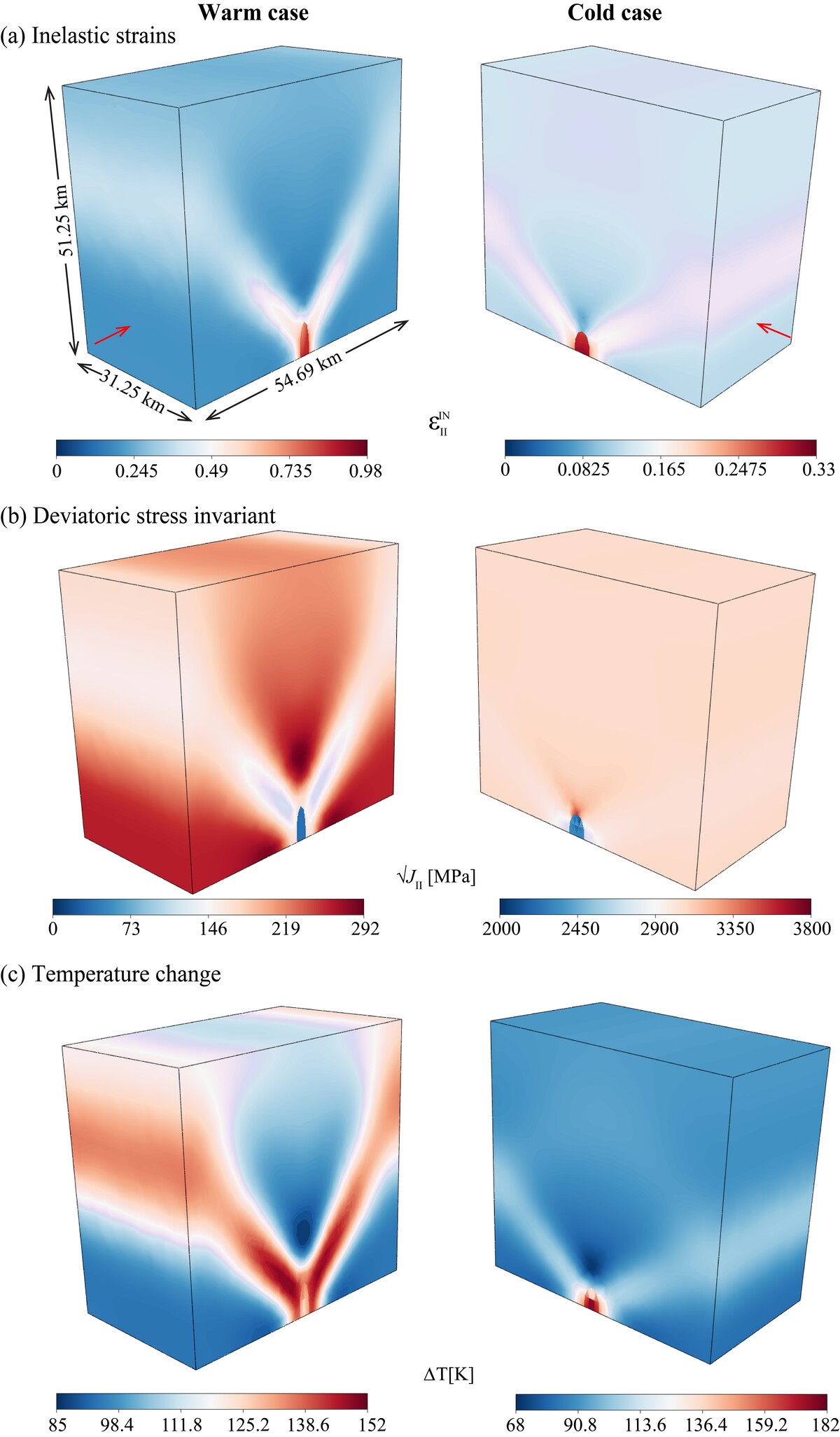}}
\caption{Simulation results of crustal dynamics for warm case, i.e., visco-elastic (left panel) and cold case, i.e., visco-plastic (right panel), respectively. The deformed dimensions are 54.69 km ${\times}$ 51.25 km ${\times}$ 31.25 km in the \textit{x}, \textit{y}, and \textit{z} directions, respectively. The red arrows indicate the starting point for a horizontal path across two deformation bands shown in \autoref{crustal_dynamics_results_1D}.}
\label{crustal_dynamics_results}
\end{figure}

\begin{figure}[htbp]
\centering{\includegraphics[width=\textwidth]{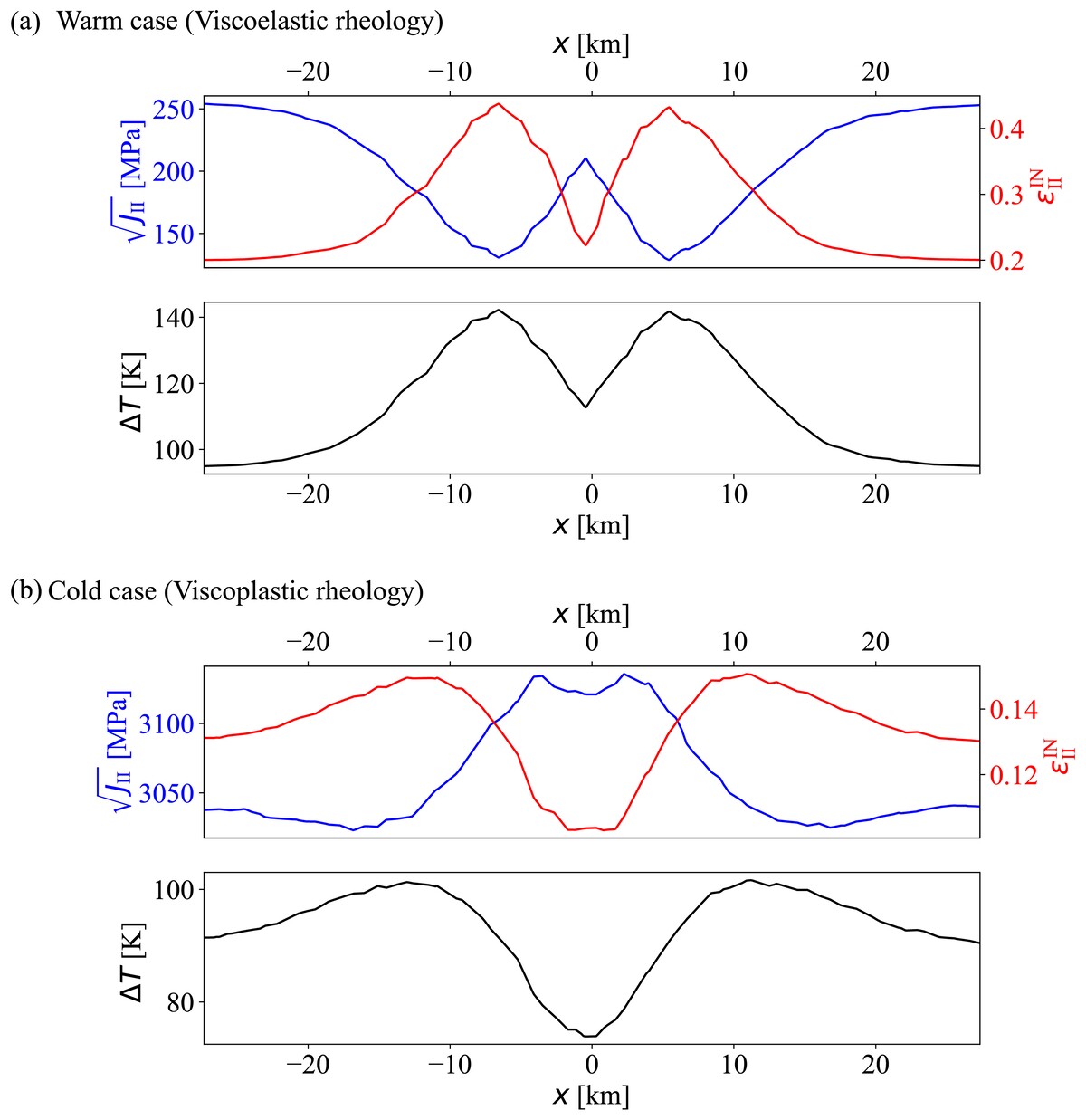}}
\caption{West-East profile across two deformation bands showing (a) inelastic strain in red and second invariant of deviatoric stress in blue and difference between final temperature and initial constant temperature for the warm case; and, (b) inelastic strain in red and second invariant of deviatoric stress in blue and difference between final temperature and initial constant temperature for the cold case.}
\label{crustal_dynamics_results_1D}
\end{figure}
\subsection{3-D Crustal Dynamics Results}
The results of the simulations are shown in \autoref{crustal_dynamics_results}. The inelastic (viscoelastic or visco-plastic) deformations are shown in \autoref{crustal_dynamics_results}a, while the second invariant of the deviatoric stress tensor is shown in \autoref{crustal_dynamics_results}b. The weak inclusion showed the most deformation and the attendant stress drop, despite the differences in the amount of internal deformation accommodated. The deformation bands nucleated within the prescribed weak zone and propagated to the boundaries at different angles within the material (${\sim}$50.5${^\circ}$ and ${\sim}$43${^\circ}$ for the visco-elastic and viscoplastic deformations, respectively). Comparing the deformation and magnitude of the deviatoric stress invariant (which can be interpreted as the magnitude of distortional stress during the loading), it was observed that while the inelastic strain magnitudes were higher for the case of visco-elastic deformation compared to the viscoplastic case, the magnitude of deviatoric stress is one order of magnitude higher in the viscoplastic case. This difference in the deviatoric stress invariant and deformation distribution led to differences in the thermal evolution with the temperature following the deformation pattern (\autoref{crustal_dynamics_results}). The maximum temperature increase is 152 K for the visco-elastic deformation, 104 K within the deformation bands and 180 K within the weak inclusion for the viscoplastic deformation (\autoref{crustal_dynamics_results}b). The maximum temperature increment is observed within the weak zone in the cold case, which can ensure a rheological change and further stress relaxation, via thermally-activated deformation. Shown in \autoref{crustal_dynamics_results_1D} are West-East profiles across two deformation bands showing peak inelastic deformation and a drop in deviatoric stress invariant for both the warm and cold cases. The temperature increase is due to deformational heating within the deformation bands.

These examples demonstrate that heat production due to irreversible deformation, whether creep or viscoplastic, is not insignificant. Zones of relatively high deformation are characterised by stress drops, which would be a self-defeating mechanism in that stress drops lead to reduced heat production, but the high rate of deformation compensates such that the temperature can increase by at least 100 K within a few hundred years. Localisation zones in nature, where faults likely nucleate or flexural regions in a buckled lithosphere with distributed deformation, may therefore be important sources of heat production. The results presented here were without gravity and extend a 2-D example in our previous benchmarking exercise \citep{momoh2025volumetric}. Gravity does impose a lithostatic stress distribution, which we note implies:
\begin{equation}
\dfrac{\partial\sigma_{zz}}{\partial z} = \rho g,
\end{equation}
 but not necessarily
 
 \begin{equation}
\dfrac{\partial P}{\partial z} = \rho g;
\end{equation}

while a temperature gradient and thermal diffusion would presumably add layers of complexity to this example.
\section{Geodynamic Simulations}
\subsection{Simulations for French Massif Central (FMC) Dynamics}
\subsubsection{Model Setup and Boundary Conditions}
\begin{figure}[htbp]
\centering
\includegraphics[width=\textwidth]{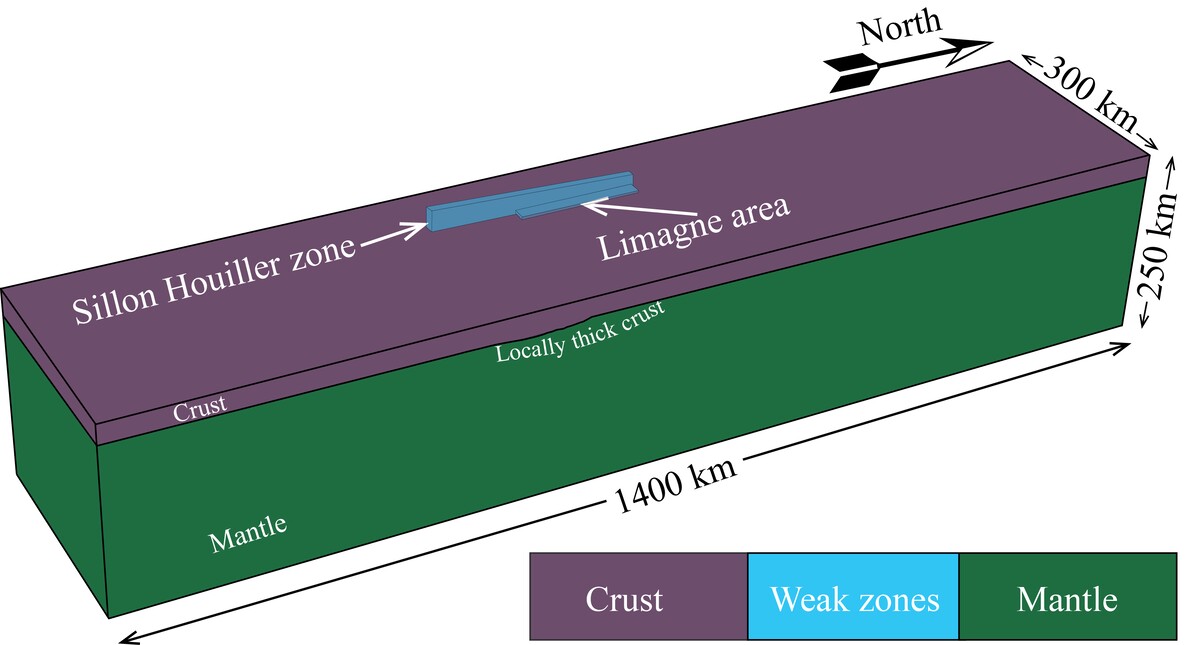}
\caption{Input for 3-D geodynamic simulation for the French Massif Central with the Hercynian Sillon Houiller Strike Slip Fault, defined by a crustal-scale damage area, and the Cenozoic Limagne area, represented by a 3-km deep weak zone, and a 26 km thick crust beneath it. Uniform initial crustal thickness of 28 km, except East of the Sillon Houiller area, where an extra 2-km-thickened crust was initialised.} 
\label{input_3D}
\end{figure}

The input model was composed of a 28 km homogeneous crustal layer with a 30 km locally thick crust, a 26 km thin crust below the graben area, a crustal-scale lithospheric fault damage area and a homogeneous mantle \autoref{input_3D}. We have utilised 2 km crustal offsets on account of observations that in some of the Cenozoic grabens in the FMC, there is crustal thinning (Limagne Graben, for example), while one of the largest Cenozoic volcanic edifices, Cantal, had 2-3 km crustal thickening as reported from seismic tomography studies of the FMC  \citep{zeyen1997refraction}. Also see \autoref{momoh_map_gravs_heat_flow}d for a zoom of the compiled crustal thicknesses. Given that the major magmatic events occurred after rifting \citep{michon2000crustal}, we used 3 km deep weak zone to characterise the graben area. For the Hercynian Sillon Houiller strike-slip fault, we defined a damaged zone (weak zone) that is 27 km deep. If the crustal offsets persisted during the time of the major magmatic event (i.e., post-rift), then Hercynian or rift-related weak zones (grabens) should be present. We found these geological inputs admissible as initial rheological conditions.

Our simulation was in 3 stages: stage 1 was defining an initial temperature field by using a 1-D geotherm, stage 2 involved defining the lithostatic stress state by gravity loading, and stage 3 was the compression stage, where one of the boundaries was compressed. In defining an initial geotherm, we have used the analytical form of continental geotherm, which assumed crustal and mantle heat fluxes, and crustal radiogenic heat production given by \cite{turcotte2002geodynamics}.

In terms of mechanical boundary conditions, the top surface was free, while the bottom surface was fixed for vertical velocity. All vertical surfaces were fixed for normal velocities during the gravity loading stage, while the Southern and Northern surfaces were imposed with normal inward velocities of 2 mm/year in the compression stage.  In terms of thermal boundary conditions, all vertical boundaries were modelled to have zero heat flux, while the upper and lower boundaries were fixed with temperatures of 273 K and 1717 K, respectively. The initial geotherm was modelled by assuming an 85 mW m\textsuperscript{-2} and 10 mW m\textsuperscript{-2}, for the crust (at the surface) and mantle (at the Moho), respectively. This resulted in an initial Moho temperature of 723 K. The material properties are shown in \autoref{matpros}.
{{\begin{table}\begin{center}
\caption{Material properties in the geodynamic model listed sequentially. \hspace{2mm}${E}$ (Young's modulus), $v $ (Poisson's ratio), $K$ (bulk modulus), $\rho_\mathrm{ref}$  (reference density), $g$ (gravitational constant), ${C_\mathrm{P}}$ (specific heat capacity), ${C_\mathrm{L}}$ (latent heat), $\alpha_\textrm{ex}$  (thermal expansivity 
coefficient), ${\lambda_\textrm{th}}$  (thermal conductivity), ${m}$ stress sensitivity exponent during creep or plasticity, ${A}$ (pre-exponential multiplier), $E_a$  (creep activation energy), $R$ (molecular gas constant), ${\varphi_i}$ (initial friction angle), ${\varphi_f}$ (final friction angle), ${\psi}$ (dilatancy angle), $c_0$ (initial cohesion), $H$  (hardening modulus), ${1/\mu}$ (relative rate of viscoplastic strain), ${\varepsilon_\mathrm{crit}}$ is the critical accumulated plastic strain at which the friction angle reaches its final value and is maintained constant, and $m$ (stress exponent during viscoplastic flow). The material parameters for dislocation creep and thermal properties were drawn from \cite{currie2015geodynamic,babeyko2008high}. Values for mechanical parameters were drawn from \cite{turcotte2002geodynamics}.  Plasticity parameters were drawn from \cite{babeyko2008high}.}
\label{matpros}
\begin{tabular}{@{}lllllll}
\hline
Material property  (unit):  &  Mantle & Continental crust & Weak zone\\
\hline
Mechanical &           &      &  \\[2pt]
\hspace{2mm}${E}$ (GPa) & 140     & 60 & 50\\[2pt]
\hspace{2mm}$v$    & 0.25     & 0.25 & 0.25 \\[2pt]
\hspace{2mm}$\rho_\mathrm{ref}$ (kg m${^\textrm{-3}}$)    & 3300     & 2800 & 2700\\[2pt]
\hspace{2mm}$ g $ (m s${^{\textrm{-2}}}$)       & 9.8& 9.8& 9.8
\\[2pt]
\hline
Thermal &            &      \\[2pt]
\hspace{2mm}${C_\mathrm{P}}$  (J kg${^\textrm{-1}}$K${^\textrm{-1}}$)       & 1000& 1000 & 1000\\[2pt]
\hspace{2mm}${C_\mathrm{L}}$  (kJ kg${^\textrm{-1}}$)       & 450& 450 & 450\\[2pt]
\hspace{2mm}$\alpha_\textrm{ex} {\;}$ (K${^\textrm{-1}}$) &  ${3.5\times{10^{-5}}}$     &  ${3.5\times{10^{-5}}}$ &  ${3.5\times{10^{-5}}}$
\\[2pt]
\hspace{2mm}${\lambda_\mathrm{th}\;\;\mathrm{{(W^{-1}\;m^{-1}\;K^{-1}}})}$ & ${0.73+\frac{1293}{T+77}}$ & ${1.18+\frac{474}{T+77}}$ & ${0.73+\frac{1293}{T+77}}$\\
\hline
Dislocation creep     &         &  \\[2pt]
\hspace{2mm}$m $  & 3     & 4    & 4 \\[2pt]
\hspace{2mm}${A}$ (Pa${^\textrm{-m}}$ s${^\textrm{-1})}$       & $3.91\times10^{-15}$   &$1.1\times10^{-28}$&$1.1\times10^{-28}$\\[2pt]
\hspace{2mm}$E_a\; $ (kJ mol${^\textrm{-1})}$    & 430     & 223 & 223 \\[2pt]
\hspace{2mm}$R\;$ (J K${^\textrm{-1}}$mol${^\textrm{{-1}}}$)  & 8.31     & 8.31 & 8.31 \\
\hline
Plasticity &     &        &      &  \\[2pt]
\hspace{2mm}${\varphi_i {\;} (^\textrm{o})}$&  28     & 15 & 2.8\\
\hspace{2mm}${\varphi_i {\;} (^\textrm{o})}$&  36.9     & 20 & 5.7\\
\hspace{2mm}${\psi {\;} (^\textrm{o})}$      & 15     & 10 & 2\\[2pt]
\hspace{2mm}$c_0 {\;} $(MPa)    & 10         & 10 & 10 \\[2pt]
\hspace{2mm}$H {\;} $ (GPa)           & 1.4     & 0.6 & 0.5\\[2pt]
\hspace{2mm}${1/\mu}$ (s${^ {-1}}$)    & ${10^{-15}}$       & ${10^{-15}}$    & ${10^{-15}}$ \\[2pt]
\hspace{2mm}${\varepsilon_\mathrm{crit}}$ (-)    & 0.005       & 0.005    & 0.005 \\[2pt]
\hline
\end{tabular}\\
\end{center}
\end{table}}}

Based on a recently published horizontal velocity model for Europe from Global Navigation Satellite Systems (GNSS) data \citep{masson2019extracting, pina20223d}, the area around the FMC (Central France) is inferred to undergo slow North-South and NorthEast-SouthWest shortening at a rate of ${\sim}$0.1 to 0.4 mm/year. Other review studies into the stress regime of France and surrounding areas indicate that, despite being far from major plate boundaries, the stress pattern within France may have been influenced by far-field plate boundary forces like ridge push force from the Mid-Atlantic Ridge and subduction/collision of the Nubia-Eurasia plates (see \cite{muller1992regional,mazzotti2020processes}). In intraplate regions like Central and Western Europe, the stress regime was found to be compressive, except in elevated areas like the FMC, where a combination of normal and strike-slip regimes is not uncommon \citep{heidbach2016world,heidbach2018world,tesauro2006analysis}. We therefore used maximum compressional velocities of 5 mm/year to study the possible influence of the Africa-Eurasia convergence on the FMC. Since the Cenozoic Era, the FMC area has undergone North-South compression, except for the period between 33.3 and 13.3 million years, where there were North-South trending grabens that were linked to a passive extensional response to the Alpine compression \citep{dezes2004evolution}.

\subsubsection{Model Shortening and Temporal Evolution}
\begin{figure}[htbp]
\centering
\includegraphics[width=\textwidth]{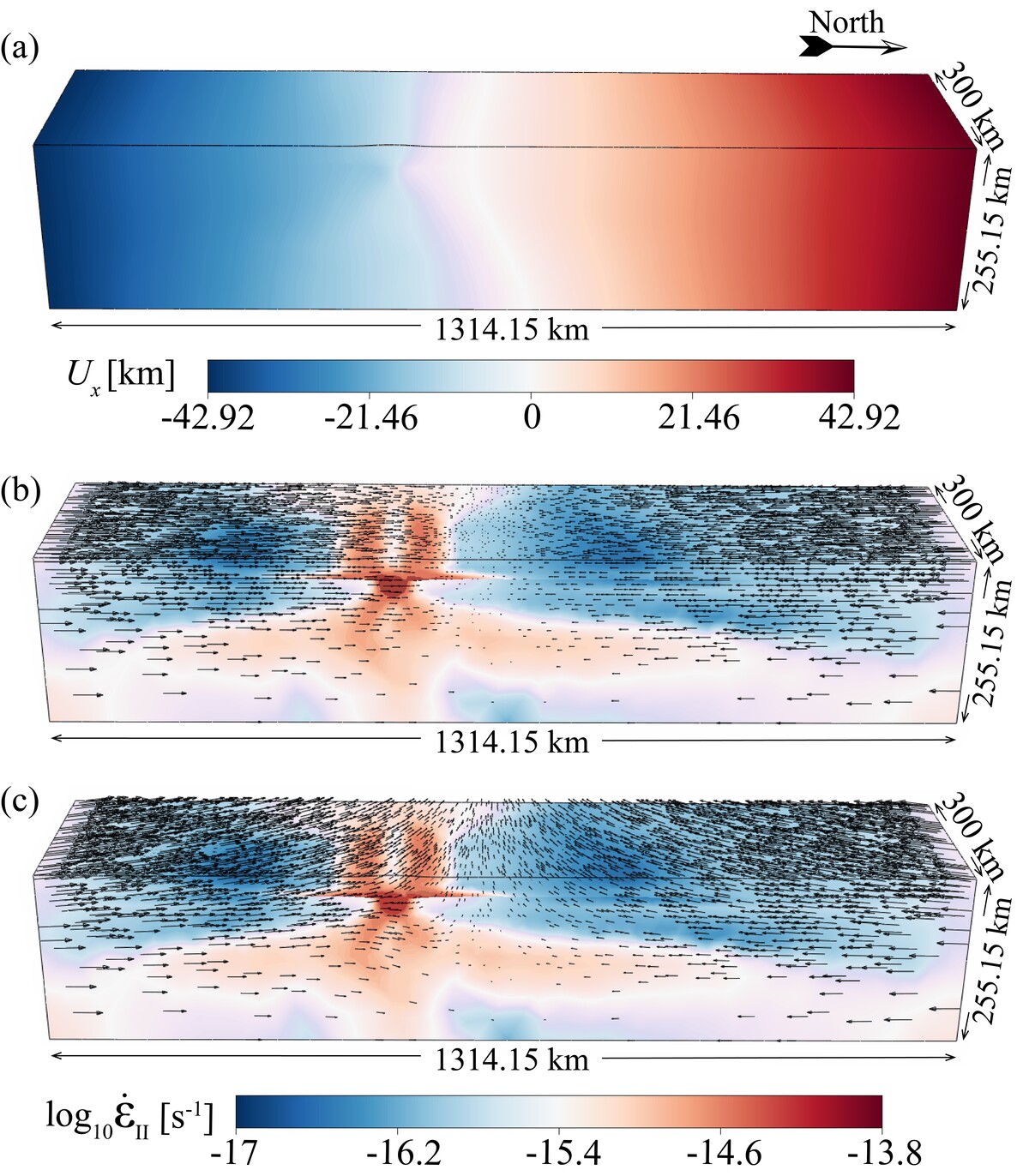}
\caption{(a) Along-compression direction horizontal displacement which can be used to infer shortening, (b) deviatoric strain rate invariant with vectors of horizontal displacement where the length of the vector represents the magnitude of horizontal displacement in the compression direction, (c) deviatoric strain rate invariant with vectors of total displacement indicating uplift in the FMC area.} 
\label{FMC_3D_displacement}
\end{figure}

Over 17.1 million years, the model underwent 6.1 per cent cumulative north-north-south shortening of 85.84 km over a domain that was initially 1,400 km long, while the local cumulative shortening around the FMC area in the model is about 20 km, representing about 1.4 to 2.5 percent local shortening. The horizontal shortening was inferred from the maximum displacements from the north and south (\autoref{FMC_3D_displacement}a), which shows 42.92 km. The change in the polarity of the displacement was observed where we had an initially thicker crust in the FMC area (\autoref{FMC_3D_displacement}b). This area shows the maximum uplift based on vectors of the magnitude of displacement (\autoref{FMC_3D_displacement}c).

\begin{figure}[htbp]
\centering
\includegraphics[width=\textwidth]{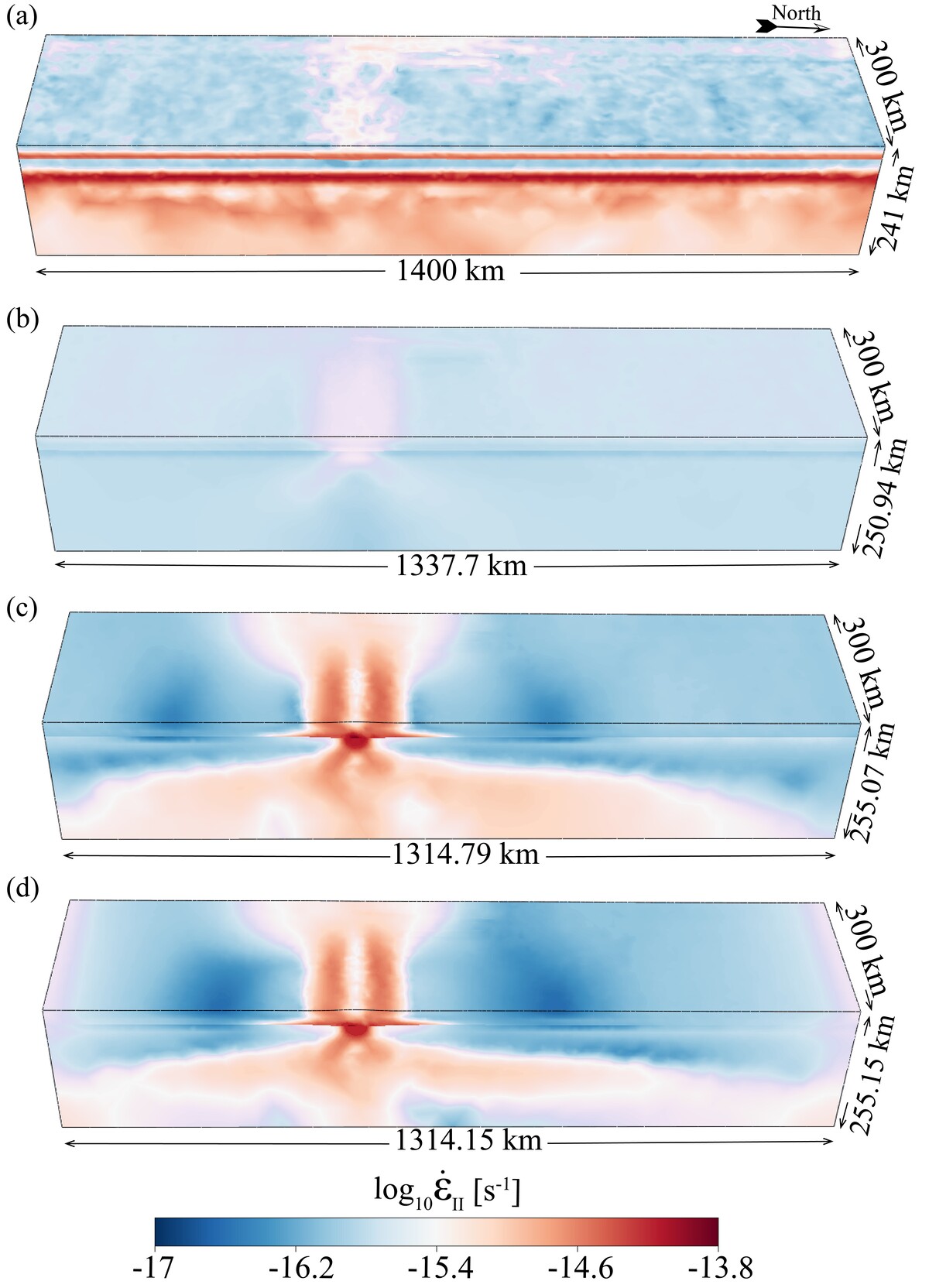}
\caption{French Massif Central results: Snapshots of deviatoric strain rate invariant taken at (a) 7,562 years, (b) 12.5 million years, (c) 17 million years and (d) 17.1 million years.} 
\label{FMC_3D_temporal}
\end{figure}

We showed the temporal evolution of the model from about 7,500 years to 17.1 million years (\autoref{FMC_3D_temporal}). The initial deformation was creep in the mantle and some crustal deformation (\autoref{FMC_3D_temporal}a). Subsequently, after creep relaxation, distributed deformation was observed in the crust and surface of the model (\autoref{FMC_3D_temporal}b). The intervening period between these two snapshots involved stress build up, stress relaxation  through creep and plastic localisation in the lithosphere. The model evolved with crustal and mantle deformation bands, where the crust in the FMC area was characterised by uplift whose shoulders were deformed by the deformation bands, while the deformation in the mantle was characterised by distributed deformation bands extending deeper into the mantle with decreasing amplitude (\autoref{FMC_3D_temporal}c and \autoref{FMC_3D_temporal}d). The intersection of these two crustal and mantle deformation bands lead to a zone of strong localisation at the base of locally thick crust. 

\subsubsection{Results of French Massif Central Dynamics}
\begin{figure}[htbp]
\centering
\includegraphics[width=\textwidth]{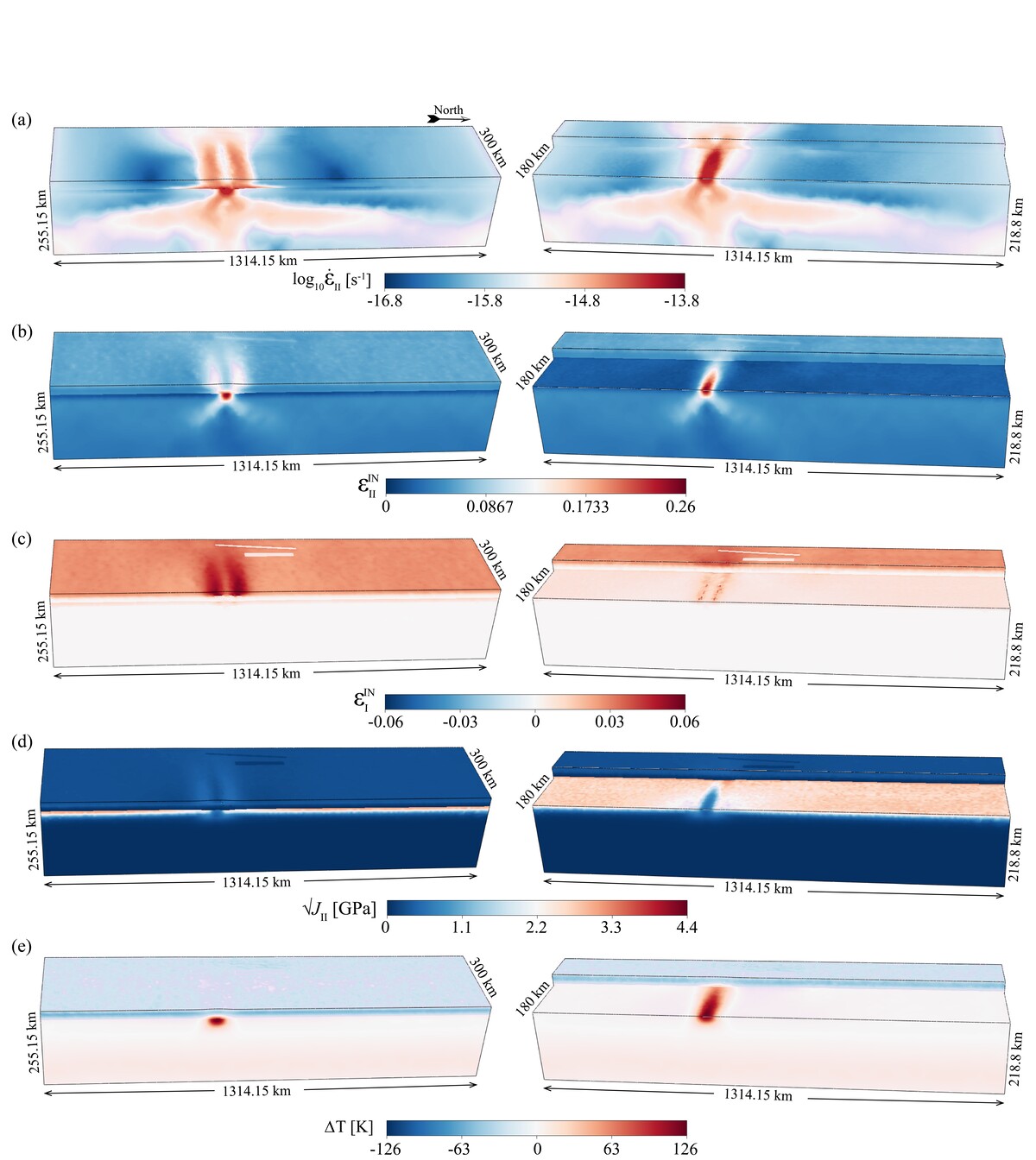}
\caption{3-D simulation results for the French Massif Central area after ${\sim}$17.1 million years of Africa-Eurasia-induced compression at a cumulative 5 mm/year. (a) Logarithm of deviatoric strain rate invariant, (b) inelastic deviatoric strain invariant, (c) inelastic volumetric strain invariant, (d) inelastic deviatoric stress invariant, (e) temperature change, i.e., difference between initial geotherm and evolved temperature with changes due to deformational heating. The figures on the right panel show cuts at a depth of 38 km from the elevated surface to show the internal deformation.} 
\label{FMC_3D}
\end{figure}

The results after 17.1 million years of 3-D simulation for the FMC are shown in \autoref{FMC_3D}. The effect of the locally thick crust in our simulations is twofold: inelastic strain localisation, which led to a drop in the deviatoric stress invariant, and localised topographic uplift (\autoref{FMC_3D}a-e). The deviatoric strain rate invariant shows localised deformation at the Moho of thick crust, with two deformation bands in the crust and two in the mantle that diffuses in two wings and one tail like a mushroom (\autoref{FMC_3D}a). The cumulative inelastic deviatoric strain shows the localised zone at the Moho forming the intersection of the crustal and mantle deformation bands (\autoref{FMC_3D}b). The distinctive contribution of plasticity is volumetric inelastic strain: this invariant (an average of the inelastic normal strains) is shown in \autoref{FMC_3D}c. Our constitutive laws only assumed dilatant plastic behaviour. Therefore the volumetric inelastic strain indicates the parts of our model where the brittle, as opposed to ductile, crust or mantle can be identified (non-zero volumetric strain and negligible volumetric strain, respectively). The magnitude of the deviatoric stress invariant (${\sqrt{J_\mathrm{II}}}$) is a measure of the deviations of normal stresses from a mean value, and of the shear stress. From our simulations, normal stresses tend to approach one another in the ductile domain, which reduces their deviations from the mean so that only the shear stress contributes to ${\sqrt{J_\mathrm{II}}}$. The localisation of deformation, whether creep or plastic, led to the normal stress components becoming equal (incompressible flow) leaving the shear stress components to dominate the deviatoric stress invariant. This explains why the deviatoric stress invariant drops at the zone of localisation (\autoref{FMC_3D}d), and indicated that our model was approaching incompressible flow there. The inelastic deformation was converted to heat and showed a temperature rise of 126 K (\autoref{FMC_3D}e). The deformation is also shown to occur internally through the model (\autoref{FMC_3D}) with peak deformation at the Moho of thick crust.

\begin{figure}[htbp]
\centering
\includegraphics[width=\textwidth]{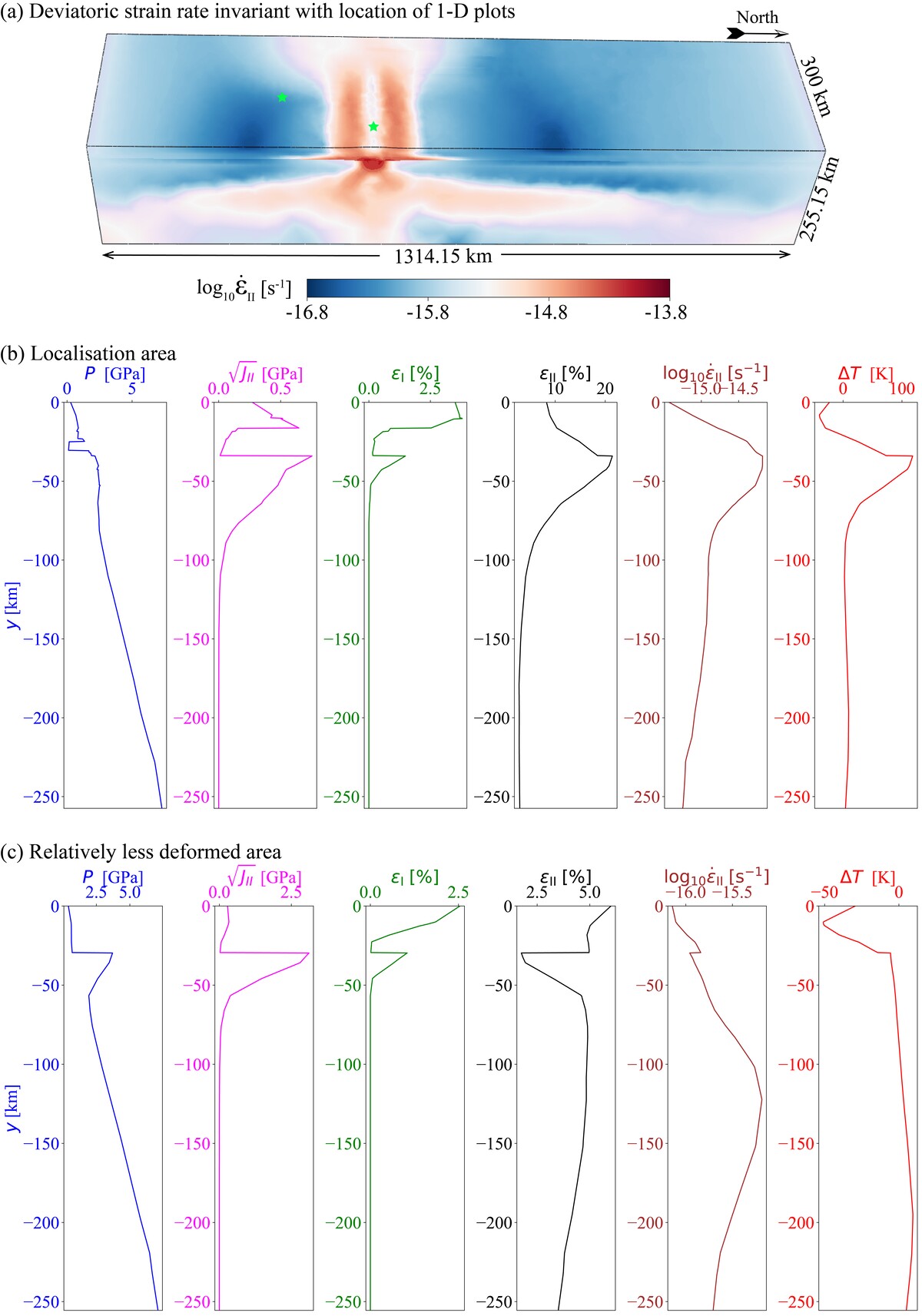}
\caption{FMC results: (a) Deviatoric strain rate invariant with green stars showing the locations for extracting 1-D profiles in the localised deformation zone (between two surface deformation bands)and a less deformed area. Profile extracts for pressure, deviatoric stress invariant, inelastic volumetric strain, inelastic deviatoric strain, deviatoric strain rate invariant and temperature change for (b) deformed area, and (c) less deformed area.} 
\label{component_extracts}
\end{figure}

We extracted illustrative 1-D vertical profiles from the localised deformation zone and less deformed zone \autoref{component_extracts}. There are strong gradients of both pressure and deviatoric stress at the Moho in both localised and non-localised areas but the profiles differ in several ways. The vertical profile in the localised area traverses a 3-D structure bisecting the deformation bands. The dominant process is upward plate buckling on a horizontal scale on the same order as the lithospheric thickness. The volumetric deformation profile indicates that while the upper crust is brittle/plastic, the lower crust is mostly ductile (where the deviatoric stress and pressure reach lower values). The mantle just below the Moho shows lithospheric (i.e., compressible) behaviour, becoming ductile incompressible below about $\sim$50 km depth (the effective lithospheric thickness under these conditions - \autoref{FMC_3D}, and \autoref{component_extracts}b). We interpret the high volumetric deformation close to the surface (\autoref{component_extracts}b) (the upper crust), as due to buckling, which induces expansion in the upper half of the plate which, given our computed plate thickness in this example, is made of crustal material. Beneath the buckled zone, both ${\sqrt{J_\mathrm{II}}}$ and ${{{\dot\varepsilon}}_{\textrm{ij}}}$ terms have their largest values at $\sim$40-50 km, where the plate is in compression and the profile crosses the intersection of the two major shear bands, thus maximising heating. The temperature anomaly which shows the cumulative effect of deformational heating and diffusion has a broad ellipsoidal shape. Ductile creep persists down to $\sim$100 km in this profile and even further in the adjacent deformation bands (\autoref{component_extracts}a). If we take ${{{\varepsilon}}_{\textrm{I}}}$ as a proxy for where seismicity may occur, we expect seismicity in the upper crust and the first $\sim$10-20 km of the uppermost mantle below the Moho but little in the ductile lower crust.

In the zone of localisation, the pressure shows decompression near the Moho (\autoref{component_extracts}b). This corresponds to the transition to ductile crust as shown in the deviatoric stress invariant and volumetric strain. In contrast, the zone of relatively less deformation shows no similar decompression, a higher deviatoric stress invariant in the mantle indicating the deformation is not localised, and heating is much less than within the zone of localisation (\autoref{component_extracts}c). {\color{blue}{}} One can hypothesise that such transition zones often involve a local change from brittle to ductile behaviour, e.g., localised thinning of the lithosphere-asthenosphere transition zone. The areas with low volumetric strains (\autoref{FMC_3D}d) highlight ductile behaviour where we made no assumptions of volumetric deformation.

\subsubsection{Heat Flow and Topography}
The FMC is in an intraplate region with high heat flow. In \autoref{temperature_heatflow}a and \autoref{temperature_heatflow}b, we showed the evolved temperature after 17.1 million years of compression for the entire model and a horizontal slice near the Moho, respectively. The highest surface heat flow is ${\sim}$70 mW m\textsuperscript{-2} (\autoref{temperature_heatflow}c). This is confined to the Sillon Houiller fault zone and the Limagne graben area. The reason for this is localised inelastic deformation within these weak zones which led to temperature changes at shallow depths to influence the surface heat flow. While it has been argued that the maximum heat flow in the Massif Central is confined to the Cenozoic grabens and volcanic area,  \citep{lucazeau1984interpretation}, our results indicate that the areas confined to the prescribed grabens produce heat flow that may be explained by localised deformational heating. This is an alternative hypothesis to the ascent of a mantle diapir commonly invoked as a heat source \citep{lucazeau1984interpretation}.

We also show an extract of the present-day topography East of the Limagne graben area (\autoref{compare_topo}a) to topography inferred from our model (\autoref{compare_topo}b) which is shown in (\autoref{compare_topo}c). The numerical results show a north-south uplifted zone comparable in horizontal scale (${\sim}$200 km) with that observed and topography similar to within 250 m to 500 m.

These results indicate the role of crustal offsets in initiating and enhancing localisation. Even when as small as 2 km, the effect is a thermal anomaly of 126 K and heat flow density comparable with observed heat flow density in the French Massif Central (\autoref{momoh_map}b). {\autoref{partial_melt} shows the profiles of temperature in comparison to potential mantle solidi taking into account the effect of pressure which is also shown. Temperatures never reach the anhydrous solidus and only reach the water-saturated solidus at $\sim$150 km. Nevertheless the likely effect of the presence, at least locally, of CO\textsubscript{2} and fusible alkali components, notably Na\textsubscript{2}O+K\textsubscript{2}O is to reduce the volatile-saturated solidus by $\sim$300 K \citep{schmidt2024origin} to around $\sim$ 1000 K or even slightly less. The strongly heated zone in the depth range $\sim$40-80 km thus may just reach such a highly depressed solidus, implying that very low degrees of partial melting may occur there, but large amounts of partial melting are very unlikely because such lithologies are not expected to be abundant. This is consistent with sparse volcanism, where individual eruptions represent low degrees of partial melting and have alkali-rich to carbonatitic compositions \citep{schmidt2024origin}, While we did not observe partial melt, mainly because of using the anhydrous solidus, we were able to observe topographic uplift, localised surface heat flow and potential change in the lithosphere-asthenosphere transition zone, which would suggest local lithospheric thinning without invoking the mantle plume hypothesis. 
\begin{figure}[htbp]
\centering
\includegraphics[width=\textwidth]{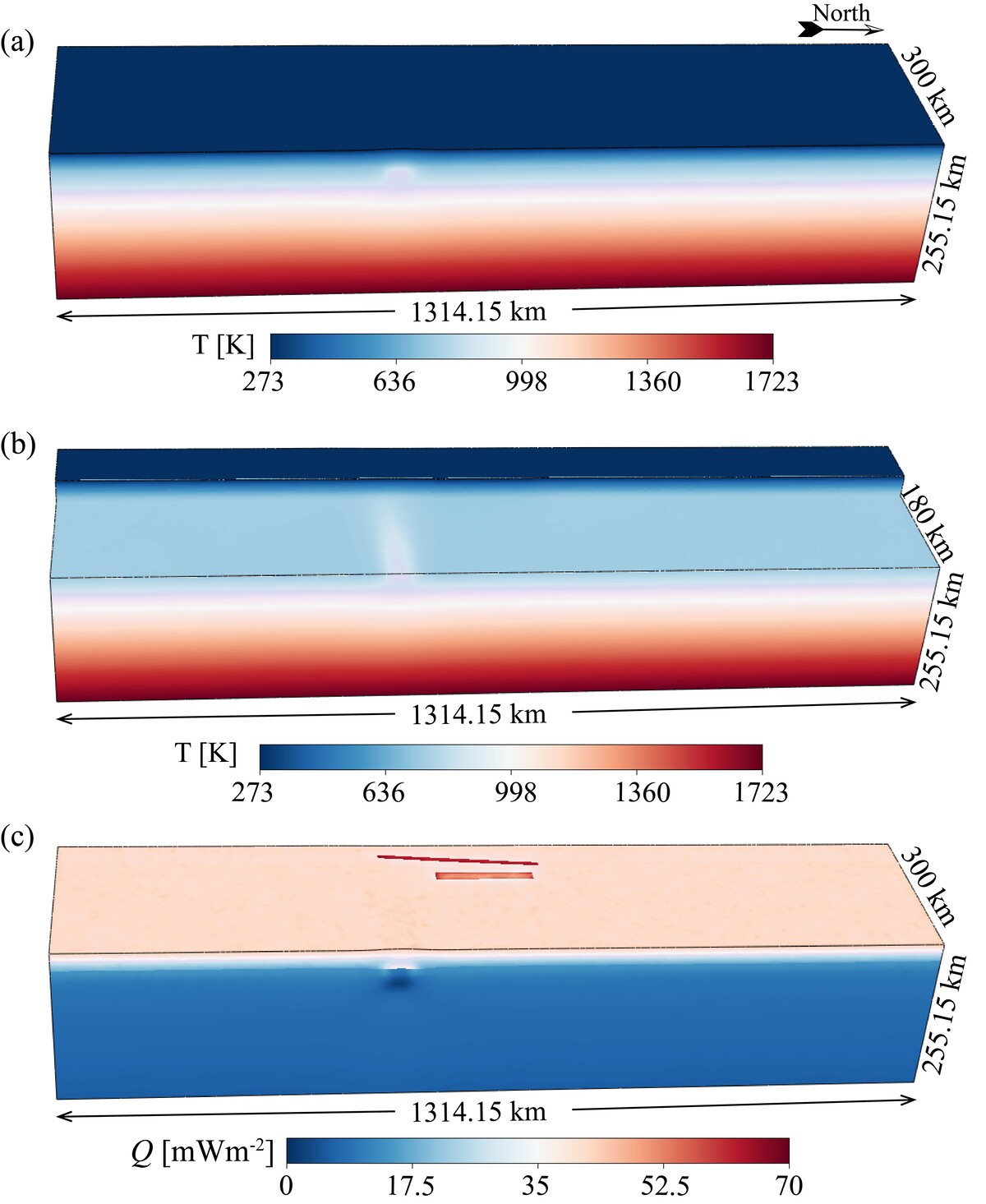}
\caption{(a) Evolved temperature showing the thermal anomaly in the localised deformation zone, (b) evolved temperature cut at 38 km depth to show the extent of the thermal anomaly, (c) heat flow showing the maximum surface heat flow at the Cenozoic grabens.} 
\label{temperature_heatflow}
\end{figure}

\begin{figure}[htbp]
\centering
\includegraphics[width=\textwidth]{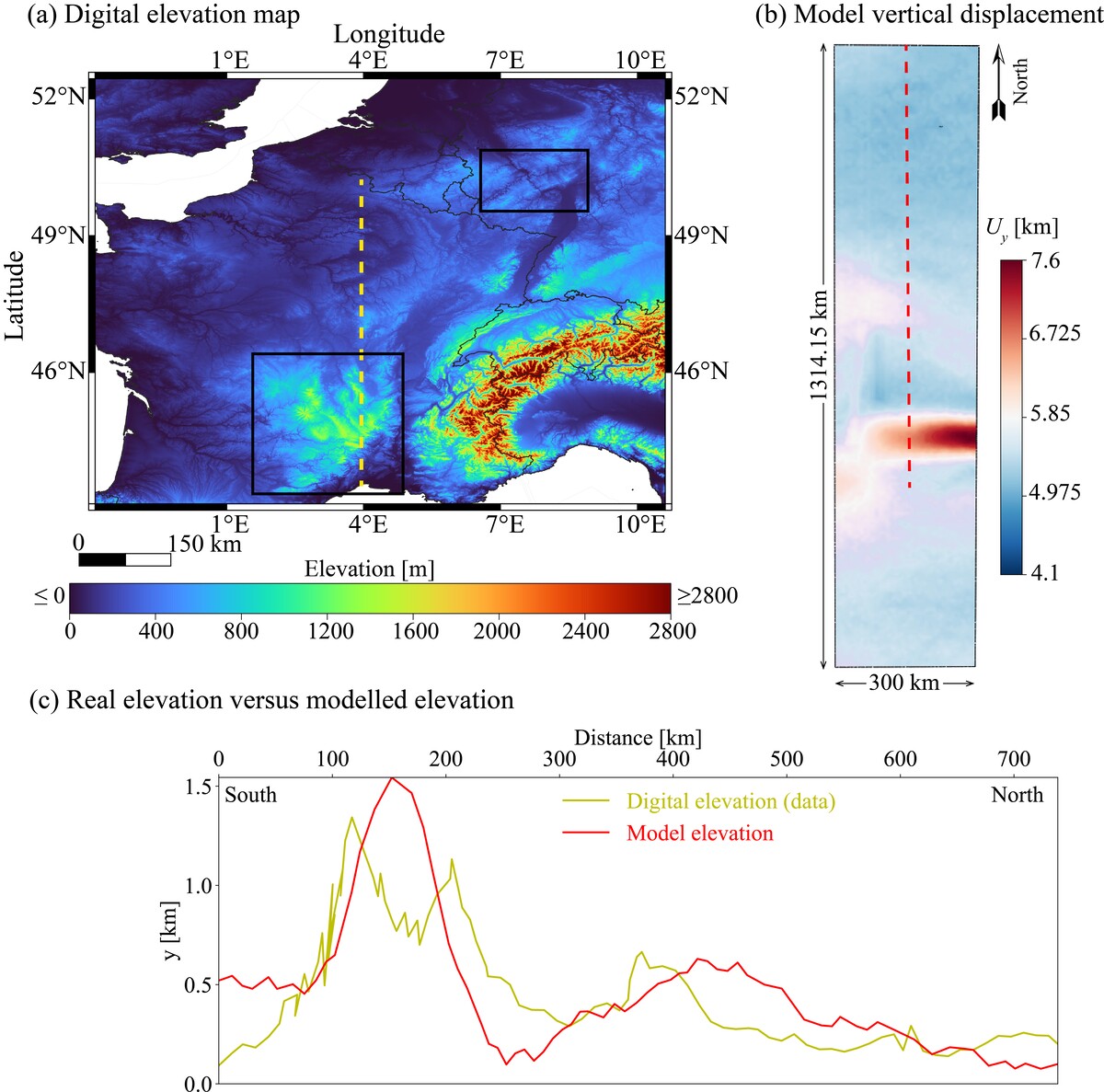}
\caption{Comparison of surface topography obtained from the digital elevation map of the FMC and our model. (a) Digital elevation map with a profile (yellow line) from South to North to extract distance-elevation profile, (b) vertical displacement shown in plan view with red line showing location of profile extract, (c) comparison of real data elevation profile with topography inferred from vertical displacements in the model. We have subtracted the displacement along the profile from the minimum displacement, i.e., $y={U_{z}-{U_{z,\mathrm{min}}}}$.} 
\label{compare_topo}
\end{figure}
\begin{figure}[htbp]
\centering
\includegraphics[width=\textwidth]{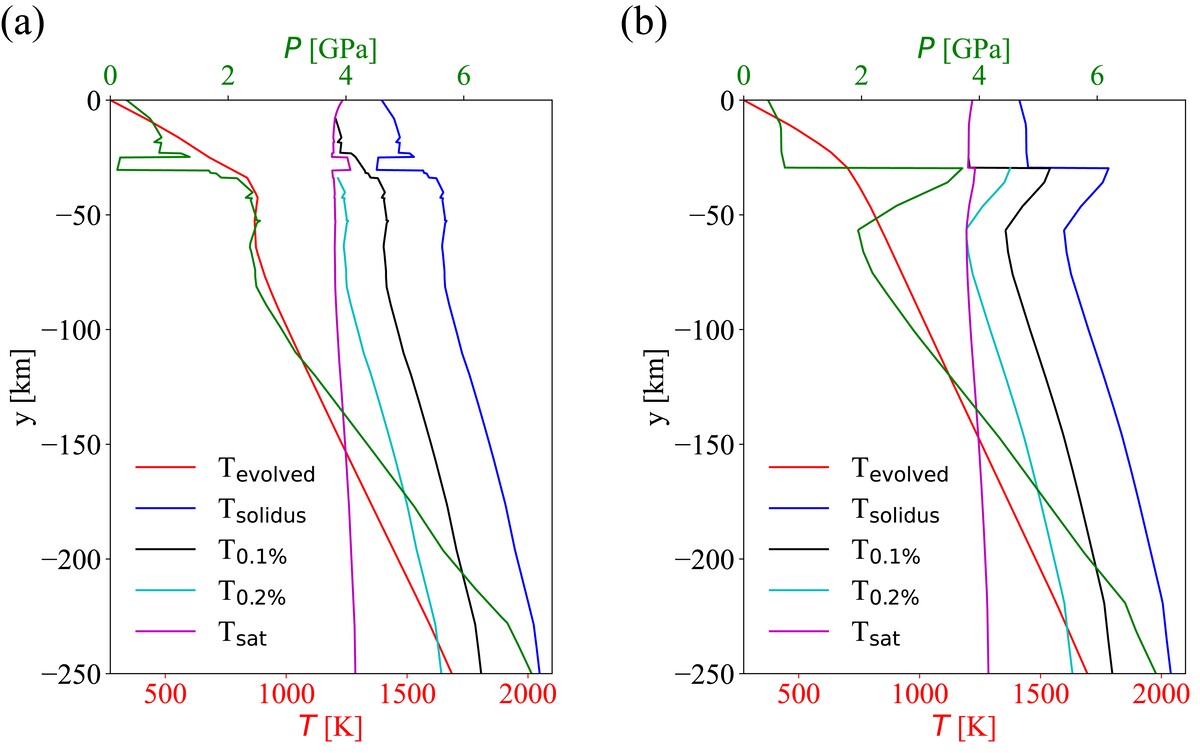}
\caption{Comparison of the evolved temperature and pressure with mantle solidi given different degrees of water saturation for (a) domed area where localisation is intense, and (b) flat area with less deformation. See \autoref{component_extracts}a for the location of profiles extracted. The green curves and top axes correspond to pressure, while the remaining coloured curves and the bottom axes correspond to temperature.} 
\label{partial_melt}
\end{figure}
\subsection{Comparison with other magmatic models of the Massif Central}
One of the earliest models of the magma source for the Massif Central highlighted crustal thinning beneath the graben area due to asthenospheric upwelling with the presence of an abnormal mantle just below the Moho of thinned crust underlying the volcanic edifices \citep{perrier1973structure,coisy1978regional}. Abnormal mantle differs from ambient mantle in terms of heterogeneities, for example, a thermal anomaly, a negative mantle Bouguer anomaly or a density change. In areas of the most recent eruptions, deep low-frequency events localised at the Moho have been reported \citep{shapiro2025deep}. The abnormal mantle could be associated with magmatic underplating or indications of the arrival of fresh magma or crystallisation of existing magma. Against this model is that the magmatic flux in the Massif Central is insufficient to be reconciled with an upwelling plume, and the spatial and temporal distribution of the magmatic fluxes seem to require ad-hoc modification of a plume model in order to make it appear viable.  

\cite{granet1995massif,granet1995imaging} proposed that low-velocity anomalies extending to 250 km depth could be interpreted as mantle "baby" plumes. A recent approach used thermomechanical models to propose a hypothesis of a subducting slab (Eurasia slab and relicts of Tethys slab) rolling back or rolling over, thereby displacing a lower mantle plume pooling at a thermal boundary layer placed at 670 km depth to generate secondary plumes which ascend and carry magma with them \citep{li2026intraplate}. For simplicity, the models of \cite{li2026intraplate} examined subducting oceanic plates to avoid complexities of continental collision. In most cases, low-velocity anomalies from tomography are interpreted in terms of mantle plumes \citep{granet1995massif,granet1995imaging,li2026intraplate}; yet these anomalies were rarely subvertical as might be expected for a classical vertical plume. An alternative explanation, suggested by our results, is that these slanted low-velocity anomalies may be due to large-scale mantle deformation bands which, given the boundary conditions, are expected to be inclined to the vertical.

Using a plate-tectonic model, \cite{chesworth1975mantle} proposed that the area between the Alps and Pyrenees had undergone vice-like compression, which led to flexural uplift with rifts formed on the flexures, and faults providing conduits for partial melt (originating from the mantle) to the surface. The Massif Central volcanism has also been interpreted as being due to the deep lithospheric root of the Alps which led to asthenospheric mantle flow leading to thermal erosion at the base of the lithosphere, volcanism and uplift in the Massif Central \citep{merle2001formation}.

The inclined nature of the velocity anomalies, which merge at a shallower depth (near the Moho) is a common feature of tomographic velocity models \citep{li2026intraplate}, and can be alternatively explained by inverted V-shaped deformation bands as suggested by our models (\autoref{FMC_3D},and \autoref{melt_hypothesis}). The deformation bands are a consequence of inelastic strain localisation due to an initially thick crust, which can in turn lead to localised heating and a local thermal anomaly (geophysical anomaly). Tectonic seismicity around Moho depth would also be expected from our results and would be expected to contain low-frequency events when small amounts of partial melt are also present.
\subsection{Simulations for Eifel Volcanic Region}
The model input for the Eifel volcanic region dynamics is shown in Figure \ref{eifel_input}. We have used a uniform crustal thickness of 28 km and included a locally thick crust beneath the intersection of Cenozoic grabens. The grabens are characterised by a three km-deep weak zone of elastically and plastically weak materials (\autoref{matpros}).
\begin{figure}[htbp]
\centering
\includegraphics[width=\textwidth]{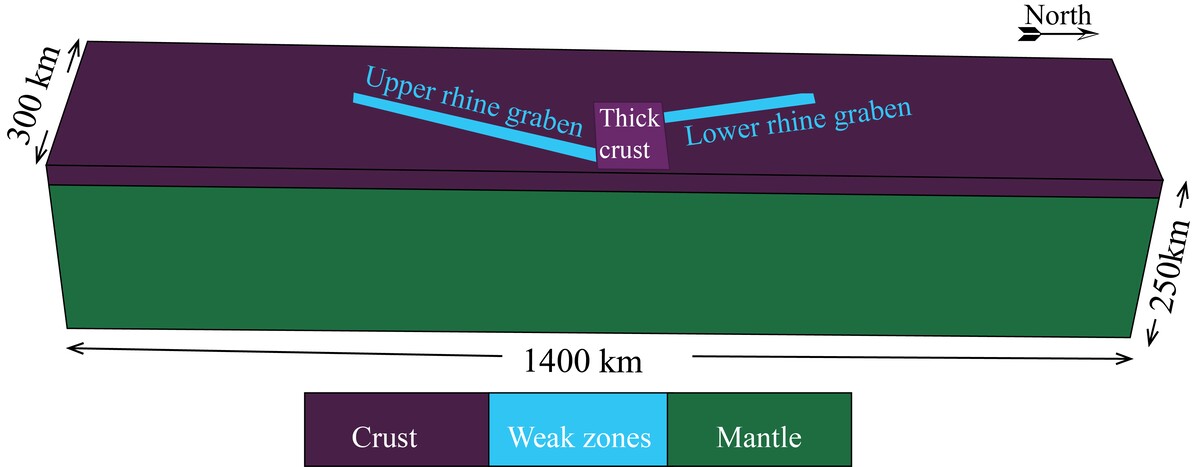}
\caption{Input model set up for the Eifel Volcanic Region. We use uniform crustal thickness of 28 km, except beneath the grabens, where the crust is thinner (26 km) and between the grabens where the crust is thicker (30 km).} 
\label{eifel_input}
\end{figure}
The Eifel volcanic area is about 300 km North of the Alpine deformation front in the Alpine foreland (Figure \ref{momoh_map}a), with Eifel Cenozoic volcanism oriented parallel to the Alpine deformation front. Given the absence of data reconciling tectonic phases and volcanic eruptions and mantle uplift \citep{schmincke1983quaternary}, we assume a shortening rate of 5 mm/year imposed Northwards from the Alpine region. The only constraint is the relative motions of the African and Eurasian plates from the Global Positioning System (GPS)-derived velocity field  \citep{reilinger2006gps,demets1994effect}. All other vertical boundaries remain fixed for normal velocities while the top surface is free.
 
\subsubsection{Results and Discussion of Eifel Volcanic Region Dynamics}
We first show the time series evolution of the simulations (\autoref{eifel_time_series}). The first few thousand years was creep deformation in the mantle and plastic deformation in the weak zones (grabens), and thicker crust between the grabens (\autoref{eifel_time_series}a and \autoref{eifel_time_series}b), while later deformation was progressive shortening, uplift, and formation of crustal and mantle deformation bands (\autoref{eifel_time_series}c). The uplifted region is flanked by two crustal-scale deformation bands, which intersect with the mantle deformation bands at the Moho (\autoref{eifel_time_series}c to \autoref{eifel_time_series}f). 
\begin{figure}[htbp]
\centering
\includegraphics[width=0.7\textwidth]{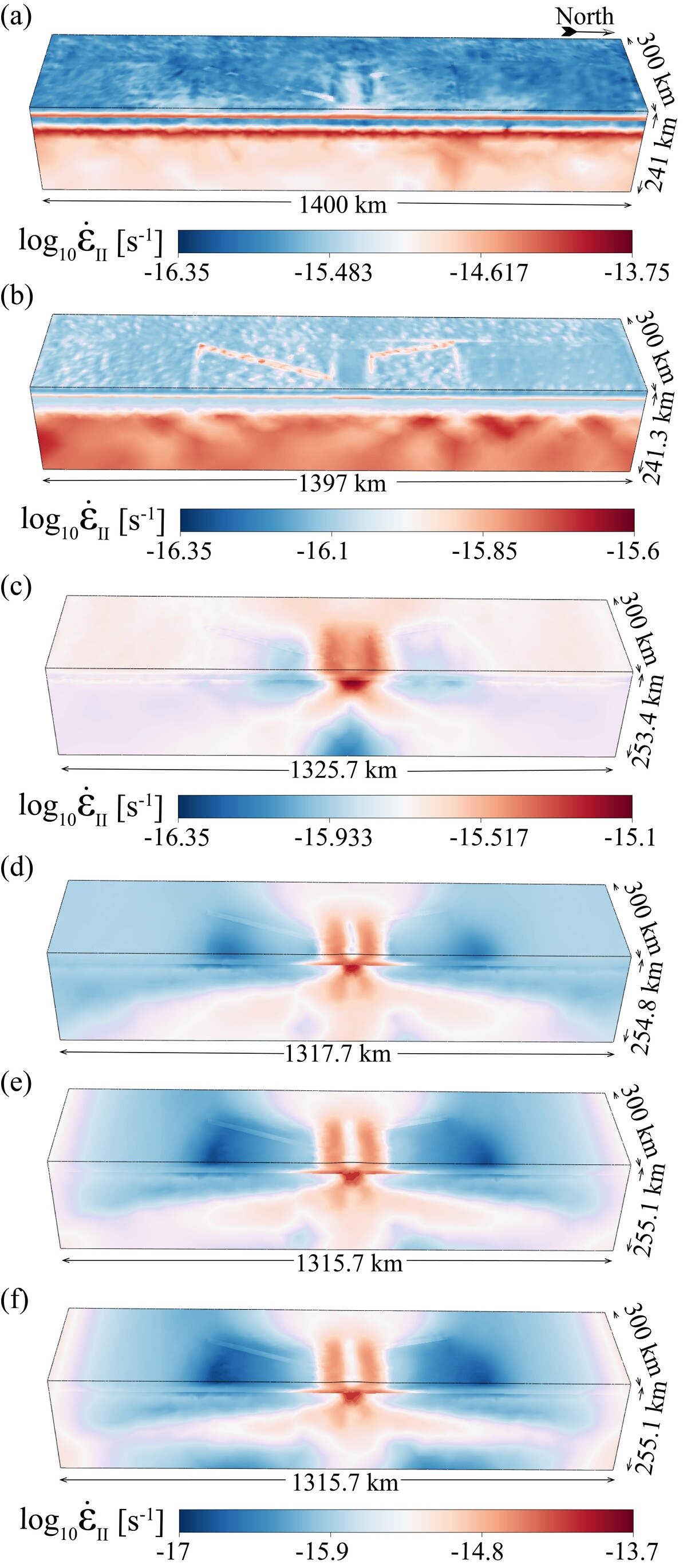}
\caption{Time series evolution for Eifel Volcanic Region showing the deviatoric strain rate invariant for (a) 6,176 years, (b) 526 thousand years, (c) 14.868 million years, (d) 16.468 million years, (e) 16.85 million years, and (f) 16.85 million years plus 320 years.} 
\label{eifel_time_series}
\end{figure}
Shown in \autoref{Eifel_3D} are the results after about 16.85 million years of shortening. First-order observations include the topographic doming at the graben junction, where we had an initially locally thick crust. As shown in \autoref{Eifel_3D}a, the deviatoric strain rate invariant essentially outlines two sets of deformation bands: the first set is in the crust, controlled by brittle plasticity, while the second one is in the mantle, controlled by ductile creep. At the intersection of these sets of deformation bands is a zone of intense deformation. The integrated inelastic deviatoric strain shows a relatively quiet region except where the crust was initially thick (\autoref{Eifel_3D}b). The surface expression of the deformation bands was also limited by the extent of the thick crust, while the deformation bands in the mantle were initially strong from the Moho and propagated into the mantle with lower amplitudes. The inelastic volumetric strain gives an indication of the parts of the model undergoing brittle plastic deformation, which is confined to the upper part of the crust and the uppermost mantle, especially at the Moho of thick crust (\autoref{Eifel_3D}c). The areas with prominent inelastic volumetric strains were dominated by brittle plasticity, which can be used as proxies for the brittle part of the crust and mantle. The zones with small volumetric strains within the crust and mantle indicate predominantly ductile deformation. The stress state and inelastic volumetric strain allow the brittle and ductile crust or mantle to emerge naturally without these rheologically different zones needing a priori specification. Due to a zone of localised inelastic deformation, a thermal anomaly was observed in the Moho of thick crust, and the maximum accumulated temperature due to deformational heating was ${\sim}$120 K (\autoref{Eifel_3D}d). The zones of highest heat flow were confined to the graben areas, highlighting a localised surface heat flow density anomaly of ${\sim}$70 mW m\textsuperscript{-2} (\autoref{Eifel_3D}e). 

To show the extent of internal deformation, we showed cuts of the deformation, stress and temperature in \autoref{Eifel_3D_Perspective_from_the_East}. The zone of instantaneous localised deformation (\autoref{Eifel_3D_Perspective_from_the_East}a) led to accumulated inelastic deformation (\autoref{Eifel_3D_Perspective_from_the_East}b), and a localised drop in the deviatoric stress invariant (\autoref{Eifel_3D_Perspective_from_the_East}c). These results further illustrate that with increasing inelastic deformation, the deviatoric stress invariant becomes lower due to the equality of normal stresses such that the contribution to the deviatoric stress invariant is mostly the shear stress. This is typical of ductile behaviour or a material approaching ductile behaviour. In general, the drop in internal stresses arises as internal resistance to the sustained inelastic deformation decreases. This sustained inelastic deformation may hint at local lithospheric thinning or transition to a thin lithosphere if stress drop can be used as a proxy for ductile mantle.  We further show the footprint of the temperature change due to deformational heating on the evolved temperature, which highlights a local temperature increase (\autoref{Eifel_3D_Perspective_from_the_East}d). Vertical profiles were also extracted for two locations in the model, one location across the most deformed region, and the other location across the less deformed region (\autoref{Eifel_3D_Perspective_from_the_East_1D}). Two distinct observations emerged. The region that had experienced the most intense deformation was associated with decompression around the Moho region (\autoref{Eifel_3D_Perspective_from_the_East_1D}b), compared to the region with less deformation (\autoref{Eifel_3D_Perspective_from_the_East_1D}c). This follows from \autoref{pressure_update}, where the pressure reduces as a fuction of volumetric plastic flow. The profile of the deviatoric stress invariant in the deformed area shows a part of the crust with brittle behaviour (characterised by high inelastic volumetric strain) followed by a part with transitory or ductile behaviour (characterised by lower inelastic volumetric strain); while the mantle (from ${\sim}$30 km showed a decay in the deviatoric stress invariant and the inelastic volumetric strain (\autoref{Eifel_3D_Perspective_from_the_East_1D}b). This region of brittle transitioning to ductile behaviour is about 50 km thick, indicating that the deformation region around the Moho is broad. Compared to the location that was less deformed (\autoref{Eifel_3D_Perspective_from_the_East_1D}c), the inelastic volumetric strain shows a progressive drop both in the crust and mantle. However, the inelastic deviatoric strain shows a bump in the mantle between 50 and 175 km depth of ${\sim}5$ per cent, which represents the large deformation band of reduced amplitude in \autoref{Eifel_3D}b. 
 
Previous tomographic images indicated a low velocity zone from 60 to 90 km with a lateral extent of approximately 100 km under the West Eifel volcanic area \citep{budweg2006eifel}. This was found to correlate with a broad reduced  \textit{S-}wave velocity anomaly extending to 180 km with a gap down to 240 km where another anomaly appears to 410 km in the mantle \citep{keyser20023d}. Previously, teleseismic \textit{P-}wave velocity inversions showed a Low Velocity Anomaly (LVA) with dip beneath the Eifel Volcanic Fields with a diameter of 100 km and equivalent to 150 to 200 K ${\pm}$100 K excess temperature \citep{ritter2000teleseismic}. In tomographic models, LVAs appear from Moho depth or in the uppermost mantle. Previous interpretations either allude to a thermal anomaly or mantle plume \citep{ritter2000teleseismic,ritter2001mantle,keyser20023d,budweg2006eifel}. Our simulations of a compression state of the Eifel would suggest a localised temperature rise of the order of 143 K beneath the Moho of thick crust, which would suggest that deformational heating in a compressional scenario, including variations in crustal thicknesses, can serve as an alternative explanation for the observed seismic velocity perturbations.  Teleseismic travel time tomography has also been used to map the thickness of the lithosphere in the Eifel region, for example, \cite{seiberlich2013topography} identified lithospheric thicknesses of 60 ${\pm}$5 km beneath the surrounding regions, while the lithosphere thins to 41 ${\pm}$5 km beneath the grabens. While our model results show the onset of ductile behaviour at 48.8 km for the location near the graben region (\autoref{Eifel_3D_Perspective_from_the_East_1D}c), if we use the inelastic volumetric strain as a discriminant between brittle and ductile behaviour, this zone begins from ${\sim}$53.6 km. 

\begin{figure}[htbp]
\centering
\includegraphics[width=0.75\textwidth]{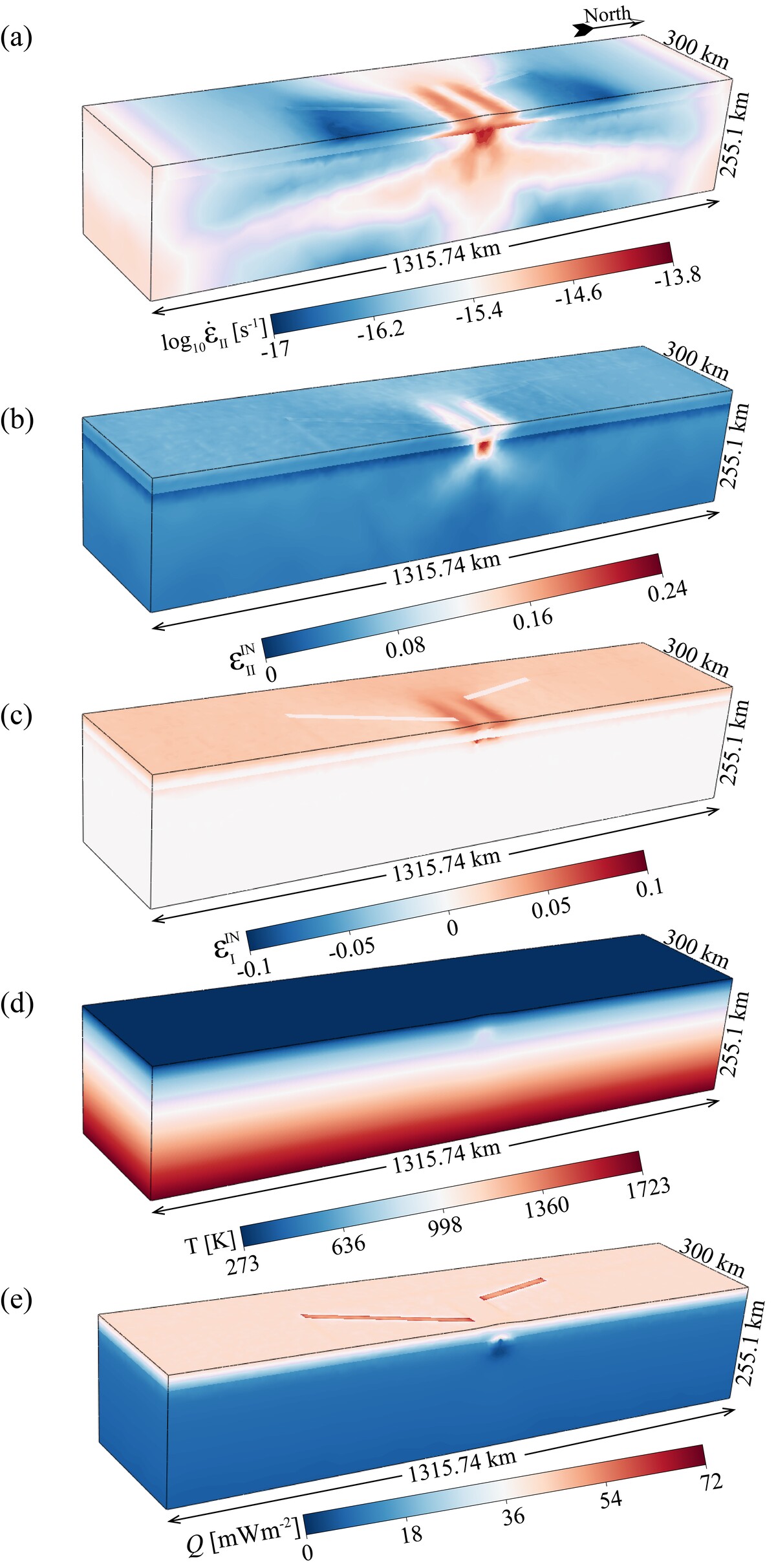}
\caption{3-D simulation results for the Eifel Volcanic Region after ${\sim}$16.85 million years of compression at a cumulative 5mm/year. (a) Logarithm of deviatoric strain rate invariant, (b) inelastic deviatoric strain invariant, (c) inelastic volumetric strain invariant, (d) evolved temperature, (e) heat flow.} 
\label{Eifel_3D}
\end{figure}
\begin{figure}[htbp]
\centering
\includegraphics[width=0.75\textwidth]{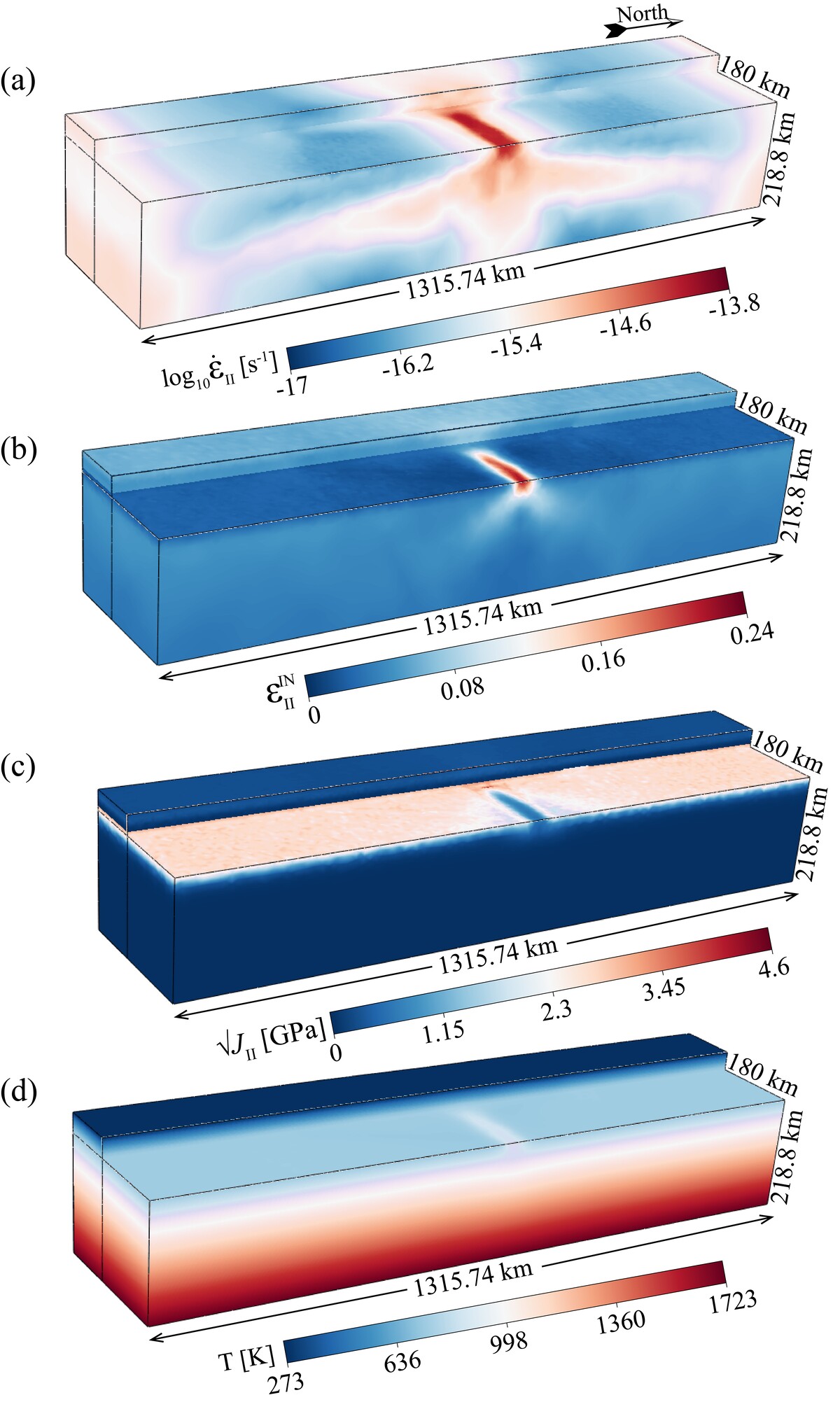}
\caption{Perspective view from the east of 3-D simulation results for the Eifel Volcanic Region after ${\sim}$16.85 million years of compression at 5mm/year, with slice near the Moho to show the extent of internal deformation. (a) Logarithm of deviatoric strain rate invariant, (b) inelastic deviatoric strain invariant, (c) deviatoric stress invariant, and (d) temperature change, (f) magnitude of heat flux. The horizontal cut at depth was made at 36.3 km from the surface.} 
\label{Eifel_3D_Perspective_from_the_East}
\end{figure}
\begin{figure}[htbp]
\centering
\includegraphics[width=0.75\textwidth]{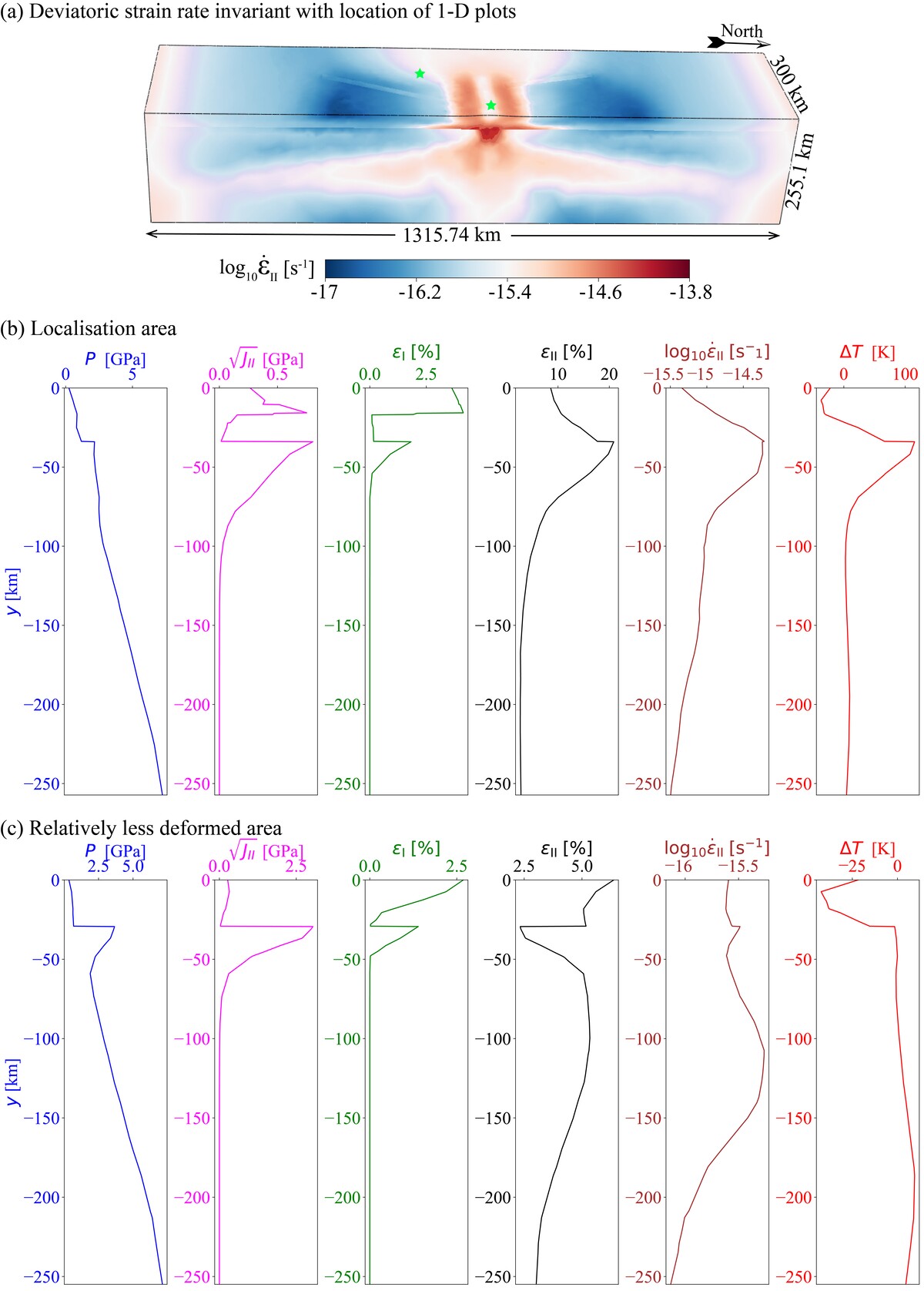}
\caption{Vertical profiles for the highly deformed region and less deformed region. (a) Logarithm of deviatoric strain rate invariant, with green stars showing the two locations in the uplifted deformed region, and less deformed region for which profiles are extracted. (b) pressure, deviatoric stress invariant, inelastic volumetric strain, inelastic deviatoric strain, deviatoric strain rate invariant and temperature change for the localisation zone, and (c) for the less deformed zone.} 
\label{Eifel_3D_Perspective_from_the_East_1D}
\end{figure}
\begin{figure}[htbp]
\centering
\includegraphics[width=\textwidth]{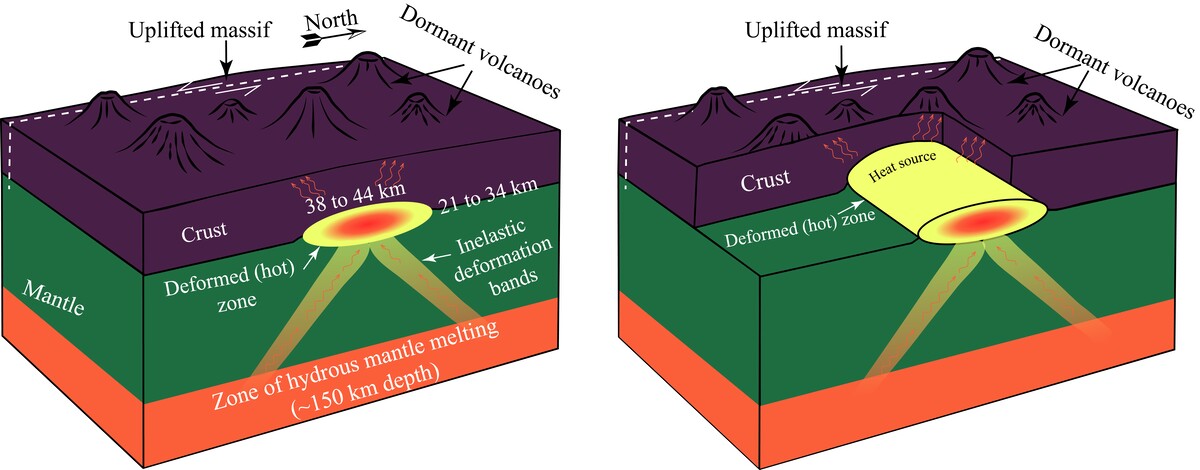}
\caption{Hypothesis for potential partial melt generation at ${\sim}$150 km depth from hydrous mantle melting, and a heat source due to deformation at the Moho. The heat source here is proposed as a cylindrical ellipsoid with ellipsoidal area oriented in the North-South direction and the longer part of the cylinder in the East-West direction. Possible melt migration direction may be along inelastic deformation bands in the mantle which intersect at the Moho and then two conjugate crustal deformation bands which introduce stress gradients and pressure changes that can enhance partial melt migration. The white lines represent the trace of the Hercynian Sillon Houiller strike-slip fault.} 
\label{melt_hypothesis}
\end{figure}
More recently, \cite{dahm2020seismological} identified a volume exhibiting low \textit{P-}wave velocity between 30 and 36 km depth, above which occurs a zone of low-magnitude (0.7 to 2.7) earthquakes with hypocenters in a vertical distribution beneath the Laacher See Volcano of the East Eifel volcanic fields. The Laacher See eruption occurred 13,000 years ago \citep{reinig2020towards}, and formed a small caldera. An interpretation of the anomalous low-velocity zone from 30 to 36 km and about 160 km wide has been made in the framework of magmatic underplating, while it has been suggested that the low-magnitude seismicity is volcanic-related \citep{dahm2020seismological}. Additionally, the Eifel Volcanic Region is undergoing ${\sim}$ 1 mm/year vertical uplift, which has been attributed to plume-related upwelling \citep{kreemer2020geodetic}. We have shown that an excess temperature of 170 K can be realised if a shortening rate of 5 mm/year induced from the Alps was sustained (\autoref{Eifel_3D}h). Moreover, the occurrence of volcano-tectonic seismicity (i.e., tectonic earthquakes occurring in a volcanic region or below a volcanic edifice) is a natural feature that emerges from our model because regional deformation has indeed localised in the zone of the thermal anomaly, whereas at shallower depths the rocks are still mostly brittle/plastic and hence seismogenic. Slow vertical uplift is also a feature of our results, as a positive topographic feature develops over the thermal anomaly during the course of the simulation, although our calculated values are quantitatively a bit less than 1 mm/yr.

The magnitude of heat flow density was found to be in excess of 60 ${\mathrm{mW\;m^{-2}}}$ in zones with intense deformation. Additionally, over the duration of the calculation, the zone of intense (localised) deformation (\autoref{Eifel_3D}e-f) was approximately geometrically and spatially stable leading to intense heating there. This heat flux is unrelated to partial melt generation, as the temperature rise did not lead to partial melting. This indicates that zones of high heat flow may not necessarily be associated with volcanism, but with local temperature changes due to deformation. These local temperature changes may also show a footprint of seismic tomography velocity perturbations and upper crustal seismicity. The maximum deformational heating occurs just below the Moho where both deviatoric stress intensity and strain rate are maximum.

\section{General Discussion of Heat Sources and Geothermal Potential of the French Massif Central and Eifel Volcanic Region}
\subsection{Potential Channels for Cenozoic Volcanism}

Driving forces for magmatic transport systems may be due to buoyancy caused by density differences between migrating melt and host rock or deformation zones, which introduce stress gradients \citep{kohlstedt2009shearing} or both. Localised deformation introduces changes in stress gradients and large-scale deformation bands which have been argued to be potential pathways for partial melt. In laboratory scales, it has been shown that shear bands influence the orientation and focusing of melt in active margins \citep{katz2006dynamics}. In a recent study, Inclined zones of localised low-velocity anomalies were imaged beneath the Laacher See Volcano in the Eifel Volcanic Region, and these zones were found to be characterised by high V\textsubscript{P}/V\textsubscript{S} ratio which have been interpreted as zones of elevated temperature or high fluid pressure which could be linked to the migration of magmatic fluids \citep{zhang2025upper}. The presence of fluids reduces the shear wave velocity which can lead to elevated V\textsubscript{P}/V\textsubscript{S} ratio. Similarly, the presence of localised deformation can also reduce the shear wave velocity through the development of fractures, damaged zones within rocks, or anisotropy, which can be pathways for fluids. 

In our conceptual model (\autoref{melt_hypothesis}), we propose a source of volatile-assisted partial melting at $\sim$150 km depth based on the temperature profile, ascending through existing inelastic deformation bands, likely pooling in a low-pressure zone and using the crustal deformation bands as preferred paths for emplacement.  Although the temperature rise of 120 K in the localised deformation zone is not substantial enough to melt a dry crust or anhydrous mantle, the solidi of lithologic heterogeneities rich in fusible components may be low enough for low degrees of partial melting to occur, possibly assisted by a local pressure minimum that our results show within the deformation zone. The pressure heterogeneity may also promote focusing and pooling of melt. In the Alps, it was reported that Oligocene magma was channeled from the base of thickened continental crust into the orogen-parallel fault system which helped to transport magma up to 40 km \citep{rosenberg2004shear}.

For the hypothesis that the origin of the latest eruptions from the FMC was the crust-mantle boundary \citep{lucazeau1984interpretation}, we can propose that localised heat sources can arise at the Moho from deformation of crust and mantle without being due to an ascending diapir or mantle plume. The major control in the deformation and associated heat sources was variation in crustal thickness, which has largely defined the FMC and Eifel Volcanic region even presently. Broad deformation zones lead to localised heat sources and changes in topography, suggesting that compressional tectonics induced by Africa-Eurasia convergence may have had an impact on the thermal footprints, deformation history, and topographic signatures of both the FMC and Eifel. Tentatively, we propose that deformational heating can contribute to overall heating of the upper mantle and the crust, to uplifting the topography, thinning the crust, could potentially contribute heat to melt parts of the mantle and the crust whose solidi are lower than the anhydrous one because of the presence of volatiles such as H\textsubscript{2}O and CO\textsubscript{2} \citep{pilet2008metasomatized} or other more fusible minor/trace components such as K\textsubscript{2}0, Na\textsubscript{2}0 consistent with the generally alkaline nature of the erupted magmas \citep{schmidt2024origin}. The colocation of volcanic and CO\textsubscript{2} vents in the Eifel region, for example, \cite{regenauer1998dilatant}, strongly suggests that a volatile-reduced solidus is appropriate. In the case of a hot and hydrous mantle at 150 km, deformation bands caused by inelastic deformation in the upper mantle may serve as channels for melt transport (\autoref{melt_hypothesis}).
\subsection{Potential Heat Sources for Cenozoic Volcanism}
It has been argued that surface heat flow density in stable continents is associated with radiogenic heat production concentrated in an enriched upper crustal layer, and that heat flow density values greater than 90 mW m\textsuperscript{-2} are typically associated with tectonically active areas possibly undergoing some crustal melting or an upper mantle that can easily deform  \citep{jaupart2007heat}. As a general matter, heat sources have been identified as radiogenic heat production within the crust and mantle, heat transfer from the earth's core, and mantle cooling \citep{morgan1985crustal,furlong2013heat}. Whether the crust melts or the upper mantle deforms, localised melting typically always involves an additional requirement, such as the addition of volatiles (mainly water and CO\textsubscript{2}) to or decompression of mantle material due to vertical ductile motions. Here, we focus on the generation of heat from inelastic deformation, which our simulations indicate has its largest contribution close to the brittle-ductile transition often referred to in shorthand as the Lithosphere-Asthenosphere boundary or LAB. The latter, of course, does move in our simulations and/or develop more complex internal structure, having a natural tendency to shallow as the temperature and deformation fields develop.

Observations and geochemistry of Cenozoic volcanism in the FMC have been taken to suggest diversity of origins. Two phases of Cenozoic eruptions were documented: pre-rift and post-rift volcanism \citep{lucazeau1984interpretation,michon2000crustal}. The pre-rift volcanism in the FMC erupted before graben opening and was composed of small volumes of undersaturated lava and may have erupted from the base of the continental lithosphere (${\sim}$150 km); while the major post-rift magmatic event was composed of lava originating from undersaturated conditions to less alkaline conditions \citep{lucazeau1984interpretation}. The most recent magmatic eruptions have been linked to the ascent of a mantle diapir - again invoking decompression melting. \cite{regenauer1998dilatant} observed the striking overlap between the surface distributions of volcanic edifices and CO\textsubscript{2} vents in the Eifel, and argued that they are related to seismogenic shear zones and plasticity, even though the work did not account for deformational heating and did not address the reason for partial melting or suggest, as we do here, that it may be part and parcel of the inelastic deformation process.

The results of our models have shown that inelastic deformation of crustal and upper mantle, especially at the base of slightly thicker crust, is an important source of heat (\autoref{FMC_3D} and Figure \autoref{Eifel_3D}). These zones concentrate not only viscous and plastic shear deformation but also volumetric (dilatant) deformation in the lithospheric domain (\autoref{FMC_3D}c and \autoref{Eifel_3D}c). This supports the view that variations in crustal thicknesses may play roles in mechanical localisation, and hence heat production and temperature changes \citep{jaupart2007heat}. The deformation localised at the base of an initially thick crust, which led to heating. The topographic uplift exposes it to erosion, which can potentially thin the crust. Explosive volcanism leading to the formation of maars, like in the Eifel or French Massif Central, is quite common and attributed to the interaction of magma with groundwater (phreatomagmatic activity). Deformation bands due to ductile mantle deformation developed, while brittle plasticity localised strain at the crust-mantle boundary and in the upper crust. Permanent brittle damage in the shallow crust should enhance permeability to groundwater due to fractured brittle crust.

Magmas in both FMC and Eifel are generally rich in alkalis which are typical relatively low melting point components in both silicate and carbonatite melts \citep{schmidt2024origin}. Alkali basalts have been inferred to originate from decarbonisation reactions caused by pressure reductions due to (hot) mantle diapirism \citep{berger1979role}. The magmas were not generated in the presence of water, but CO\textsubscript{2}, and in the Rhine graben we have the Kaiserstuhl example of a carbonatite magma. While it is far from simple
to propose a representative solidus for such volatile-rich exotic magma compositions, something on the order of 1000 K and relatively insensitive to pressure in the range  $\sim$ 0.5 - 4 GPa \citep{schmidt2024origin} seems reasonable. Our results offer a new perspective on the possible lithospheric origin of carbonatitic and associated alkaline magmas as discussed by \citep{schmidt2024origin}, who nevertheless assume a lithostatic pressure equal to hydrostatic pressure. From our results, a heated zone was generated in the aforementioned pressure range, but we found that pressure is not a simple proportionality with depth, nor constant at any given depth throughout the deformation cycle. The coincidence in surface distribution of volcanic and CO\textsubscript{2} vents in the Eifel noted by \cite{regenauer1998dilatant} hints at the release of CO\textsubscript{2} by dolomite breakdown in some depth range as laid out in \cite{schmidt2024origin} and would also constrain some of the P-T-t curves that our results point towards.

\subsection{Source of High Heat Flow Anomalies in the French Massif Central and Eifel Volcanic Region}
 A key observation presently from the French Massif Central and Eifel Volcanic region is a halo of SSW-NNE trending high heat flow density from the South of France, terminating near the graben junction in the Eifel volcanic region. While the heat flow is high for an intraplate region, the volcanic flux in the two regions is insufficient to explain the observations. Taking orders of magnitude for the Eifel to illustrate this point: a cumulative volume of $\sim$15 km\textsuperscript{3} erupted since $\sim$700,000 years \citep{ritter2001mantle}. If we take the amount of heat required to melt this amount of material as a measure of that which the magma transports towards the surface, and which can potentially heat the crust, using latent heat $\sim$10\textsuperscript{6} J kg\textsuperscript{-1}, gives a total energy of $\sim 4.5 \times 10\textsuperscript{19}$ J. Spreading this energy over 700,000 years and over an area $\sim 40\times40$ km, i.e., ${\sim 1.6\; \times}$ 10\textsuperscript{9} m\textsuperscript{2} is equivalent to a surface heat flux of $\sim$ 1 mW m\textsuperscript{-2}. These numbers do not support magma as a major transporter of heat to fuel the crustal geothermal system. Invoking two orders of magnitude more intrusive magma to fuel a geothermal system is always possible but not very plausible in a province where low to very low degrees of partial melting are the norm. A common explanation for the high heat flow density, hitherto, has been a short-lived thermal mantle diapir rising from 150 km depth since ${\sim}$34 million years, since crustal heat production was ruled out as the sole explanation for the high heat flux observation \citep{lucazeau1984interpretation}.

The high heat flux reported in the FMC \citep{vcermak1979heat,lucazeau1984interpretation,lucazeau1989heat}
has been broken down as the aggregate of radiogenic crustal heat production \citep{vigneresse1987heat} of the order of 50 to 60 mW m\textsuperscript{-2}, with a potential localised mantle contribution of 25 - 30 mW m\textsuperscript{-2} in the proximity of the Cenozoic grabens and volcanic areas attributed to a transient mantle diapir \citep{lucazeau1984interpretation,lucazeau1989heat}, and 35 mW m\textsuperscript{-2} in areas without Cenozoic rifts or grabens, like Brittany in the stable Hercynian Paris Basin, Northern France \citep{lucazeau1984interpretation, lucazeau1989heat}. In Central Europe, anomalous heat flux (${>}$70 mW m\textsuperscript{-2}) has been correlated with areas with thick crust, where positive correlations were particularly strong for tectonically-induced thicker crustal domains \citep{bodri1985correlation}.

In our simulations, we included radiogenic heat as a contributor to crustal heat production (Equation \ref{eqn:thermal}) as well as deformational heating. Surface heat flow exceeded 65 mW m\textsuperscript{-2} near the specified Cenozoic grabens and fault systems (Figure \ref{temperature_heatflow}c and Figure \ref{Eifel_3D}d). Based on similar heat flux values in the Eifel volcanic region and the timing of the volcanism and Oligocene grabens, we therefore suggest that dominantly deformational heating (which exceeded radiogenic heat production) may account for most of the elevated surface heat flux on the graben edges and faults. Arguably, additional heat flux that has led to the hypothesis of a mantle thermal diapir could in fact be attributed to a localised deformational feature. This could also be a reason why there is a gap in shear wave velocity anomalies, which has led to the interpretation that \enquote{hanging} mantle baby plumes do not appear to show a connection with deeper super plumes \citep{budweg2006eifel}.
 
\subsection{Geothermal Energy Budget for the French Massif Central and Eifel Volcanic Region}
Areas that were volcanically active in recent geological history are proposed as reasonable targets for geothermal exploration, but the technology for assessing the heat stored in place is not at a practical stage hitherto \citep{glassley2014geothermal}. Moreover, important questions about the level of risk to infrastructure in the event of renewed volcanic activity are difficult to answer as long as fundamental questions about the energy sources and underlying geodynamic processes remain open.  While currently inactive or extinct volcanoes that have undergone their eruptive phase in recent geological history may be associated with intrusions or even the possible formation of a caldera, the attendant heat source is not extinguished rapidly due to the low thermal conductivity of crustal rocks. For example, the Larderello geothermal system, which currently accounts for 10\% of global geothermal energy supply, is thought to have a granitic intrusion as a heat source at ${\sim}$3 km to 6 km depth which supplies heat into aquifers which in turn expel supercritical fluids as hot springs  \citep{gherardi2025geochemistry,rochira2018regional,minissale1991larderello}. Nevertheless, volcanism is but one manifestation of the geological activity that underlies the current state of the FMC. If one takes the view that magma is the essential heat provider in a geothermal system, then its role must be seen as one of advecting heat into the crust from the region in the mantle undergoing partial melting. In analysing the results of our simulations, we have shown that the attendant tectonic activity, a very important part of the FMC's history, is ultimately also a considerable source of heat. The parts of France where the FMC is located were said to be in a state of compression and moderate seismicity \citep{masson2019extracting,pina20223d,ferhat1998geodetic,baroux2001analyses,sylvander2021seismicity}. Indeed, localised deformational heating has dissipated enough energy to heat up what was initially lithospheric mantle to a temperature somewhere between its anhydrous solidus and its plausible hydrous and/or CO\textsubscript{2}-depressed solidus. That is, small amounts of the most fusible material will melt, but high amounts of partial melting would not be expected. In this view, the deformational heating is the cause of melting and so also the ultimate heat source feeding an overlying geothermal system. Magma is the symptom rather than the cause. Indeed, much of the heat advected by the magma that is produced is erupted to the surface, and therefore, its heat does not, in fact, feed the geothermal system; only a possible intrusive counterpart to the volcanic flux can be invoked, but this is hard to quantify. Finally, we observe that invoking a large volume of magma intruded into the crust is not particularly plausible given that volcanism in the FMC and Eifel is essentially a sparse collection of monogenetic edifices. As is well documented, moreover, many lavas forming these edifices bear mantle xenoliths which is more consistent with rapid uninterrupted magma ascent than intrusive crustal storage. Presently, mantle-derived volatile fluxes have been documented for the Eifel region \citep{aeschbach1996quantification, brauer2013indications}, indicating that magmatic activity may be associated with mantle volatiles in the Laacher See maar lake in East Eifel \citep{zhang2026seismological}. 

Similar to petroleum exploration,  a complete geothermal system requires several essential elements: a source rock and a reservoir rock. Additionally, a network of fractures for fluid injection and extraction is essential for geothermal exploration. Common sources of geothermal systems are wet and dry steam, hot brine and hot rocks \citep{shepherd2014energy}. 

In estimating the geothermal potential due to deformational heating, we utilise a conceptual sketch shown in \autoref{melt_hypothesis} where dimensions have been extracted based on the thermal anomaly shown in \autoref{FMC_3D}e and \autoref{Eifel_3D}e. The shape is an elongated cylindrical ellipsoid (with volume ${V_D}$) inferred from the thermal anomaly. We observed stresses of the order of 0.1 to 1 GPa, and inelastic strain rates of 10\textsuperscript{-14} to 10\textsuperscript{-16} s\textsuperscript{-1} within the deformed zone. Using the formula of ${P_D=V_D.\sigma.\dot{\varepsilon}^{\mathrm{IN}}}$ to obtain potential geothermal power of 1.5675 MW to 2.938 GW. This leads to 13.73 GW h and 25.74 TWh, respectively, for one year. Using an average consumption of 22.982 MWh for a typical house in rural France \citep{belaid2016understanding}, the potential contribution from geothermal energy at source can supply energy to 598 and 1.12 million rural houses. Uncertainties can arise from the size of the anomalies, heat conduction efficiency, and energy conversion efficiency. Assuming only 10\% of the heat in place is recoverable, this works out to powering between 60 houses and 112,000 houses.

It is currently impossible to drill to access heat sources at depths where heat sources are present. However, heat in place can be conducted to colder regions, and the presence of faults, which characterise the FMC and Eifel, can serve as permeability channels for fluids and mining geothermal heat. The framework presented here inherently involves spatial association of a deep heat source with brittle deformation at shallower depths to produce permeability in the upper crust.

While we take cognizance of the current technological limitations to access geothermal resources at the scale of the FMC Moho (${>}$25 km), our recommendation is to identify areas with thick crust and low absolute velocities or velocity anomalies from seismic tomography, dormant volcanoes on uplifted massifs from observational geology, high surface high heat flow and deep fault systems like the Sillon Houiller strike slip fault and the Great Limagne fault for permeability zones for fluid circulation.  These areas are suitable for drilling targets for geothermal exploration. The heat in place may be sufficient to heat the overlying crust or aquifers, where steam can be extracted for powering geothermal plants.
\section{Conclusions and Recommendations}\label{Conclusions}
Our numerical implementation of a 3-D constitutive update for the FMC and Eifel volcanic region has enabled us to explain some of the observations in the two regions. Based on our studies, we conclude as follows:
\begin{enumerate}
\item Imposing inward velocities from the north and south to mimic the far-field shortening rate due to the Africa-Eurasia convergence and using a local variation of crustal thickness, our models showed localisation of inelastic strain at the base of initially thickened crust more than 300 km from the boundaries. The localised inelastic strain led to deformational heating, which raised the temperature locally by over 120 K at below the Moho. The localisation zones were at the base of the thicker crust and were associated with a topographic bulge, whose shoulders were formed by viscoplastic deformation bands. Although deformational heating has not been a popular source of partial melt generation, which has been modelled by decompression or flux melting, we have not attained partial melt here either. However, the local temperature increase can be a potential source of heat for partial melt generation near the Moho, if we take into account the likely reduction of the solidus due to the presence of volatiles and other fusible components, rather than the strictly anhydrous solidus that we have taken into account in our calculations.
\item The unusually high surface heat flow on the Cenozoic grabens in the Eifel and French Massif Central (${>70 \;\mathrm{mW/m^2}}$) can be explained by the contribution from deformational heating near the crust-mantle boundary even in the absence of partial melting, and radiogenic heat generation. This observation arose from areas where the highest heat flow was localised in the region with significant inelastic deformation, even in the absence of melt flux.
\item We recommend zones of tectonic terminations, like graben junctions; increased crustal thickness and high surface heat flow as sites for potential targets for geothermal exploration. Additionally, the presence of faults, especially deep-penetration faults like the Great Limagne Fault or the Sillon Houiller Fault, can be investigated as permeability zones for deep geothermal studies. Para-seismic protection of permanent infrastructure is clearly appropriate and dependent on the intensity of deformation.
\item While we do not preclude upper mantle flow processes contributing heat sources and partial melt, we propose that a mantle plume (if present) may not be the only mechanism to supply heat and cause local high heat flow density. Without including a buoyant mantle plume, we obtained in our models a heat flow density comparable with that observed, with deformation zones contributing more than 50 mW m\textsuperscript{-2}. The deformation bands of areas with thicker crust mimic plume-heads, which can locally raise the temperature and uplift the topography and potentially supply enough heat for partial melting.
\item Our simulations have shown that uplifted areas are associated with a thermal anomaly if the crust is initially thick. It is hypothesised that sites for possible future eruptions may be areas currently undergoing uplift.
\end{enumerate}
\section*{Acknowledgements}
This work was supported by AXA Research Fund, for which EM is a beneficiary. We acknowledge partial support from the European Research Council grant PERSISMO (grant number 865411). Computing facilities and foundational funding support were also provided by the Institut Physique du Globe de Paris (IPGP), INTERREG V Cara\"{i}bes program, the European Regional Development Fund (FEDER) through the PREST project. We acknowledge Val{\'e}rie Clouard for her support. EM appreciates the directorate of the G\'{e}oscience Environment Toulouse/Observatoire Midi-Pyr\'{e}n\'{e}es (GET/OMP) for supporting and hosting EM during his AXA fellowship.


\setcounter{section}{0}
\renewcommand\thesection{Appendix \Alph{section}}
\renewcommand\thesubsection{\Alph{section}\arabic{subsection}}
\renewcommand{\theequation}{\Alph{section}\arabic{equation}}
\setcounter{equation}{0}
\renewcommand{\thefigure}{\Alph{section}\arabic{figure}}
\setcounter{figure}{0}

\renewcommand{\theHsection}{appendix.\Alph{section}}
\renewcommand{\theHsubsection}{appendix.\Alph{section}.\arabic{subsection}}
\renewcommand{\theHequation}{appendix.\Alph{section}.\arabic{equation}}

\section{Synthetic case for potential geothermal output}
Using observations of orders of magnitude of stresses and inelastic strain rates from our simulations, we investigated two scenarios: (1) one for which stresses increase with inelastic strain rates, and (2) one for which stresses reduce with inelastic strain rates. Using the mean semi-major and semi-minor axes and a length of 250 km for the ellipsoidal cylinder, we estimate potential power output and the upper and lower bounds for the two cases. We also used a 100 km-long ellipsoidal cylinder to estimate the power output. These estimates are shown in \autoref{powergeneration}. The powers were converted to power-hours per year, and using a mean consumption of 22,982 kWh with a standard deviation of 13,256 kWh per year for a French rural house \citep{belaid2016understanding}, we estimated the number of rural houses that could be supplied, assuming the heat in place can be extracted. 
\begin{figure}[htbp]
\centering{\includegraphics[width=\textwidth]{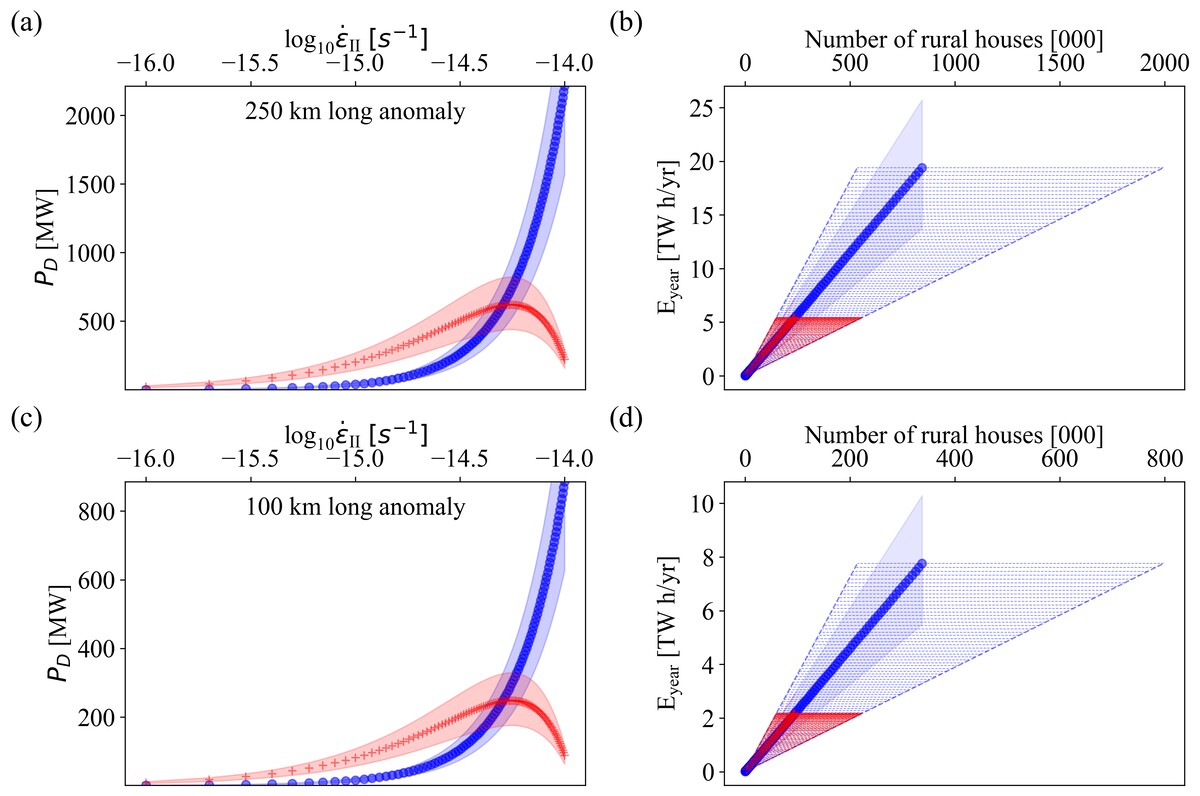}}
\caption{(a) Power generation for an ellipsoidal cylinder of dimensions 19 km and 10.5 km for semi-major and semi-minor axes, respectively  (red plot), and 22 km and 17 km for semi-major and semi-minor axes, respectively  (blue plot),(b) estimated power hours supplied per year for a mean consumption of 22,982 kW h for a standard deviation of 13,256 kW h for a typical French rural house. Both (a) and (b) used a 250 km long ellipsoidal cylinder to calculate the volume for the mean, upper and lower bounds, which represent the geometrical uncertainties from the deformed zone. The uncertainties in (b) are from the standard deviations of the power consumption in a rural house. (c) and (d) are similar to (a) and (b), except that we have used a 100 km long ellipsoidal cylinder.}
\label{powergeneration}
\end{figure}

\clearpage
\renewcommand*{\bibfont}{\scriptsize}
\printbibliography

@book{glassley2014geothermal,
  title={Geothermal energy: renewable energy and the environment},
  author={Glassley, William E},
  year={2014},
  publisher={CRC press}
}

@article{brauer2013indications,
  title={Indications for the existence of different magmatic reservoirs beneath the Eifel area (Germany): A multi-isotope (C, N, He, Ne, Ar) approach},
  author={Br{\"a}uer, Karin and K{\"a}mpf, Horst and Niedermann, Samuel and Strauch, Gerhard},
  journal={Chemical Geology},
  volume={356},
  pages={193--208},
  year={2013},
  publisher={Elsevier}
}

@article{reinig2020towards,
  title={Towards a dendrochronologically refined date of the Laacher See eruption around 13,000 years ago},
  author={Reinig, Frederick and Cherubini, Paolo and Engels, Stefan and Esper, Jan and Guidobaldi, Giulia and J{\"o}ris, Olaf and Lane, Christine and Nievergelt, Daniel and Oppenheimer, Clive and Park, Cornelia and others},
  journal={Quaternary Science Reviews},
  volume={229},
  pages={106128},
  year={2020},
  publisher={Elsevier}
}

@article{rosenberg2004shear,
  title={Shear zones and magma ascent: A model based on a review of the Tertiary magmatism in the Alps},
  author={Rosenberg, CL},
  journal={Tectonics},
  volume={23},
  number={3},
  year={2004},
  publisher={Wiley Online Library}
}

@book{perrier1973structure,
  title={Structure profonde du Massif Central fran{\c{c}}ais},
  author={Perrier, G and Ruegg, JC},
  year={1973}
}

@article{li2026intraplate,
  title={Intraplate volcanism driven by slab-plume interaction: Numerical modeling and its application to the Eifel, Massif Central and Hainan volcanic areas},
  author={Li, Yingying and Steinberger, Bernhard and Brune, Sascha and Le Breton, Eline and Glerum, Anne and Pons, Micha{\"e}l},
  journal={Journal of Geophysical Research: Solid Earth},
  volume={131},
  number={1},
  pages={e2025JB032799},
  year={2026},
  publisher={Wiley Online Library}
}

@article{coisy1978regional,
  title={Regional structure and geodynamics of the upper mantle beneath the Massif Central},
  author={Coisy, Ph and Nicolas, A},
  journal={Nature},
  volume={274},
  number={5670},
  pages={429--432},
  year={1978},
  publisher={Nature Publishing Group UK London}
}

@article{berger1979role,
  title={The role on partial melting of mantle diapirism, CO2 and H2O from the study of lherzolite nodules of intracontinental alkali basalts: example of the French Massif Central},
  author={Berger, E},
  journal={Physics and Chemistry of the Earth},
  volume={11},
  pages={619--629},
  year={1979},
  publisher={Elsevier}
}

@article{leloup1999shear,
  title={Shear heating in continental strike-slip shear zones: model and field examples},
  author={Leloup, Philippe Herv{\'e} and Ricard, Yannick and Battaglia, Jean and Lacassin, Robin},
  journal={Geophysical Journal International},
  volume={136},
  number={1},
  pages={19--40},
  year={1999},
  publisher={Blackwell Publishing Ltd Oxford, UK}
}

@article{regenauer1998rapid,
  title={Rapid conversion of elastic energy into plastic shear heating during incipient necking of the lithosphere},
  author={Regenauer-Lieb, Klaus and Yuen, David A},
  journal={Geophysical Research Letters},
  volume={25},
  number={14},
  pages={2737--2740},
  year={1998},
  publisher={Wiley Online Library}
}

@article{jaquet2018spontaneous,
  title={Spontaneous ductile crustal shear zone formation by thermal softening and related stress, temperature and strain rate evolution},
  author={Jaquet, Yoann and Schmalholz, Stefan M},
  journal={Tectonophysics},
  volume={746},
  pages={384--397},
  year={2018},
  publisher={Elsevier}
}

@article{burg2005role,
  title={The role of viscous heating in Barrovian metamorphism of collisional orogens: thermomechanical models and application to the Lepontine Dome in the Central Alps},
  author={Burg, J-P and Gerya, TV},
  journal={Journal of Metamorphic Geology},
  volume={23},
  number={2},
  pages={75--95},
  year={2005},
  publisher={Wiley Online Library}
}

@article{merle2001formation,
  title={The formation of the West European Rift; a new model as exemplified by the Massif Central area},
  author={Merle, Olivier and Michon, Laurent},
  journal={Bulletin de la Societ{\'e} g{\'e}ologique de France},
  volume={172},
  number={2},
  pages={213--221},
  year={2001},
  publisher={Societe Geologique de France}
}

@article{katz2006dynamics,
  title={The dynamics of melt and shear localization in partially molten aggregates},
  author={Katz, Richard F and Spiegelman, Marc and Holtzman, Benjamin},
  journal={Nature},
  volume={442},
  number={7103},
  pages={676--679},
  year={2006},
  publisher={Nature Publishing Group UK London}
}

@article{kohlstedt2009shearing,
  title={Shearing melt out of the Earth: An experimentalist's perspective on the influence of deformation on melt extraction},
  author={Kohlstedt, David L and Holtzman, Benjamin K},
  journal={Annual Review of Earth and Planetary Sciences},
  volume={37},
  number={1},
  pages={561--593},
  year={2009},
  publisher={Annual Reviews}
}

@article{zhang2026seismological,
  title={Seismological evidence for fluid accumulation beneath the dormant East Eifel volcanic region, Germany},
  author={Zhang, Hao and Dahm, Torsten and B{\"u}y{\"u}kakp{\i}nar, P{\i}nar and Cesca, Simone and Isken, Marius Paul and Laumann, Patrick and Gei{\ss}ler, Wolfram Hartmut},
  journal={Communications Earth \& Environment},
  volume={7},
  number={1},
  pages={571},
  year={2026},
  publisher={Nature Publishing Group UK London}
}

@article{aeschbach1996quantification,
  title={Quantification of gas fluxes from the subcontinental mantle: The example of Laacher See, a maar lake in Germany},
  author={Aeschbach-Hertig, Werner and Kipfer, R and Hofer, M and Wieler, R and Signer, P and others},
  journal={Geochimica et cosmochimica acta},
  volume={60},
  number={1},
  pages={31--41},
  year={1996},
  publisher={Elsevier}
}

@article{belaid2016understanding,
  title={Understanding the spectrum of domestic energy consumption: Empirical evidence from France},
  author = {Belaïd, F.},
  journal={Energy Policy},
  volume={92},
  pages={220--233},
  year={2016},
  publisher={Elsevier}
}

@article{poutanen2020geodesist,
  title={The geodesist’s handbook 2020},
  author={Poutanen, Markku and R{\'o}zsa, Szabolcs},
  journal={Journal of Geodesy},
  volume={94},
  number={11},
  pages={109},
  year={2020},
  publisher={Springer}
}

@article{rochira2018regional,
  title={Regional thermo-rheological field related to granite emplacement in the upper crust: implications for the Larderello area (Tuscany, Italy)},
  author={Rochira, Federica and Caggianelli, Alfredo and de Lorenzo, Salvatore},
  journal={Geodinamica Acta},
  volume={30},
  number={1},
  pages={225--240},
  year={2018},
  publisher={Taylor \& Francis}
}

@article{minissale1991larderello,
  title={The Larderello geothermal field: a review},
  author={Minissale, Angelo},
  journal={Earth-Science Reviews},
  volume={31},
  number={2},
  pages={133--151},
  year={1991},
  publisher={Elsevier}
}

@article{gherardi2025geochemistry,
  title={Geochemistry of geothermal well gases from superhot zones of Larderello, Italy},
  author={Gherardi, Fabrizio and Magro, Gabriella},
  journal={Applied Geochemistry},
  pages={106480},
  year={2025},
  publisher={Elsevier}
}

@article{drucker1952soil,
  title={Soil mechanics and plastic analysis or limit design},
  author={Drucker, Daniel Charles and Prager, William},
  journal={Quarterly of Applied Mathematics},
  volume={10},
  number={2},
  pages={157--165},
  year={1952}
}

@article{vigneresse1987heat,
  title={Heat flow, heat production and granite depth in western France},
  author={Vigneresse, JL and Jolivet, J and Cuney, M and Bienfait, G},
  journal={Geophysical Research Letters},
  volume={14},
  number={3},
  pages={275--278},
  year={1987},
  publisher={Wiley Online Library}
}

@article{njinju2023instantaneous,
  title={Instantaneous 3{D} tomography-based convection beneath the {R}ungwe {V}olcanic {P}rovince, {E}ast {A}frica: implications for melt generation},
  author={Njinju, Emmanuel A and Stamps, D Sarah and Atekwana, Estella A and Rooney, Tyrone O and Rajaonarison, Tahiry A},
  journal={Geophysical Journal International},
  volume={235},
  number={1},
  pages={296--311},
  year={2023},
  publisher={Oxford University Press}
}

@incollection{vcermak1979heat,
  title={Heat flow map of {E}urope},
  author={{\v{C}}erm{\'a}k, Vladim{\'\i}r},
  booktitle={Terrestrial heat flow in {E}urope},
  pages={3--40},
  year={1979},
  publisher={Springer}
}

@article{saunders2007regional,
  title={Regional uplift associated with continental large igneous provinces: The roles of mantle plumes and the lithosphere},
  author={Saunders, AD and Jones, SM and Morgan, Lisa A and Pierce, Kenneth Lee and Widdowson, M and Xu, YG},
  journal={Chemical Geology},
  volume={241},
  number={3-4},
  pages={282--318},
  year={2007},
  publisher={Elsevier}
}

@article{cloetingh1999lithosphere,
  title={Lithosphere folding: Primary response to compression? (from {C}entral {A}sia to {P}aris basin)},
  author={Cloetingh, SAPL and Burov, E and Poliakov, A},
  journal={Tectonics},
  volume={18},
  number={6},
  pages={1064--1083},
  year={1999},
  publisher={Wiley Online Library}
}

@article{masson2019extracting,
  title={Extracting small deformation beyond individual station precision from dense Global Navigation Satellite System {(GNSS)} networks in {F}rance and {W}estern {E}urope},
  author={Masson, Christine and Mazzotti, Stephane and Vernant, Philippe and Doerflinger, Erik},
  journal={Solid Earth},
  volume={10},
  number={6},
  pages={1905--1920},
  year={2019},
  publisher={Copernicus Publications G{\"o}ttingen, Germany}
}

@article{babeyko2008high,
  title={High-resolution numerical modeling of stress distribution in visco-elasto-plastic subducting slabs},
  author={Babeyko, AY and Sobolev, Stephan V},
  journal={Lithos},
  volume={103},
  number={1-2},
  pages={205--216},
  year={2008},
  publisher={Elsevier}
}

@incollection{sobolev1997temperature,
  title={Temperature and dynamics of the upper mantle beneath the {F}rench {M}assif {C}entral},
  author={Sobolev, Stephan V and Babeyko, A Yu and Christensen, U and Granet, M},
  booktitle={Upper mantle heterogeneities from active and passive seismology},
  pages={269--275},
  year={1997},
  publisher={Springer}
}

@article{shapiro2025deep,
  title={Deep long period earthquakes beneath volcanoes of the {F}rench {M}assif {C}entral},
  author={Shapiro, Nikolai M and Aubert, Coralie and Chevrot, Sebastien and Mordret, Aur{\'e}lien and Paul, Anne and Sylvander, Matthieu and Laumonier, Mickael and Boudoire, Guillaume and Laporte, Didier and Cluzel, Nicolas and others},
  journal={Geophysical Research Letters},
  volume={52},
  number={12},
  pages={e2025GL114904},
  year={2025},
  publisher={Wiley Online Library}
}

@article{pina20223d,
  title={3{D GNSS} velocity field sheds light on the deformation mechanisms in {E}urope: Effects of the vertical crustal motion on the distribution of seismicity},
  author={Pi{\~n}a-Vald{\'e}s, Jes{\'u}s and Socquet, Anne and Beauval, C{\'e}line and Doin, Marie-Pierre and D’Agostino, Nicola and Shen, Zheng-Kang},
  journal={Journal of Geophysical Research: Solid Earth},
  volume={127},
  number={6},
  pages={e2021JB023451},
  year={2022},
  publisher={Wiley Online Library}
}

@article{granet1995imaging,
  title={Imaging a mantle plume beneath the {F}rench {M}assif {C}entral},
  author={Granet, Michel and Wilson, Marjorie and Achauer, Ulrich},
  journal={Earth and Planetary Science Letters},
  volume={136},
  number={3-4},
  pages={281--296},
  year={1995},
  publisher={Elsevier}
}

@article{granet1995massif,
  title={Massif {C}entral ({F}rance): new constraints on the geodynamical evolution from teleseismic tomography},
  author={Granet, M and Stoll, G and Dorel, J and Achauer, U and Poupinet, G and Fuchs, K},
  journal={Geophysical Journal International},
  volume={121},
  number={1},
  pages={33--48},
  year={1995},
  publisher={Blackwell Publishing Ltd Oxford, UK}
}

@article{momoh2025volumetric,
	title={Volumetric (dilatant) plasticity in geodynamic models and implications on thermal dissipation and strain localization},
	author={Momoh, Ekeabino and Bhat, Harsha S and Tait, Stephen and Gerbault, Muriel},
	journal={Geophysical Journal International},
	volume={240},
	number={3},
	pages={1551--1578},
	year={2025},
	publisher={Oxford Academic}
}

@article{vermeer1984non,
  title={Non-associated plasticity for soils, concrete and rock},
  author={Vermeer, Pieter A and de Borst, R},
  journal={HERON, 29 (3), 1984},
  year={1984},
  publisher={Delft University of Technology}
}

@article{white1995mantle,
  title={Mantle plumes and flood basalts},
  author={White, RS and McKenzie, D},
  journal={Journal of Geophysical Research: Solid Earth},
  volume={100},
  number={B9},
  pages={17543--17585},
  year={1995},
  publisher={Wiley Online Library}
}

@article{richards1989flood,
  title={Flood basalts and hot-spot tracks: plume heads and tails},
  author={Richards, Mark A and Duncan, Robert A and Courtillot, Vincent E},
  journal={Science},
  volume={246},
  number={4926},
  pages={103--107},
  year={1989},
  publisher={American Association for the Advancement of Science}
}

@article{hill1993mantle,
  title={Mantle plumes and continental tectonics},
  author={Hill, Robert I},
  journal={Lithos},
  volume={30},
  number={3-4},
  pages={193--206},
  year={1993},
  publisher={Elsevier}
}

@article{dongmo2023imaging,
  title={Imaging deep-mantle plumbing beneath {L}a {R}{\'e}union and {C}omores hot spots: Vertical plume conduits and horizontal ponding zones},
  author={Dongmo Wamba, Mathurin and Montagner, Jean-Paul and Romanowicz, Barbara},
  journal={Science Advances},
  volume={9},
  number={4},
  pages={eade3723},
  year={2023},
  publisher={American Association for the Advancement of Science}
}

@article{davies2013global,
  title={Global map of solid Earth surface heat flow},
  author={Davies, J Huw},
  journal={Geochemistry, Geophysics, Geosystems},
  volume={14},
  number={10},
  pages={4608--4622},
  year={2013},
  publisher={Wiley Online Library}
}

@book{turcotte2002geodynamics,
  title={Geodynamics},
  author={Turcotte, Donald L and Schubert, Gerald},
  year={2002},
  publisher={Cambridge university press}
}

@article{sylvander2021seismicity,
  title={Seismicity patterns in {S}outhwestern {F}rance},
  author={Sylvander, Matthieu and Rigo, Alexis and S{\'e}n{\'e}chal, Guy and Battaglia, Jean and Benahmed, S{\'e}bastien and Calvet, Marie and Chevrot, S{\'e}bastien and Douchain, Jean-Michel and Grimaud, Frank and Letort, Jean and others},
  journal={Comptes Rendus. G{\'e}oscience},
  volume={353},
  number={S1},
  pages={79--104},
  year={2021}
}

@article{baroux2001analyses,
  title={Analyses of the stress field in southeastern {F}rance from earthquake focal mechanisms},
  author={Baroux, Emmanuel and B{\'e}thoux, Nicole and Bellier, Olivier},
  journal={Geophysical Journal International},
  volume={145},
  number={2},
  pages={336--348},
  year={2001},
  publisher={Blackwell Publishing Ltd Oxford, UK}
}

@article{ferhat1998geodetic,
  title={Geodetic measurement of tectonic deformation in the southern {A}lps and {P}rovence, {F}rance, 1947--1994},
  author={Ferhat, Gilbert and Feigl, Kurt L and Ritz, Jean-Fran{\c{c}}ois and Souriau, Annie},
  journal={Earth and Planetary Science Letters},
  volume={159},
  number={1-2},
  pages={35--46},
  year={1998},
  publisher={Elsevier}
}

@article{hirschmann2000mantle,
  title={Mantle solidus: Experimental constraints and the effects of peridotite composition},
  author={Hirschmann, Marc M},
  journal={Geochemistry, Geophysics, Geosystems},
  volume={1},
  number={10},
  year={2000},
  publisher={Wiley Online Library}
}

@article{katz2003new,
  title={A new parameterization of hydrous mantle melting},
  author={Katz, Richard F and Spiegelman, Marc and Langmuir, Charles H},
  journal={Geochemistry, Geophysics, Geosystems},
  volume={4},
  number={9},
  year={2003},
  publisher={Wiley Online Library}
}

@article{jacquey2020multiphysics,
  title={Multiphysics modeling of a brittle-ductile lithosphere: 1. Explicit visco-elasto-plastic formulation and its numerical implementation},
  author={Jacquey, Antoine B and Cacace, Mauro},
  journal={Journal of Geophysical Research: Solid Earth},
  volume={125},
  number={1},
  pages={e2019JB018474},
  year={2020},
  publisher={Wiley Online Library}
}

@article{gere2009mechanics,
  title={Mechanics of materials. Cengage learning},
  author={Gere, James M and Goodno, Barry J},
  journal={Inc.: Independence, KY},
  year={2009}
}

@article{leroy1989finite,
  title={Finite element analysis of strain localization in frictional materials},
  author={Leroy, Y and Ortiz, M},
  journal={International Journal for Numerical and Analytical Methods in Geomechanics},
  volume={13},
  number={1},
  pages={53--74},
  year={1989},
  publisher={Wiley Online Library}
}

@article{leroy1990finite,
  title={Finite element analysis of transient strain localization phenomena in frictional solids},
  author={Leroy, Y and Ortiz, M},
  journal={International Journal for Numerical and Analytical Methods in Geomechanics},
  volume={14},
  number={2},
  pages={93--124},
  year={1990},
  publisher={Wiley Online Library}
}

@article{njinju2021lithospheric,
  title={Lithospheric control of melt generation beneath the {R}ungwe volcanic province, {E}ast {A}frica: Implications for a plume source},
  author={Njinju, Emmanuel A and Stamps, D Sarah and Neumiller, Kodi and Gallager, James},
  journal={Journal of Geophysical Research: Solid Earth},
  volume={126},
  number={5},
  pages={e2020JB020728},
  year={2021},
  publisher={Wiley Online Library}
}

@article{dannberg2019new,
  title={A new formulation for coupled magma/mantle dynamics},
  author={Dannberg, Juliane and Gassm{\"o}ller, Rene and Grove, Ryan and Heister, Timo},
  journal={Geophysical Journal International},
  volume={219},
  number={1},
  pages={94--107},
  year={2019},
  publisher={Oxford University Press}
}

@article{ball2022coupled,
  title={A coupled geochemical-geodynamic approach for predicting mantle melting in space and time},
  author={Ball, PW and Duvernay, T and Davies, DR},
  journal={Geochemistry, Geophysics, Geosystems},
  volume={23},
  number={4},
  pages={e2022GC010421},
  year={2022},
  publisher={Wiley Online Library}
}

@article{jaupart2007heat,
  title={Heat flow and thermal structure of the lithosphere},
  author={Jaupart, C and Mareschal, JC and Schubert, G},
  journal={Treatise on geophysics},
  volume={6},
  pages={217--252},
  year={2007}
}

@article{furlong2013heat,
  title={Heat flow, heat generation, and the thermal state of the lithosphere},
  author={Furlong, Kevin P and Chapman, David S},
  journal={Annual Review of Earth and Planetary Sciences},
  volume={41},
  number={1},
  pages={385--410},
  year={2013},
  publisher={Annual Reviews}
}

@article{morgan1985crustal,
  title={Crustal radiogenic heat production and the selective survival of ancient continental crust},
  author={Morgan, Paul},
  journal={Journal of Geophysical Research: Solid Earth},
  volume={90},
  number={S02},
  pages={C561--C570},
  year={1985},
  publisher={Wiley Online Library}
}

@article{lucazeau1984interpretation,
  title={Interpretation of heat flow data in the French Massif Central},
  author={Lucazeau, Francis and Vasseur, Guy and Bayer, Roger},
  journal={Tectonophysics},
  volume={103},
  number={1-4},
  pages={99--119},
  year={1984},
  publisher={Elsevier}
}

@article{lucazeau1989heat,
  title={Heat flow density data from {F}rance and surrounding margins},
  author={Lucazeau, Francis and Vasseur, Guy},
  journal={Tectonophysics},
  volume={164},
  number={2-4},
  pages={251--258},
  year={1989},
  publisher={Elsevier}
}

@article{vasseur1980trend,
  title={Trend of heat flow in {F}rance: relation with deep structures},
  author={Vasseur, Guy and Nouri, Yamina and Fluxchaf, Groupe},
  journal={Tectonophysics},
  volume={65},
  number={3-4},
  pages={209--223},
  year={1980},
  publisher={Elsevier}
}

@article{zeyen1997refraction,
  title={Refraction-seismic investigations of the northern Massif Central ({F}rance)},
  author={Zeyen, Hermann and Novak, Olaf and Landes, Michael and Prodehl, Claus and Driad, Lynda and Hirn, Alfred},
  journal={Tectonophysics},
  volume={275},
  number={1-3},
  pages={99--117},
  year={1997},
  publisher={Elsevier}
}

@incollection{schmincke2007quaternary,
  title={The {Q}uaternary volcanic fields of the {E}ast and {W}est {E}ifel ({G}ermany)},
  author={Schmincke, Hans-Ulrich},
  booktitle={Mantle plumes: A multidisciplinary approach},
  pages={241--322},
  year={2007},
  publisher={Springer}
}

@inproceedings{schmincke1983quaternary,
  title={The {Q}uaternary {E}ifel volcanic fields},
  author={Schmincke, H-U and Lorenz, V and Seck, HA},
  booktitle={Plateau Uplift: The Rhenish Shield—A Case History},
  pages={139--151},
  year={1983},
  organization={Springer}
}

@article{michon2000crustal,
  title={Crustal structures of the {R}hinegraben and the {M}assif {C}entral grabens: An experimental approach},
  author={Michon, Laurent and Merle, Olivier},
  journal={Tectonics},
  volume={19},
  number={5},
  pages={896--904},
  year={2000},
  publisher={Wiley Online Library}
}

@article{dezes2004evolution,
  title={Evolution of the {E}uropean {C}enozoic {R}ift {S}ystem: interaction of the {A}lpine and {P}yrenean orogens with their foreland lithosphere},
  author={D{\`e}zes, Pierre and Schmid, Stefan M and Ziegler, Peter A},
  journal={Tectonophysics},
  volume={389},
  number={1-2},
  pages={1--33},
  year={2004},
  publisher={Elsevier}
}

@article{michon2001evolution,
  title={The evolution of the {M}assif {C}entral {R}ift; spatio-temporal distribution of the volcanism},
  author={Michon, Laurent and Merle, Olivier},
  journal={Bulletin de la Soci{\'e}t{\'e} g{\'e}ologique de France},
  volume={172},
  number={2},
  pages={201--211},
  year={2001},
  publisher={Societe Geologique de France}
}

@article{deves2014strain,
  title={Strain heating in process zones; implications for metamorphism and partial melting in the lithosphere},
  author={Dev{\`e}s, Maud H and Tait, Stephen R and King, Geoffrey CP and Grandin, Rapha{\"e}l},
  journal={Earth and Planetary Science Letters},
  volume={394},
  pages={216--228},
  year={2014},
  publisher={Elsevier}
}

@article{lefort1981kinematic,
  title={A kinematic model for the collision and complete suturing between Gondwanaland and Laurussia in the Carboniferous},
  author={Lefort, Jean-Pierre and Van der Voo, Rob},
  journal={The Journal of Geology},
  volume={89},
  number={5},
  pages={537--550},
  year={1981},
  publisher={University of Chicago Press}
}

@article{sobolev1997upper,
  title={Upper mantle temperatures and lithosphere-asthenosphere system beneath the {F}rench {M}assif {C}entral constrained by seismic, gravity, petrologic and thermal observations},
  author={Sobolev, Stephan V and Zeyen, Hermann and Granet, Michel and Achauer, Ulrich and Bauer, Christian and Werling, Friederike and Altherr, Rainer and Fuchs, Karl},
  journal={Tectonophysics},
  volume={275},
  number={1-3},
  pages={143--164},
  year={1997},
  publisher={Elsevier}
}

@article{mazzotti2020processes,
  title={Processes and deformation rates generating seismicity in metropolitan {F}rance and conterminous {W}estern {E}urope},
  author={Mazzotti, St{\'e}phane and Jomard, Herv{\'e} and Masson, Fr{\'e}d{\'e}ric},
  journal={Bulletin de la Soci{\'e}t{\'e} G{\'e}ologique de France},
  volume={191},
  number={1},
  year={2020},
  publisher={GeoScienceWorld}
}

@article{faure2009review,
  title={A review of the pre-{P}ermian geology of the {V}ariscan {F}rench {M}assif {C}entral},
  author={Faure, Michel and Lardeaux, Jean-Marc and Ledru, Patrick},
  journal={Comptes rendus g{\'e}oscience},
  volume={341},
  number={2-3},
  pages={202--213},
  year={2009},
  publisher={Elsevier}
}

@book{de2011computational,
  title={Computational methods for plasticity: theory and applications},
  author={de Souza Neto, Eduardo A and Peric, Djordje and Owen, David RJ},
  year={2011},
  publisher={John Wiley \& Sons}
}

@book{lemaitre1994mechanics,
  title={Mechanics of solid materials},
  author={Lemaitre, Jean and Chaboche, Jean-Louis},
  year={1994},
  publisher={Cambridge university press}
}

@book{de2012nonlinear,
  title={Nonlinear finite element analysis of solids and structures},
  author={de Borst, Ren{\'e} and Crisfield, Mike A and Remmers, Joris JC and Verhoosel, Clemens V},
  year={2012},
  publisher={John Wiley \& Sons}
}

@misc{bower2009applied,
  title={Applied Mechanics of Solids},
  author={Bower, AF},
  year={2009},
  publisher={CRC Press}
}

@article{werling1997thermal,
  title={Thermal evolution of the lithosphere beneath the French Massif Central as deduced from geothermobarometry on mantle xenoliths},
  author={Werling, Friederike and Altherr, Rainer},
  journal={Tectonophysics},
  volume={275},
  number={1-3},
  pages={119--141},
  year={1997},
  publisher={Elsevier}
}

@article{plomerova2000temporary,
  title={Temporary Array Data for Studying Seismic Anisotropy of {V}ariscan {M}assifs--The {A}rmorican {M}assif, {F}rench {M}assif {C}entral and {B}ohemian {M}assif},
  author={Plomerov{\'a}, Jaroslava and Granet, Michel and Judenherc, Sebasten and Achauer, Ulrich and Babu{\v{s}}ka, Vladislav and Jedli{\v{c}}ka, Petr and Kouba, Daniel and Vecsey, Lud{\v{e}}k},
  journal={Studia Geophysica et Geodaetica},
  volume={44},
  number={2},
  pages={195--209},
  year={2000},
  publisher={Springer}
}

@article{duretz2014physics,
  title={Physics-controlled thickness of shear zones caused by viscous heating: Implications for crustal shear localization},
  author={Duretz, T and Schmalholz, SM and Podladchikov, YY and Yuen, DA},
  journal={Geophysical Research Letters},
  volume={41},
  number={14},
  pages={4904--4911},
  year={2014},
  publisher={Wiley Online Library}
}

@article{bessat2020stress,
  title={Stress and deformation mechanisms at a subduction zone: insights from 2-D thermomechanical numerical modelling},
  author={Bessat, Annelore and Duretz, Thibault and Het{\'e}nyi, Gy{\"o}rgy and Pilet, S{\'e}bastien and Schmalholz, Stefan M},
  journal={Geophysical Journal International},
  volume={221},
  number={3},
  pages={1605--1625},
  year={2020},
  publisher={Oxford University Press}
}

@article{afonso2005thermal,
  title={Thermal expansivity and elastic properties of the lithospheric mantle: results from mineral physics of composites},
  author={Afonso, Juan Carlos and Ranalli, Giorgio and Fern{\`a}ndez, Manel},
  journal={Physics of the Earth and Planetary Interiors},
  volume={149},
  number={3-4},
  pages={279--306},
  year={2005},
  publisher={Elsevier}
}

@article{gerya2004thermomechanical,
  title={Thermomechanical modelling of slab detachment},
  author={Gerya, Taras V and Yuen, David A and Maresch, Walter V},
  journal={Earth and Planetary Science Letters},
  volume={226},
  number={1-2},
  pages={101--116},
  year={2004},
  publisher={Elsevier}
}

@article{barth2007crustal,
  title={Crustal and upper mantle structure of the {F}rench {M}assif {C}entral plume},
  author={Barth, Andreas and Jordan, Michael and Ritter, Joachim RR},
  journal={Mantle Plumes: A Multidisciplinary Approach},
  pages={159--184},
  year={2007},
  publisher={Springer}
}

@article{thielmann2012shear,
  title={Shear heating induced lithospheric-scale localization: Does it result in subduction?},
  author={Thielmann, Marcel and Kaus, Boris JP},
  journal={Earth and Planetary Science Letters},
  volume={359},
  pages={1--13},
  year={2012},
  publisher={Elsevier}
}

@article{budweg2006eifel,
  title={The {E}ifel Plume—imaged with converted seismic waves},
  author={Budweg, Martin and Bock, G{\"u}nter and Weber, Michael},
  journal={Geophysical Journal International},
  volume={166},
  number={2},
  pages={579--589},
  year={2006},
  publisher={Blackwell Publishing Ltd Oxford, UK}
}

@article{seiberlich2013topography,
  title={Topography of the lithosphere--asthenosphere boundary below the {U}pper {R}hine {G}raben {R}ift and the volcanic {E}ifel region, {C}entral {E}urope},
  author={Seiberlich, CKA and Ritter, JRR and Wawerzinek, B},
  journal={Tectonophysics},
  volume={603},
  pages={222--236},
  year={2013},
  publisher={Elsevier}
}

@article{ritter2001mantle,
  title={A mantle plume below the {E}ifel volcanic fields, {G}ermany},
  author={Ritter, Joachim RR and Jordan, Michael and Christensen, Ulrich R and Achauer, Ulrich},
  journal={Earth and Planetary Science Letters},
  volume={186},
  number={1},
  pages={7--14},
  year={2001},
  publisher={Elsevier}
}

@article{ritter2000teleseismic,
  title={The teleseismic tomography experiment in the {E}ifel region, {C}entral {E}urope: design and first results},
  author={Ritter, Joachim RR and Achauer, Ulrich and Christensen, Ulrich R and Eifel Plume Team},
  journal={Seismological Research Letters},
  volume={71},
  number={4},
  pages={437--443},
  year={2000},
  publisher={Seismological Society of America}
}

@article{guillou2007deciphering,
  title={Deciphering plume--lithosphere interactions beneath {E}urope from topographic signatures},
  author={Guillou-Frottier, Laurent and Burov, Evgenii and Nehlig, Pierre and Wyns, Robert},
  journal={Global and Planetary Change},
  volume={58},
  number={1-4},
  pages={119--140},
  year={2007},
  publisher={Elsevier}
}

@article{keyser20023d,
  title={3{D} shear-wave velocity structure of the {E}ifel plume, {G}ermany},
  author={Keyser, Matthias and Ritter, Joachim RR and Jordan, Michael},
  journal={Earth and Planetary Science Letters},
  volume={203},
  number={1},
  pages={59--82},
  year={2002},
  publisher={Elsevier}
}

@article{zhang2025upper,
  title={The upper crustal structure of the Eifel volcanic region (southwest Germany) from local earthquake tomography using large-N seismic network data},
  author={Zhang, Hao and Dahm, Torsten and Haberland, Christian and Isken, Marius Paul and Laumann, Patrick and B{\"u}y{\"u}kakp{\i}nar, P{\i}nar},
  journal={Journal of Geophysical Research: Solid Earth},
  volume={130},
  number={11},
  pages={e2025JB031338},
  year={2025},
  publisher={Wiley Online Library}
}

@article{goes1999lower,
  title={A lower mantle source for {C}entral {E}uropean volcanism},
  author={Goes, Saskia and Spakman, Wim and Bijwaard, Harmen},
  journal={Science},
  volume={286},
  number={5446},
  pages={1928--1931},
  year={1999},
  publisher={American Association for the Advancement of Science}
}

@article{hetenyi2009anomalously,
  title={Anomalously deep mantle transition zone below {C}entral {E}urope: evidence of lithospheric instability},
  author={Het{\'e}nyi, Gy{\"o}rgy and Stuart, Graham W and Houseman, Gregory A and Horv{\'a}th, Frank and Heged{\H{u}}s, Endre and Br{\"u}ckl, Ewald},
  journal={Geophysical Research Letters},
  volume={36},
  number={21},
  year={2009},
  publisher={Wiley Online Library}
}

@article{chesworth1975mantle,
  title={Mantle plumes, plate tectonics, and the Cenozoic volcanism of the {M}assif {C}entral},
  author={Chesworth, Ward},
  journal={The Journal of Geology},
  volume={83},
  number={5},
  pages={579--588},
  year={1975},
  publisher={University of Chicago Press}
}

@article{demets1994effect,
  title={Effect of recent revisions to the geomagnetic reversal time scale on estimates of current plate motions},
  author={DeMets, Charles and Gordon, Richard G and Argus, Donald F and Stein, Seth},
  journal={Geophysical research letters},
  volume={21},
  number={20},
  pages={2191--2194},
  year={1994},
  publisher={Wiley Online Library}
}

@article{reilinger2006gps,
  title={GPS constraints on continental deformation in the Africa-Arabia-Eurasia continental collision zone and implications for the dynamics of plate interactions},
  author={Reilinger, Robert and McClusky, Simon and Vernant, Philippe and Lawrence, Shawn and Ergintav, Semih and Cakmak, Rahsan and Ozener, Haluk and Kadirov, Fakhraddin and Guliev, Ibrahim and Stepanyan, Ruben and others},
  journal={Journal of Geophysical Research: Solid Earth},
  volume={111},
  number={B5},
  year={2006},
  publisher={Wiley Online Library}
}

@article{ziegler2006crustal,
  title={Crustal evolution of Western and Central Europe},
  author={Peter A. Ziegler and P. D{\`e}zes},
  journal={Geological Society, London, Memoirs},
  year={2006},
  volume={32},
  pages={43 - 56},
  url={https://api.semanticscholar.org/CorpusID:73602279}
}

@article{fichtner2015crust,
  title={Crust and upper mantle of the western Mediterranean--Constraints from full-waveform inversion},
  author={Fichtner, Andreas and Villase{\~n}or, Antonio},
  journal={Earth and Planetary Science Letters},
  volume={428},
  pages={52--62},
  year={2015},
  publisher={Elsevier}
}

@book{shepherd2014energy,
  title={Energy studies},
  author={Shepherd, William and Shepherd, David William},
  year={2014},
  publisher={World Scientific Publishing Company}
}

@article{plomerova2016cenozoic,
  title={Cenozoic volcanism in the {B}ohemian {M}assif in the context of {P}-and {S}-velocity high-resolution teleseismic tomography of the upper mantle},
  author={Plomerov{\'a}, Jaroslava and Munzarov{\'a}, Helena and Vecsey, Lud{\v{e}}k and Kissling, Eduard and Achauer, Ulrich and Babu{\v{s}}ka, Vladislav},
  journal={Geochemistry, Geophysics, Geosystems},
  volume={17},
  number={8},
  pages={3326--3349},
  year={2016},
  publisher={Wiley Online Library}
}

@article{heidbach2016world,
  title={World stress map database release 2016},
  author={Heidbach, Oliver and Rajabi, Mojtaba and Reiter, Karsten and Ziegler, Moritz and Wsm Team and others},
  journal={GFZ Data Services},
  volume={10},
  number={1},
  year={2016}
}

@article{heidbach2018world,
  title={The World Stress Map database release 2016: Crustal stress pattern across scales},
  author={Heidbach, Oliver and Rajabi, Mojtaba and Cui, Xiaofeng and Fuchs, Karl and M{\"u}ller, Birgit and Reinecker, John and Reiter, Karsten and Tingay, Mark and Wenzel, Friedemann and Xie, Furen and others},
  journal={Tectonophysics},
  volume={744},
  pages={484--498},
  year={2018},
  publisher={Elsevier}
}

@article{tesauro2006analysis,
  title={Analysis of {C}entral {W}estern {E}urope deformation using {GPS} and seismic data},
  author={Tesauro, Magdala and Hollenstein, Christine and Egli, Ramon and Geiger, Alain and Kahle, Hans-Gert},
  journal={Journal of Geodynamics},
  volume={42},
  number={4-5},
  pages={194--209},
  year={2006},
  publisher={Elsevier}
}

@incollection{currie2015geodynamic,
  title={Geodynamic models of Cordilleran orogens: Gravitational instability of magmatic arc roots},
  author={Currie, Claire A and Ducea, Mihai N and DeCelles, Peter G and Beaumont, Christopher},
  booktitle={Geodynamics of a {C}ordilleran {O}rogenic {S}ystem: {T}he {C}entral {A}ndes of {A}rgentina and {N}orthern {C}hile},
editor={DeCelles, Peter G and Ducea, Mihai N and Carrapa, Barbara and Kapp, Paul A},
  volume  = {212},
  pages= {1--22},
  year={2015},
  publisher={Memoir of the Geological Society of America},
doi= {10.1130/2015.1212(01)}
}

@article{schmidt2024origin,
  title={The origin of carbonatites—combining the rock record with available experimental constraints},
  author={Schmidt, Max W and Giuliani, Andrea and Poli, Stefano},
  journal={Journal of Petrology},
  volume={65},
  number={10},
  pages={egae105},
  year={2024},
  publisher={Oxford University Press}
}

@article{pilet2008metasomatized,
  title={Metasomatized lithosphere and the origin of alkaline lavas},
  author={Pilet, S{\'e}bastien and Baker, Michael B and Stolper, Edward M},
  journal={Science},
  volume={320},
  number={5878},
  pages={916--919},
  year={2008},
  publisher={American Association for the Advancement of Science}
}

@article{regenauer1998dilatant,
  title={Dilatant plasticity applied to Alpine collision: Ductile void growth in the intraplate area beneath the Eifel volcanic field},
  author={Regenauer-Lieb, Klaus},
  journal={Journal of Geodynamics},
  volume={27},
  number={1},
  pages={1--21},
  year={1998},
  publisher={Elsevier}
}

@article{bodri1985correlation,
  title={On the correlation between heat flow and crustal thickness},
  author={Bodri, L and Bodri, B},
  journal={Tectonophysics},
  volume={120},
  number={1-2},
  pages={69--81},
  year={1985},
  publisher={Elsevier}
}

@article{muller1992regional,
  title={Regional patterns of tectonic stress in {E}urope},
  author={M{\"u}ller, Birgit and Zoback, Mary Lou and Fuchs, Karl and Mastin, Larry and Gregersen, Soren and Pavoni, Nazario and Stephansson, Ove and Ljunggren, Christer},
  journal={Journal of Geophysical Research: Solid Earth},
  volume={97},
  number={B8},
  pages={11783--11803},
  year={1992},
  publisher={Wiley Online Library}
}

@Manual{QGIS_software,
        title = {QGIS Geographic Information System},
        author = {{QGIS Development Team}},
        organization = {QGIS Association},
        year = {2025},
        url = {https://www.qgis.org}
      }

@article{grad2009moho,
  title={The {M}oho depth map of the {E}uropean Plate},
  author={Grad, Marek and Tiira, Timo and ESC Working Group},
  journal={Geophysical Journal International},
  volume={176},
  number={1},
  pages={279--292},
  year={2009},
  publisher={Blackwell Publishing Ltd Oxford, UK}
}

@article{aubert2025maciv,
  title={The {MACIV} multiscale seismic experiments in the French Massif Central (2023-2027): deployment, data quality and availability},
  author={Aubert, Coralie and Scheiblin, Guilhem and Paul, Anne and Pauchet, H{\'e}l{\`e}ne and Mordret, Aur{\'e}lien and Baudot, Vincent and Chevrot, S{\'e}bastien and Cluzel, Nicolas and Douste-Bacqu{\'e}, Isabelle and Grimaud, Franck and Jung, Axel and Mercier, St\`{}phane and Pawlowski, Piel and Roussel, Sandrine and Souriot, Thierry and Shapiro, Nikola\"i and Sylvander, Matthieu and Vial, Benjamin and Wolyniec, David},
  journal={Annals of Geophysics},
 volume={68},
  year={2025}
}

@article{dahm2020seismological,
  title={Seismological and geophysical signatures of the deep crustal magma systems of the {C}enozoic volcanic fields beneath the {E}ifel, {G}ermany},
  author={Dahm, Torsten and Stiller, Manfred and Mechie, James and Heimann, Sebastian and Hensch, Martin and Woith, Heiko and Schmidt, Bernd and Gabriel, Gerald and Weber, Michael},
  journal={Geochemistry, Geophysics, Geosystems},
  volume={21},
  number={9},
  pages={e2020GC009062},
  year={2020},
  publisher={Wiley Online Library}
}

@article{walker2005shear,
  title={Shear-wave splitting around the Eifel hotspot: evidence for a mantle upwelling},
  author={Walker, KT and Bokelmann, GHR and Klemperer, SL and Bock, G},
  journal={Geophysical Journal International},
  volume={163},
  number={3},
  pages={962--980},
  year={2005},
  publisher={Blackwell Publishing Ltd Oxford, UK}
}

@article{grunthal1992recent,
  title={The recent crustal stress field in {C}entral {E}urope: trajectories and finite element modeling},
  author={Gr{\"u}nthal, Gottfried and Stromeyer, Dietrich},
  journal={Journal of Geophysical Research: Solid Earth},
  volume={97},
  number={B8},
  pages={11805--11820},
  year={1992},
  publisher={Wiley Online Library}
}

@article{wilson1999tertiary,
  title={Tertiary-Quaternary magmatism within the {M}editerranean and surrounding regions},
  author={Wilson, Marjorie and Bianchini, Gianluca},
  journal={Geological Society, London, Special Publications},
  volume={156},
  number={1},
  pages={141--168},
  year={1999},
  publisher={The Geological Society of London}
}

@article{kreemer2020geodetic,
  title={Geodetic evidence for a buoyant mantle plume beneath the {E}ifel volcanic area, {NW} {E}urope},
  author={Kreemer, Corn{\'e} and Blewitt, Geoffrey and Davis, Paul M},
  journal={Geophysical Journal International},
  volume={222},
  number={2},
  pages={1316--1332},
  year={2020},
  publisher={Oxford University Press}
}

@article{hensch2019deep,
  title={Deep low-frequency earthquakes reveal ongoing magmatic recharge beneath {L}aacher {S}ee Volcano ({E}ifel, {G}ermany)},
  author={Hensch, Martin and Dahm, Torsten and Ritter, Joachim and Heimann, Sebastian and Schmidt, Bernd and Stange, Stefan and Lehmann, Klaus},
  journal={Geophysical Journal International},
  volume={216},
  number={3},
  pages={2025--2036},
  year={2019},
  publisher={Oxford University Press}
}

@article{silverii2023lithospheric,
  title={Lithospheric sill intrusions and present-day ground deformation at {R}henish {M}assif, {C}entral {E}urope},
  author={Silverii, Francesca and Mantiloni, Lorenzo and Rivalta, Eleonora and Dahm, Torsten},
  journal={Geophysical Research Letters},
  volume={50},
  number={23},
  pages={e2023GL105824},
  year={2023},
  publisher={Wiley Online Library}
}

@article{eickhoff2024seismic,
  title={Seismic reflection imaging of fluid-filled sills in the west {E}ifel volcanic field, {G}ermany},
  author={Eickhoff, Dario and Ritter, Joachim RR and Hlou{\v{s}}ek, Felix and Buske, Stefan},
  journal={Geophysical Research Letters},
  volume={51},
  number={24},
  pages={e2024GL111425},
  year={2024},
  publisher={Wiley Online Library}
}

@article{eickhoff2025seismic,
  title={Seismic reprocessing of {BELCORP-DEKORP 1A} reveals deep fault reflections of the {P}aleozoic {E}ifel {F}old and {T}hrust {B}elt in {G}ermany},
  author={Eickhoff, D and Back, S and Reicherter, K and Ritter, JRR},
  journal={Tectonophysics},
  volume={903},
  pages={230702},
  year={2025},
  publisher={Elsevier}
}

@article{chevrot2014high,
  title={High-resolution imaging of the {P}yrenees and {M}assif {C}entral from the data of the {PYROPE} and {IBERARRAY} portable array deployments},
  author={Chevrot, S{\'e}bastien and Villase{\~n}or, Antonio and Sylvander, Matthieu and Benahmed, S{\'e}bastien and Beucler, Eric and Cougoulat, Glenn and Delmas, Philippe and De Saint Blanquat, Michel and Diaz, Jordi and Gallart, Josep and others},
  journal={Journal of Geophysical Research: Solid Earth},
  volume={119},
  number={8},
  pages={6399--6420},
  year={2014},
  publisher={Wiley Online Library}
}
\end{document}